\documentclass[11pt]{article}
\usepackage{amssymb}
\usepackage{amsmath}
\usepackage{tcolorbox}
\usepackage{graphics}
\usepackage[final]{pdfpages}
\usepackage{slashed}
\usepackage{amssymb}
\usepackage{epstopdf}
\catcode`@11
\def\seceqaa{\@addtoreset{equation}{section}
\def\theequation{A\arabic{equation}}}
\def\seceqbb{\@addtoreset{equation}{section}
\def\theequation{B\arabic{equation}}}
\def\seceqcc{\@addtoreset{equation}{section}
\def\theequation{C\arabic{equation}}}
\def\seceqdd{\@addtoreset{equation}{section}
\def\theequation{D\arabic{equation}}}
\def\seceqee{\@addtoreset{equation}{section}
\def\theequation{E\arabic{equation}}}
\def\seceqff{\@addtoreset{equation}{section}
\def\theequation{F\arabic{equation}}}
\def\seceqgg{\@addtoreset{equation}{section}
\def\theequation{G\arabic{equation}}}
\def\seceqhh{\@addtoreset{equation}{section}
\def\theequation{H\arabic{equation}}}
\catcode`@11

\date{\today}

\begin{document}

\begin{titlepage}
\begin{center}
{\large \bf Delocalized SYZ Mirrors and Confronting Top-Down  $SU(3)$-Structure Holographic Meson Masses at  Finite $g$ and $N_c$ with P(article) D(ata) G(roup) Values}\\
Vikas Yadav$^{(a)}$\footnote{email: viitr.dph2015@iitr.ac.in}, Aalok Misra$^{(a),(b)}$\footnote{e-mail: aalokfph@iitr.ac.in} and Karunava Sil$^{(a)}$\footnote{e-mail: krusldph@iitr.ac.in}\\
(a) Department of Physics, Indian Institute of Technology,
Roorkee - 247 667, Uttarakhand, India\\
(b) Physics Department, McGill University, 3600 University St, Montr\'eal, QC H3A 2T8, Canada

\date{\today}
\end{center}
\thispagestyle{empty}
\begin{abstract}
Meson spectroscopy at {\it finite} gauge coupling - whereat any perturbative QCD computation would break down -  and {\it finite number of colors}, from a {\it top-down holographic string model}, has thus far been entirely missing in the literature. This paper fills in this gap. Using the delocalized type IIA SYZ mirror (with $SU(3)$ structure) of the  holographic type IIB dual of large-$N$ thermal QCD of \cite{metrics} as constructed in \cite{MQGP} at finite coupling and number of colors ($N_c =$ Number of $D5(\overline{D5}$)-branes wrapping a vanishing two-cycle in the top-down holographic construct of \cite{metrics}  = ${\cal O}(1)$ in the IR in the MQGP limit of \cite{MQGP} at the end of a Seiberg duality cascade), we obtain analytical (not just numerical) expressions for the vector and scalar meson spectra and compare our results with previous calculations of \cite{Sakai-Sugimoto-1}, \cite{Dasgupta_et_al_Mesons}, and obtain a closer match with the Particle Data Group (PDG) results \cite{PDG}. Through explicit computations, we verify that the vector and scalar meson spectra obtained by the gravity dual with a black hole for all temperatures (small and large) are nearly isospectral with the spectra obtained by a thermal gravity dual valid for only low temperatures; the isospectrality is much closer for vector mesons than scalar mesons. The black hole gravity dual (with a horizon radius smaller than the deconfinement scale) also provides the expected large-$N$ suppressed decrease in vector meson mass with increase of temperature.
\end{abstract}
\end{titlepage}


\section{Introduction}
The AdS/CFT \cite{maldacena} correspondence and its non-conformal generalizations, conjecture the equivalence between string theory on a ten dimensional space-time and gauge theory living on the boundary of space-time. A generalization of the AdS/CFT correspondence is necessary to explore more realistic theories (less supersymmetric and non conformal) such as QCD. The original AdS/CFT conjecture \cite{maldacena} proposed a duality between maximally supersymmetric ${\cal N}=4$ SU(N) SYM gauge theory and type IIB supergravity on $AdS_{5}\times S^{5}$ in the low enegry limit. Different generalized versions of the AdS/CFT have thus far been proposed to study non-supersymmetric field theories. One way of constructing gauge theories with less supersymmetry is to consider stacks of Dp branes at the singular tip of a Calabi-Yau cone. 
 In this paper we use a large N top down holographic dual of QCD\cite{metrics} to obtain the meson spectrum from type IIA perspective. %
Embedding of additional D-branes({\it flavor branes})  in the near-horizon limit gives rise to a modification of the original AdS/CFT correspondence which involves field theory degrees of freedom that transform in the {\it fundamental} representation of the gauge group. This is useful for describing  field theories like QCD, where quark fields transform in the fundamental representation. Mesons operator or a gauge invariant bilinear operator corresponds to the bound state of anti-fundamental and fundamental field.

In the past decade, (glueballs and) mesons have been studied extensively to gain new insight into the non-perturbative regime of QCD. Various holographic setups such as soft-wall model, hard wall model, modified soft wall model, etc. have been used to obtain the glueballs' and mesons' spectra and obtain interaction between them. In the following two paragraphs a brief summary of the work is given that has been done in past decades.


Most of existing literature on holographic meson spectroscopy is of the bottom-up variety based often on soft/hard wall AdS/QCD models. Here is a short summary of some of the relevant works.
Soft-wall holographic QCD model was used in \cite{Bellantuono:2014lra} and \cite{Colangelo:2009ra} to obtain spectrum and decay constants for $1^{-+}$ hybrid mesons and to study the scalar glueballs and scalar mesons at $T\neq 0$ respectively. In \cite{Bellantuono:2014lra} no states with exotic quantum numbers were observed in the heavy quark sector. Comparison of the computed mass with the experimental mass of the $1^{-+}$ candidates $\pi_{1}(1400)$, $\pi_{1}(1600)$ and $\pi_{1}(2015)$, favored $\pi_{1}(1400)$ as the lightest hybrid state.
%
In \cite{Cui:2013xva} an IR-improved soft-wall AdS/QCD model in good agreement with linear confinement and chiral symmetry breaking was constructed to study the mesonic spectrum. The model was constructed to rectify inconsistencies associated with both simple soft-wall and hard-wall models. The hard-wall model gave a good realization for the chiral symmetry breaking, but the mass spectra obtained for the excited mesons didn't match up with the experimental data well. The soft-wall model with a quadratic dilaton background field showed the Regge behaviour for excited vector mesons but chiral symmetry breaking phenomena cannot be realized consistently in the simple soft-wall AdS/QCD model.
%
A hard wall holographic model of QCD was used in \cite{Alvares:2011wb}, \cite{Domokos:2007kt} and \cite{Kim:2009bp} to analyze the mesons.
In \cite{Li:2013oda} a two-flavor quenched dynamical holographic QCD(hQCD) model was constructed in the graviton-dilaton framework by adding two light flavors.
%
In  \cite{Sakai-Sugimoto-1} the mesonic spectrum was obtained for a $D4/D8(-\overline{D8})$-brane configuration in type IIA string theory; in  \cite{Imoto:2010ef}  massive excited states in the open string spectrum were used to obtain the spectrum for higher spin mesons $J\geq 2$. NLO terms were obtained by taking into account the effect of the curved background perturbatively which led to corrections in formula  $J=\alpha_{0}+\alpha{'}M^{2}$. The results obtained for the meson spectrum were compared with the experimental data to identify $a_{2}(1320),b_{1}(1235),\pi(1300),a_{0}(1450)$ etc. first and second excited states. In \cite{Kruczenski:2004me} a holographic model was constructed with extremal $N_{c}$ $D4$-branes and $D6$-flavor branes in the probe approximation. The model gave a good approximation for Regge behaviour of glueballs but failed to explain mesonic spinning strings because the dual theory did not include quark in the fundamental representations.

To our knowledge, the only top-down holographic dual of large-$N$ thermal QCD which is IR confining, UV conformal and UV-complete (e.g. the holographic Sakai-Sugimoto model \cite{Sakai-Sugimoto-1} does not address the UV) with fundamental quarks is the one given in \cite{metrics} involving $N\ D3$-branes, $M\ D5/(\overline{D5})$ branes wrapping a vanishing two-cycle and $N_f\ D7(\overline{D7})$ flavor-branes in a warped resolved conifold at finite temperature in the brane picture (and $M\ \overline{D5}$-branes and $D7(\overline{D7})$-branes with a black-hole and fluxes in a resolved warped deformed conifold gravitational dual).
In \cite{Dasgupta_et_al_Mesons}, the authors (some also part of \cite{metrics}) obtained the vector and scalar mesonic spectra by taking a single T-dual of the holographic type IIB background of \cite{metrics}. Comparison of the  (pseudo-)vector mesons with PDG results, provided a reasonable agreement.  One of the main objectives of our work is to see if by taking a mirror of the type IIB  background of \cite{metrics} via delocalized Strominger-Yau-Zaslow's triple-T-duality prescription - a new tool in this field - at finite gauge coupling and with finite number of colors - a new limit and one which is closest to realistic strongly coupled thermal QCD - one can obtain a better agreement between the mesonic spectra so obtained and PDG results than previously obtained in \cite{Sakai-Sugimoto-1} and \cite{Dasgupta_et_al_Mesons}, and in the process gain new insights into a holographic understanding of thermal QCD.

In \cite{Sil_Yadav_Misra-EPJC-17}, we initiated top-down $G$-structure holographic large-$N$ thermal QCD phenomenology at {\it finite} gauge coupling and {\it finite} number of colors, in particular from the vantage point of the M theory uplift of the delocalized SYZ type IIA mirror of the top-down UV complete holographic dual of large-$N$ thermal QCD of \cite{metrics}, as constructed in \cite{MQGP}. We calculated up to (N)LO in $N$, masses of $0^{++}, 0^{--}, 0^{-+}, 1^{++}$ and $2^{++}$ glueballs, and found very good agreement with some of the lattice results on the same. In this paper, we continue exploring top-down $G$-structure holographic large-$N$ thermal QCD phenomenology at {\it finite} gauge coupling by evaluating the spectra of (pseudo-)vector and (pseudo-)scalar mesons, and in particular comparing their ratios for both types with P(artcile) D(ata) G(roup) results.

The rest of the paper is organized as follows. In Section {\bf 2}, via four sub-sections, we briefly review a UV-complete top-down type IIB holographic dual of large-$N$ thermal QCD (subsection {\bf 2.1}) as given in \cite{metrics} and its M theory uplift in the `MQGP limit' as worked out in \cite{MQGP} (sub-section {\bf 2.2});  sub-section {\bf 2.3} has a discussion on the construction in \cite{MQGP} of the delocalized Strominger-Yau-Zaslow (SYZ) type IIA mirror of the aforementioned type IIB background of \cite{metrics} and sub-section {\bf 2.4} has a brief review of $SU(3)$ and $G_2$ structures relevant to \cite{metrics} (type IIB) and \cite{MQGP, NPB} (type IIA and M theory). Section {\bf 3} is on the construction of the embedding of $D6$-branes via delocalized SYZ type IIA mirror of the embedding of $D7$-branes of type IIB. Via five sub-sections, Section {\bf 4} is on obtaining the (pseudo-)vector meson spectra in the framework of \cite{MQGP} at finite coupling assuming a black hole gravity dual for all temperatures, small and large. The (pseudo-)vector mesons correspond to gauge fluctuations about a background gauge field along the world volume of the $D6$ branes. Unlike \cite{Dasgupta_et_al_Mesons}, the gravity dual involves a black-hole ($r_h\neq0$) and consequently, while factorizing the gauge fluctuations along $\mathbb{R}^3\times \mathbb{S}^1$-radial direction into fluctuations along $\mathbb{R}^3\times\mathbb{S}^1$ and eigenmode fluctuations along the radial direction, there are two types of eigenmodes along the radial direction - one denoted by $\alpha^{\left\{i\right\}}_n(Z)$ which is coupled to gauge fluctuations along the space-like $\mathbb{R}^3$  and the other denoted by $\alpha^{\left\{0\right\}}_n(Z)$ which is coupled to the compact time-like $\mathbb{S}^1$ (metric along which includes the black-hole function). After obtaining the EOMs for $\alpha^{\left\{i\right\}}_n$ and $\alpha^{\left\{0\right\}}_n$, the following is the outline of what is done in subsections {\bf 4.1} - {\bf 4.5}. First, in {\bf 4.1}, assuming an IR-valued vector meson spectrum, the same is obtained by solving the EOMs near the horizon. Next, by converting the EOMs to a Schr\"{o}dinger-like EOMs, the vector meson spectra are worked out for $\alpha^{\left\{i\right\}}_n$ eigenmodes in {\bf 4.2} (in the IR limit in {\bf 4.2.1} and the UV limit in {\bf 4.2.2}) and $\alpha^{\left\{0\right\}}_n$ eigenmodes in {\bf 4.3} (in the IR limit in {\bf 4.3.1} and the UV limit in {\bf 4.3.2}). Finally, using the WKB quantization prescription, the vector meson spectrum corresponding to the $\alpha^{\left\{i\right\}}_n$ eigenmodes was worked out (in the small and large mass-limits) in {\bf 4.4}, and that corresponding to the $\alpha^{\left\{0\right\}}_n$ eigenmodes (in the small and large mass-limits) in {\bf 4.5}. In Section {\bf 5}, we obtain the scalar meson spectrum by considering fluctuations of the $D6$-branes orthogonal to their world-volume in the absence of any background gauge fields in a black hole background for all temperatures, small and large. In the same vein as vector meson spectroscopy, after obtaining the EOM for the radial eigenfunction mode, the following is an outline of what is done in section {\bf 5}, devoted to scalar meson spectroscopy.  First, in {\bf 5.1}, assuming an IR-valued scalar meson spectrum, the same is obtained by solving the EOMs near the horizon. Next, by converting the EOMs to a Schr\"{o}dinger-like EOMs, the scalar meson spectrum is worked out for in {\bf 5.2} (in the IR limit in {\bf 5.2.1} and the UV limit in {\bf 5.2.2}). Finally, using the WKB quantization prescription, the scalar meson spectrum  was worked out (in the small and large mass-limits) in {\bf 5.3}. In Section {\bf 6}, we obtain the (pseudo-)vector meson spectrum in {\bf 6.1} (and the three sub-sub-sections therein) and (pseudo-)scalar meson spectrum in {\bf 6.2} (and the three sub-sub-sections therein) using a thermal background, and hence verify that the mesonic spectra of Sections {\bf 4} and {\bf 5} are nearly isospectral with {\bf 6}. Section {\bf 7} has a discussion on the new insights and results obtained in this work and some future directions. There are three supplementary appendices.

\section{Background: A Top-Down Type IIB Holographic Large-$N$ Thermal QCD and its M-Theory Uplift in the `MQGP' Limit}

Via four sub-sections, in this section, we will:

\begin{itemize}
\item
provide a short review of the type IIB background of \cite{metrics},  a UV complete holographic dual of large-$N$ thermal QCD, in subsection {\bf 2.1},

\item
discuss the 'MQGP' limit of \cite{MQGP} and the motivation for considering this limit in subsection {\bf 2.2},

\item
briefly review issues as discussed in \cite{MQGP}, \cite{transport-coefficients}, \cite{NPB} and \cite{EPJC-2}, pertaining to construction of delocalized S(trominger) Y(au) Z(aslow) mirror and approximate supersymmetry, in subsection {\bf 2.3},

\item
briefly review the new results of \cite{NPB} and \cite{EPJC-2} pertaining to construction of explicit $SU(3)$ and $G_2$ structures respectively of type IIB/IIA, and M-theory uplift,

\end{itemize}

\subsection{Type IIB Dual of Large-$N$ Thermal QCD}

In this subsection, we will discuss a UV complete holographic dual of large-$N$ thermal QCD as given in  Dasgupta-Mia et al \cite{metrics}.
In order to include fundamental quarks at non-zero temperature in the context of type IIB string theory, \cite{metrics} considered at finite temperature,  $N$  $D3$-branes, $M\ D5$-branes wrapping a vanishing two-cycle and $M\ \overline{D5}$-branes  distributed along a resolved two-cycle and placed at the outer boundary  of the IR-UV interpolating region/inner boundary of the UV region. The $D5/\overline{D5}$ separation is given by ${\cal R}_{D5/\overline{D5}}$. The radial space, in \cite{metrics} is divided into the IR, the IR-UV interpolating region and the UV. $N_f\ D7$-branes, via Ouyang embedding,  are holomorphically embedded in the UV, the IR-UV interpolating region and dipping into the (confining) IR (up to a certain minimum value of $r$ corresponding to the lightest quark)  and $N_f\ \overline{D7}$-branes present in the UV and the UV-IR interpolating (not the confining IR). This is to ensure turning off of three-form fluxes. The resultant ten-dimensional geometry is given by a  resolved warped deformed conifold. In the gravity dual $D3$-branes and the $D5$-branes are replaced by fluxes in the IR. The finite temperature resolves \footnote{The non-zero resolution parameter `$a$' is also there to introduce a separation ${\cal R}_{D5/\overline{D5}}$ between the $D5$ and $\overline{D5}$ branes, which as in \cite{EPJC-2} we assume is $\sqrt{3}a$ (as $r>\sqrt{3}a$ lies in the large-$r$ region in a resolved conifold \cite{Candelas+Ossa-Conifolds}).} and IR confinement deforms the conifold. Back-reactions are included in the warp factor and fluxes.

One has $SU(N+M)\times SU(N+M)$ color gauge group and $SU(N_f)\times SU(N_f)$ flavor gauge group, in the UV. It is expected that there will be a partial Higgsing of $SU(N+M)\times SU(N+M)$ to $SU(N+M)\times SU(N)$ at $r={\cal R}_{D5/\overline{D5}}$  \cite{K. Dasgupta et al [2012]}. The two gauge couplings, $g_{SU(N+M)}$ and $g_{SU(N)}$ flow  logarithmically  and oppositely in the IR: $4\pi^2\left(\frac{1}{g_{SU(N+M)}^2} + \frac{1}{g_{SU(N)}^2}\right)e^\phi \sim \pi;\
 4\pi^2\left(\frac{1}{g_{SU(N+M)}^2} - \frac{1}{g_{SU(N)}^2}\right)e^\phi \sim \frac{1}{2\pi\alpha^\prime}\int_{S^2}B_2$
.Had it not been for $\int_{S^2}B_2$, in the UV, one could have set $g_{SU(M+N)}^2=g_{SU(N)}^2=g_{YM}^2\sim g_s\equiv$ constant (implying conformality) which is the reason for inclusion of $M$ $\overline{D5}$-branes at the common boundary of the UV-IR interpolating and the UV regions, to annul this contribution. In fact, the running also receives a contribution from the $N_f$ flavor $D7$-branes which needs to be annulled via $N_f\ \overline{D7}$-branes. Under an NVSZ RG flow, the gauge coupling $g_{SU(N+M)}$ - having a larger rank - flows towards strong coupling and the $SU(N)$ gauge coupling flows towards weak coupling. Upon application of Seiberg duality, $SU(N+M)_{\rm strong}\stackrel{\rm Seiberg\ Dual}{\longrightarrow}SU(N-(M - N_f))_{\rm weak}$ in the IR;  assuming after duality cascade, $N$ decreases to 0 and there is a finite $M$, {one will be left with $SU(M)$ gauge theory with $N_f$ flavors that confines in the IR - the finite temperature version of the same is what was looked at by \cite{metrics}}.

  So, in the IR, at the end of the duality cascade, number of colors $N_c$ is identified with $M$, which in the `MQGP limit' can be tuned to equal 3. One can identify $N_c$ with $N_{\rm eff}(r) + M_{\rm eff}(r)$, where $N_{\rm eff}(r) = \int_{\rm Base\ of\ Resolved\ Warped\ Deformed\ Conifold}F_5$ and $M_{\rm eff} = \int_{S^3}\tilde{F}_3$ (the $S^3$ being dual to $\ e_\psi\wedge\left(\sin\theta_1 d\theta_1\wedge d\phi_1 - B_1\sin\theta_2\wedge d\phi_2\right)$, wherein $B_1$ is an asymmetry factor defined in \cite{metrics}, and $e_\psi\equiv d\psi + {\rm cos}~\theta_1~d\phi_1 + {\rm cos}~\theta_2~d\phi_2$) where $\tilde{F}_3 (\equiv F_3 - \tau H_3)\propto M(r)\equiv\frac{1}{1 + e^{\alpha\left(r-{\cal R}_{D5/\overline{D5}}\right)}}, \alpha\gg1$  \cite{IR-UV-desc_Dasgupta_etal}. 
The number of colors $N_c$ varies between $M$ in the deep IR and a large value [even in the MQGP limit of (\ref{limits_Dasguptaetal-ii}) (for a large value of $N$)] in the UV.  {Hence, at very low energies, the number of colors $N_c$ can be approximated by $M$, which in the MQGP limit is taken to be finite and can hence be taken to be equal to three. } In \cite{metrics}, the effective number of $D3$-branes, $D5$-branes wrapping the vanishing two-cycle and the flavor $D7$-branes, denoted respectively by $ N_{\rm eff}(r)$, $M_{\rm eff}(r)$ and $N^{\rm eff}_f(r)$, are given as:
\begin{eqnarray}
\label{NeffMeffNfeff}
& & N_{\rm eff}(r) = N\left[ 1 + \frac{3 g_s M_{\rm eff}^2}{2\pi N}\left(\log r + \frac{3 g_s N_f^{\rm eff}}{2\pi}\left(\log r\right)^2\right)\right],\nonumber\\
& & M_{\rm eff}(r) = M + \frac{3g_s N_f M}{2\pi}\log r + \sum_{m\geq1}\sum_{n\geq1} N_f^m M^n f_{mn}(r),\nonumber\\
& & N^{\rm eff}_f(r) = N_f + \sum_{m\geq1}\sum_{n\geq0} N_f^m M^n g_{mn}(r).
\end{eqnarray}
It was argued in \cite{NPB} that  the length scale of the OKS-BH metric in the IR after Seiberg-duality cascading away almost the whole of $N_{\rm eff}$, will be given by:
\begin{eqnarray}
\label{length-IR}
& & L_{\rm OKS-BH}\sim\sqrt{M}N_f^{\frac{3}{4}}\sqrt{\left(\sum_{m\geq0}\sum_{n\geq0}N_f^mM^nf_{mn}(\Lambda)\right)}\left(\sum_{l\geq0}\sum_{p\geq0}N_f^lM^p g_{lp}(\Lambda)\right)^{\frac{1}{4}}g_s^{\frac{1}{4}}\sqrt{\alpha^\prime}\nonumber\\
& & \equiv N_f^{\frac{3}{4}}\left.\sqrt{\left(\sum_{m\geq0}\sum_{n\geq0}N_f^mM^nf_{mn}(\Lambda)\right)}\left(\sum_{l\geq0}\sum_{p\geq0}N_f^lM^p g_{lp}(\Lambda)\right)^{\frac{1}{4}} L_{\rm KS}\right|_{\Lambda:\log \Lambda{<}{\frac{2\pi}{3g_sN_f}}},
\end{eqnarray}
which implies that  {in the IR, relative to KS, there is a color-flavor enhancement of the length scale in the OKS-BH metric}. Hence,  in the IR, even for $N_c^{\rm IR}=M=3$ and $N_f=2$ (light flavors) upon inclusion of $n,m>1$  terms in
$M_{\rm eff}$ and $N_f^{\rm eff}$ in (\ref{NeffMeffNfeff}), $L_{\rm OKS-BH}\gg L_{\rm KS}(\sim L_{\rm Planck})$ in the MQGP limit involving $g_s\stackrel{\sim}{<}1$, implying that {the stringy corrections are suppressed and one can trust supergravity calculations}. Further, the global  flavor group $SU(N_f)\times SU(N_f)$, is broken in the IR to $SU(N_f)$ as the IR has only $N_f$ $D7$-branes.

Hence, the type IIB model of \cite{metrics} makes it an ideal holographic dual of thermal QCD because, it is UV conformal and IR confining with required chiral symmetry breaking in the IR. The quarks present in the theory transform in the fundamental representation, plus theory is defined for full range of temperature both low and high.\\







\noindent (d) \underline{Supergravity solution on resolved warped deformed conifold}

The metric in the gravity dual of the resolved warped deformed conifold  with $g_i$'s:
$ g_{1,2}(r,\theta_1,\theta_2)= 1-\frac{r_h^4}{r^4} + {\cal O}\left(\frac{g_sM^2}{N}\right)$ is given by :
\begin{equation}
\label{metric}
ds^2 = \frac{1}{\sqrt{h}}
\left(-g_1 dt^2+dx_1^2+dx_2^2+dx_3^2\right)+\sqrt{h}\biggl[g_2^{-1}dr^2+r^2 d{\cal M}_5^2\biggr].
\end{equation}
The compact five dimensional metric in (\ref{metric}), is given as:
\begin{eqnarray}
\label{RWDC}
& & d{\cal M}_5^2 =  h_1 (d\psi + {\rm cos}~\theta_1~d\phi_1 + {\rm cos}~\theta_2~d\phi_2)^2 +
h_2 (d\theta_1^2 + {\rm sin}^2 \theta_1 ~d\phi_1^2)  + h_4 (h_3 d\theta_2^2 + {\rm sin}^2 \theta_2 ~d\phi_2^2) \nonumber\\
& & + h_5\left[~{\rm cos}~\psi \left(d\theta_1 d\theta_2 -
{\rm sin}~\theta_1 {\rm sin}~\theta_2 d\phi_1 d\phi_2\right) + ~{\rm sin}~\psi \left({\rm sin}~\theta_1~d\theta_2 d\phi_1 +
{\rm sin}~\theta_2~d\theta_1 d\phi_2\right)\right],
\end{eqnarray}
wherein we will assume $r\gg a, h_5\sim\frac{({\rm deformation\ parameter})^2}{r^3}\ll  1$ for $r \gg({\rm deformation\ parameter})^{\frac{2}{3}}$.  The $h_i$'s appearing in internal metric up to linear order depend on $g_s, M, N_f$ are given as below:
\begin{eqnarray}
\label{h_i}
& & \hskip -0.45in h_1 = \frac{1}{9} + {\cal O}\left(\frac{g_sM^2}{N}\right),\  h_2 = \frac{1}{6} + {\cal O}\left(\frac{g_sM^2}{N}\right),\ h_4 = h_2 + \frac{a^2}{r^2},\nonumber\\
& & h_3 = 1 + {\cal O}\left(\frac{g_sM^2}{N}\right),\ h_5\neq0,\
 L=\left(4\pi g_s N\right)^{\frac{1}{4}}.
\end{eqnarray}
One sees from (\ref{RWDC}) and (\ref{h_i}) that one has a non-extremal resolved warped deformed conifold involving
an $S^2$-blowup (as $h_4 - h_2 = \frac{a^2}{r^2}$), an $S^3$-blowup (as $h_5\neq0$) and squashing of an $S^2$ (as $h_3$ is not strictly unity). The horizon (being at a finite $r=r_h$) is warped squashed $S^2\times S^3$. In the deep IR, in principle one ends up with a warped squashed $S^2(a)\times S^3(\epsilon),\ \epsilon$ being the deformation parameter. Assuming $\epsilon^{\frac{2}{3}}>a$ and given that $a={\cal O}\left(\frac{g_s M^2}{N}\right)r_h$ \cite{K. Dasgupta  et al [2012]}, in the IR and in the MQGP limit, $N_{\rm eff}(r\in{\rm IR})=\int_{{\rm warped\ squashed}\ S^2(a)\times S^3(\epsilon)}F_5(r\in{\rm IR})\ll   M = \int_{S^3(\epsilon)}F_3(r\in{\rm IR})$; we have a confining $SU(M)$ gauge theory in the IR.

 The warp factor that includes the back-reaction, in the IR is given as:
\begin{eqnarray}
\label{eq:h}
&& \hskip -0.45in h =\frac{L^4}{r^4}\Bigg[1+\frac{3g_sM_{\rm eff}^2}{2\pi N}{\rm log}r\left\{1+\frac{3g_sN^{\rm eff}_f}{2\pi}\left({\rm
log}r+\frac{1}{2}\right)+\frac{g_sN^{\rm eff}_f}{4\pi}{\rm log}\left({\rm sin}\frac{\theta_1}{2}
{\rm sin}\frac{\theta_2}{2}\right)\right\}\Biggr],
\end{eqnarray}
where, in principle, $M_{\rm eff}/N_f^{\rm eff}$ are not necessarily the same as $M/N_f$; we however will assume that up to ${\cal O}\left(\frac{g_sM^2}{N}\right)$, they are. Proper UV behavior requires \cite{K. Dasgupta et al [2012]}:
\begin{eqnarray}
\label{h-large-small-r}
& & h = \frac{L^4}{r^4}\left[1 + \sum_{i=1}\frac{{\cal H}_i\left(\phi_{1,2},\theta_{1,2},\psi\right)}{r^i}\right],\ {\rm large}\ r;
\nonumber\\
& & h = \frac{L^4}{r^4}\left[1 + \sum_{i,j; (i,j)\neq(0,0)}\frac{h_{ij}\left(\phi_{1,2},\theta_{1,2},\psi\right)\log^ir}{r^j}\right],\ {\rm small}\ r.
\end{eqnarray}


  In the IR, up to ${\cal O}(g_s N_f)$ and setting $h_5=0$, the three-forms are as given in \cite{metrics}:
\begin{eqnarray}
\label{three-form-fluxes}
& & \hskip -0.4in (a) {\widetilde F}_3  =  2M { A_1} \left(1 + \frac{3g_sN_f}{2\pi}~{\rm log}~r\right) ~e_\psi \wedge
\frac{1}{2}\left({\rm sin}~\theta_1~ d\theta_1 \wedge d\phi_1-{ B_1}~{\rm sin}~\theta_2~ d\theta_2 \wedge
d\phi_2\right)\nonumber\\
&& \hskip -0.4in -\frac{3g_s MN_f}{4\pi} { A_2}~\frac{dr}{r}\wedge e_\psi \wedge \left({\rm cot}~\frac{\theta_2}{2}~{\rm sin}~\theta_2 ~d\phi_2
- { B_2}~ {\rm cot}~\frac{\theta_1}{2}~{\rm sin}~\theta_1 ~d\phi_1\right)\nonumber \\
&& \hskip -0.3in -\frac{3g_s MN_f}{8\pi}{ A_3} ~{\rm sin}~\theta_1 ~{\rm sin}~\theta_2 \left(
{\rm cot}~\frac{\theta_2}{2}~d\theta_1 +
{ B_3}~ {\rm cot}~\frac{\theta_1}{2}~d\theta_2\right)\wedge d\phi_1 \wedge d\phi_2, \nonumber\\
& & \hskip -0.4in (b) H_3 =  {6g_s { A_4} M}\Biggl(1+\frac{9g_s N_f}{4\pi}~{\rm log}~r+\frac{g_s N_f}{2\pi}
~{\rm log}~{\rm sin}\frac{\theta_1}{2}~
{\rm sin}\frac{\theta_2}{2}\Biggr)\frac{dr}{r}\wedge \frac{1}{2}\Biggl({\rm sin}~\theta_1~ d\theta_1 \wedge d\phi_1\nonumber \\
&& \hskip -0.4in
- { B_4}~{\rm sin}~\theta_2~ d\theta_2 \wedge d\phi_2\Biggr)
+ \frac{3g^2_s M N_f}{8\pi} { A_5} \Biggl(\frac{dr}{r}\wedge e_\psi -\frac{1}{2}de_\psi \Biggr) \wedge \Biggl({\rm cot}~\frac{\theta_2}{2}~d\theta_2
-{ B_5}~{\rm cot}~\frac{\theta_1}{2} ~d\theta_1\Biggr). \nonumber\\
\end{eqnarray}
The asymmetry factors in (\ref{three-form-fluxes}) are given by: $ A_i=1 +{\cal O}\left(\frac{a^2}{r^2}\ {\rm or}\ \frac{a^2\log r}{r}\ {\rm or}\ \frac{a^2\log r}{r^2}\right) + {\cal O}\left(\frac{{\rm deformation\ parameter }^2}{r^3}\right),$ $  B_i = 1 + {\cal O}\left(\frac{a^2\log r}{r}\ {\rm or}\ \frac{a^2\log r}{r^2}\ {\rm or}\ \frac{a^2\log r}{r^3}\right)+{\cal O}\left(\frac{({\rm deformation\ parameter})^2}{r^3}\right)$.    As in the UV, $\frac{({\rm deformation\ parameter})^2}{r^3}\ll  \frac{({\rm resolution\ parameter})^2}{r^2}$, we will assume the same three-form fluxes for $h_5\neq0$.  With ${\cal R}_{D5/\overline{D5}}$ denoting the boundary common to the UV-IR interpolating region and the UV region, $\tilde{F}_{lmn}, H_{lmn}=0$ for $r\geq {\cal R}_{D5/\overline{D5}}$ is required to ensure conformality in the UV.  Near the $\theta_1=\theta_2=0$-branch, assuming: $\theta_{1,2}\rightarrow0$ as $\epsilon^{\gamma_\theta>0}$ and $r\rightarrow {\cal R}_{\rm UV}\rightarrow\infty$ as $\epsilon^{-\gamma_r <0}, \lim_{r\rightarrow\infty}\tilde{F}_{lmn}=0$ and  $\lim_{r\rightarrow\infty}H_{lmn}=0$ for all components except $H_{\theta_1\theta_2\phi_{1,2}}$; in the MQGP limit and near $\theta_{1,2}=\pi/0$-branch, $H_{\theta_1\theta_2\phi_{1,2}}=0/\left.\frac{3 g_s^2MN_f}{8\pi}\right|_{N_f=2,g_s=0.6, M=\left({\cal O}(1)g_s\right)^{-\frac{3}{2}}}\ll  1.$ So, the UV nature too is captured near $\theta_{1,2}=0$-branch in the MQGP limit. This mimics addition of $\overline{D5}$-branes in \cite{metrics} to ensure cancellation of $\tilde{F}_3$.

Further, to ensure UV conformality, it is important to ensure that the axion-dilaton modulus approaches a constant implying a vanishing beta function in the UV. This was discussed in detail in  appendix B of \cite{NPB}, wherein in particular, assuming the F-theory uplift involved, locally, an elliptically fibered $K3$, it was shown that UV conformality and the Ouyang embedding are mutually consistent.

\subsection{The `MQGP Limit'}

In \cite{MQGP}, we had considered the following two limits:
\begin{eqnarray}
\label{limits_Dasguptaetal-i}
&   & \hskip -0.17in (i) {\rm weak}(g_s){\rm coupling-large\ t'Hooft\ coupling\ limit}:\nonumber\\
& & \hskip -0.17in g_s\ll  1, g_sN_f\ll  1, \frac{g_sM^2}{N}\ll  1, g_sM\gg1, g_sN\gg1\nonumber\\
& & \hskip -0.17in {\rm effected\ by}: g_s\sim\epsilon^{d}, M\sim\left({\cal O}(1)\epsilon\right)^{-\frac{3d}{2}}, N\sim\left({\cal O}(1)\epsilon\right)^{-19d}, \epsilon\ll  1, d>0
 \end{eqnarray}
(the limit in the first line  though not its realization in the second line, considered in \cite{metrics});
\begin{eqnarray}
\label{limits_Dasguptaetal-ii}
& & \hskip -0.17in (ii) {\rm MQGP\ limit}: \frac{g_sM^2}{N}\ll  1, g_sN\gg1, {\rm finite}\
 g_s, M\ \nonumber\\
& & \hskip -0.17in {\rm effected\ by}:  g_s\sim\epsilon^d, M\sim\left({\cal O}(1)\epsilon\right)^{-\frac{3d}{2}}, N\sim\left({\cal O}(1)\epsilon\right)^{-39d}, \epsilon\lesssim 1, d>0.
\end{eqnarray}

The motivation for considering the MQGP limit which was discussed in detail in \cite{NPB} is:
\begin{enumerate}
\item
Unlike the AdS/CFT limit wherein $g_{\rm YM}\rightarrow0, N\rightarrow\infty$ such that $g_{\rm YM}^2N$ is large, for strongly coupled thermal systems like sQGP, what is relevant is $g_{\rm YM}\sim{\cal O}(1)$ and $N_c=3$. From the discussion in the previous paragraphs one sees that in the IR after the Seiberg duality cascade, effectively $N_c=M$ which in the MQGP limit of (\ref{limits_Dasguptaetal-ii})  can be tuned to 3. Further, in the same limit, the string coupling $g_s\stackrel{<}{\sim}1$. The finiteness of the string coupling necessitates addressing the same from an M theory perspective. This is the reason for coining: `MQGP limit'. In fact this is the reason why one is required to first construct a type IIA mirror, which was done in \cite{MQGP} \`{a} la delocalized Strominger-Yau-Zaslow mirror prescription, and then take its M-theory uplift.

\item
The second set of reasons for looking at the MQGP limit of (\ref{limits_Dasguptaetal-ii}) is calculational simplification in supergravityy:
\begin{itemize}
\item
In the UV-IR interpolating region and the UV,
$(M_{\rm eff}, N_{\rm eff}, N_f^{\rm eff})\stackrel{\rm MQGP}{\approx}(M, N, N_f)$
\item
Asymmetry Factors $A_i, B_j$(in three-form fluxes)$\stackrel{MQGP}{\rightarrow}1$  in the UV-IR interpolating region and the UV.

\item
Simplification of ten-dimensional warp factor and non-extremality function in MQGP limit
\end{itemize}
\end{enumerate}


\subsection{Approximate Supersymmetry, Construction of  the Delocalized SYZ IIA Mirror and Its M-Theory Uplift in the MQGP Limit}


{ To implement the quantum mirror symmetry a la SYZ \cite{syz}, one needs a special Lagrangian (sLag) $T^3$ fibered over a large base. Defining delocalized T-duality coordinates, $(\phi_1,\phi_2,\psi)\rightarrow(x,y,z)$ valued in $T^3(x,y,z)$ \cite{MQGP}:
\begin{equation}
\label{xyz defs}
x = \sqrt{h_2}h^{\frac{1}{4}}sin\langle\theta_1\rangle\langle r\rangle \phi_1,\ y = \sqrt{h_4}h^{\frac{1}{4}}sin\langle\theta_2\rangle\langle r\rangle \phi_2,\ z=\sqrt{h_1}\langle r\rangle h^{\frac{1}{4}}\psi,
\end{equation}
using the results of \cite{M.Ionel and M.Min-OO (2008)} it was shown in \cite{transport-coefficients,EPJC-2} that the following conditions are satisfied:
\begin{eqnarray}
\label{sLag-conditions}
& & \left.i^* J\right|_{\rm RC/DC} \approx 0,\nonumber\\
& & \left.\Im m\left( i^*\Omega\right)\right|_{\rm RC/DC} \approx 0,\nonumber\\
& & \left.\Re e\left(i^*\Omega\right)\right|_{\rm RC/DC}\sim{\rm volume \ form}\left(T^3(x,y,z)\right),
\end{eqnarray}
for the $T^2$-invariant sLag of \cite{M.Ionel and M.Min-OO (2008)} for a deformed conifold $\sum_{i=1}^4z_i^2 = 1$:
\begin{eqnarray}
\label{T2sLag-1}
& & K'(r^2) \Im m(z_1{\bar z}_2) = c_1,  K'(r^2) \Im m(z_3{\bar z}_4) = c_2,\Im m (z_1^2 + z_2^2) = c_3,
\end{eqnarray}
and the $T^2$-invariant sLag of \cite{M.Ionel and M.Min-OO (2008)} of a resolved conifold:
\begin{eqnarray}
\label{sLagRC-I}
& & \frac{K^\prime}{2}\left(|x|^2-|y|^2\right) + 4 a^2\frac{|\lambda_2|^2}{|\lambda_1|^2+|\lambda_2|^2} = c_1,\nonumber\\
& & \frac{K^\prime}{2}\left(|v|^2-|u|^2\right) + 4 a^2\frac{|\lambda_2|^2}{|\lambda_1|^2+|\lambda_2|^2} = c_2,\nonumber\\
& & \Im m\left(xy\right) = c_3,
\end{eqnarray}
wherein one uses the following complex structure for a resolved conifold \cite{Knauf+Gwyn[2007]}:
\begin{eqnarray}
\label{resolvedconifold-compl-struc}
  x & =&  \left ( 9 a^2 r^4 + r ^6 \right ) ^{1/4} e^{i/2(\psi-\phi_1-\phi_2)}\,\sin\frac{\theta_1}{2}\,\sin\frac{\theta_2}{2}  \nonumber\\
  y & =&  \left ( 9 a^2 r^4 + r ^6 \right ) ^{1/4} e^{i/2(\psi+\phi_1+\phi_2)}\,\cos\frac{\theta_1}{2}\,\cos\frac{\theta_2}{2}  \nonumber\\
  u & =&  \left ( 9 a^2 r^4 + r ^6 \right ) ^{1/4} e^{i/2(\psi+\phi_1-\phi_2)}\,\cos\frac{\theta_1}{2}\,\sin\frac{\theta_2}{2}  \nonumber\\
  v & =&  \left ( 9 a^2 r^4 + r ^6 \right ) ^{1/4} e^{i/2(\psi-\phi_1+\phi_2)}\,\sin\frac{\theta_1}{2}\,\cos\frac{\theta_2}{2}\,
\nonumber\\
& & \left(\begin{array}{c}
x\\ y\\ u\\ v \end{array}\right) = \frac{1}{\sqrt{2}}\left(\begin{array}{cccc}
1 & - i & 0 & 0 \\
1 & i & 0 & 0 \\
0 & 0 & -i & 1 \\
0 & 0 & -i & -1
\end{array}\right).
\end{eqnarray}
In (\ref{sLagRC-I}), $[\lambda_1:\lambda_2]$ are the homogeneous coordinates of the blown-up $\mathbb{CP}^1=S^2$; $\frac{\lambda_2}{\lambda_1}=\frac{x}{-u}=\frac{v}{-y}=-e^{-i\phi_1}\tan\frac{\theta_1}{2}$. In (\ref{sLagRC-I}),
$\gamma(r^2)\equiv r^2 K^\prime(r^2)= - 2 a^2 + 4 a^4 N^{-\frac{1}{3}}(r^2) + N^{\frac{1}{3}}(r^2)$, where $N(r^2)\equiv\frac{1}{2}\left(r^4 - 16 a^6 + \sqrt{r^8 - 32 a^6 r^4}\right)$.
  Hence, if the resolved warped deformed conifold is predominantly either resolved or deformed, the local $T^3$ of (\ref{xyz defs}) is the required sLag to effect SYZ mirror construction.}

{Interestingly, in the `delocalized limit' \cite{M. Becker et al [2004]}  $\psi=\langle\psi\rangle$, under the coordinate transformation:
\begin{equation}
\label{transformation_psi}
\left(\begin{array}{c} sin\theta_2 d\phi_2 \\ d\theta_2\end{array} \right)\rightarrow \left(\begin{array}{cc} cos\langle\psi\rangle & sin\langle\psi\rangle \\
- sin\langle\psi\rangle & cos\langle\psi\rangle
\end{array}\right)\left(\begin{array}{c}
sin\theta_2 d\phi_2\\
d\theta_2
\end{array}
\right),
\end{equation}
and $\psi\rightarrow\psi - \cos\langle{\bar\theta}_2\rangle\phi_2 + \cos\langle\theta_2\rangle\phi_2 - \tan\langle\psi\rangle ln\sin{\bar\theta}_2$, the $h_5$ term becomes $h_5\left[d\theta_1 d\theta_2 - sin\theta_1 sin\theta_2 d\phi_1d\phi_2\right]$, $e_\psi\rightarrow e_\psi$, i.e.,  one introduces an local (not global) isometry along $\psi$ in addition to the isometries along $\phi_{1,2}$.

{ To enable use of SYZ-mirror duality via three T dualities, remembering that SYZ mirror symmetry is in fact a quantum mirror symmetry, one also needs to ensure a large base (implying large complex structures of the aforementioned two two-tori) of the $T^3(x,y,z)$ fibration, ensuring the disc instantons' contribution is very small \cite{syz}. This is effected via
\cite{F. Chen et al [2010]}:
\begin{eqnarray}
\label{SYZ-large base}
& & d\psi\rightarrow d\psi + f_1(\theta_1)\cos\theta_1 d\theta_1 + f_2(\theta_2)\cos\theta_2d\theta_2,\nonumber\\
& & d\phi_{1,2}\rightarrow d\phi_{1,2} - f_{1,2}(\theta_{1,2})d\theta_{1,2},
\end{eqnarray}
for appropriately chosen large values of $f_{1,2}(\theta_{1,2}) = \pm \cot\theta_{1,2}$ \cite{NPB}. The three-form fluxes
 remain invariant. The guiding principle behind choosing such large values of $f_{1,2}(\theta_{1,2})$, as given in \cite{MQGP}, is that one requires the metric obtained after SYZ-mirror transformation applied to the non-K\"{a}hler  resolved warped deformed conifold to be  like a non-K\"{a}hler warped resolved conifold at least locally. 
For completenes, we summarize the Buscher triple-T duality rules \cite{SYZ 3 Ts},\cite{MQGP} in appendix A.


A single T-duality along a direction orthogonal to the $D3$-brane world volume, e.g., $z$ of (\ref{xyz defs}), yields $D4$ branes straddling a pair of $NS5$-branes consisting of world-volume coordinates $(\theta_1,x)$ and $(\theta_2,y)$. Further, T-dualizing along $x$ and then $y$ would yield a Taub-NUT space  from each of the two $NS5$-branes \cite{T-dual-NS5-Taub-NUT-Tong}. The $D7$-branes yield $D6$-branes which get uplifted to Kaluza-Klein monopoles in M-theory \cite{KK-monopoles-A-Sen} which too involve Taub-NUT spaces. Globally, probably the eleven-dimensional uplift would involve a seven-fold of $G_2$-structure, analogous to the uplift of $D5$-branes wrapping a two-cycle in a resolved warped conifold \cite{Dasguptaetal_G2_structure}. We obtained a local $G_2$ structure in \cite{NPB}, which is summarized in {\bf 2.4}.

\subsection{$G$-Structures}

In this sub-section, we give a quick overview of $G=SU(3), G_2$-structures and how the same appear in the holographic type IIB dual of \cite{metrics}, its delocalized type IIA SYZ mirror and its M-theory uplift constructed in \cite{MQGP}.

Any metric-compatible connection can be written in terms of the Levi-Civita connection and the contorsion tensor $\kappa$ (\cite{Louis_et_al} and references therein). Metric compatibility requires $\kappa\in\Lambda^1\otimes\Lambda^2$, $\Lambda^n$ being the space of $n$-forms. Alternatively, in $d$ complex dimensions, since $\Lambda^2\cong so(d)$, $\kappa$  also be thought of as $\Lambda^1\otimes so(d)$. Given the existence of a $G$-structure, one can decompose $so(d)$
into a part in the Lie algebra $g$ of $G \subset SO(d)$ and its orthogonal complement $g^\perp = so(d)/g$.  The contorsion $\kappa$ splits accordingly into
$\kappa = \kappa^0 + \kappa^g$, where $\kappa^0$ - the intrinsinc torsion - is the part in $\Lambda^1\otimes g^\perp$. One can decompose $\kappa^0$ into irreducible $G$ representations providing a classification of $G$-structures in terms of which representations appear in the decomposition. Let us consider the decomposition of $T^0$ in the case of $SU(3)$-structure. The relevant
representations are
$\Lambda^1\sim 3\oplus\bar{3}, g \sim 8, g^\perp\sim  1 \oplus 3 \oplus \bar{3}.$
Thus the intrinsic torsion, an element of $\Lambda^1\oplus su(3)^\perp$, can be decomposed into the following $SU(3)$ representations \cite{Louis_et_al} :
\begin{eqnarray}
& & \Lambda^1 \otimes su(3)^\perp = (3 \oplus \bar{3}) \otimes (1 \oplus 3 \oplus \bar{3)}
\nonumber\\
& & = (1 \oplus 1) \oplus (8 \oplus 8) \oplus (6 \oplus \bar{6}) \oplus (3 \oplus \bar{3}) \oplus (3 \oplus \bar{3})^\prime\equiv W_1\oplus W_2\oplus W_3\oplus W_4\oplus W_5.
\end{eqnarray}
The $SU(3)$ structure torsion classes \cite{torsion} can be defined in terms of  J, $ \Omega $, dJ, $ d{\Omega}$ and
the contraction operator  $\lrcorner : {\Lambda}^k T^{\star} \otimes {\Lambda}^n
T^{\star} \rightarrow {\Lambda}^{n-k} T^{\star}$.
The torsion classes are then defined in the following way:
\begin{itemize}
\item
$W_1 \leftrightarrow [dJ]^{(3,0)}$, given by real numbers
$W_1=W_1^+ + W_1^-$
with $ d {\Omega}_+ \wedge J = {\Omega}_+ \wedge dJ = W_1^+ J\wedge J\wedge J$
and $ d {\Omega}_- \wedge J = {\Omega}_- \wedge dJ = W_1^- J \wedge J \wedge J$;

\item
$W_2 \leftrightarrow [d \Omega]_0^{(2,2)}$ :
$(d{\Omega}_+)^{(2,2)}=W_1^+ J \wedge J + W_2^+ \wedge J$
and $(d{\Omega}_-)^{(2,2)}=W_1^- J \wedge J + W_2^- \wedge J$;

\item
 $W_3 \leftrightarrow [dJ]_0^{(2,1)}$ is defined
as $W_3=dJ^{(2,1)} -[J \wedge W_4]^{(2,1)}$;

\item
  $W_4 =\frac{1}{2} J\lrcorner dJ$;

 \item
  $W_5 = \frac{1}{2} {\Omega}_+\lrcorner d{\Omega}_+$
(the subscript 0 indicative of the primitivity of the respective forms).
\end{itemize}
 In \cite{transport-coefficients}, it was shown that the five $SU(3)$ structure torsion classes, in the MQGP limit, satisfied (schematically):
\begin{eqnarray}
\label{T-IIB-i}
& & T_{SU(3)}^{\rm IIB}\in W_1 \oplus W_2 \oplus W_3 \oplus W_4 \oplus W_5 \sim \frac{e^{-3\tau}}{\sqrt{g_s N}} \oplus \left(g_s N\right)^{\frac{1}{4}} e^{-3\tau}\oplus \sqrt{g_s N}e^{-3\tau}\oplus -\frac{2}{3} \oplus -\frac{1}{2}\nonumber\\
\end{eqnarray}
$(r\sim e^{\frac{\tau}{3}})$, such that
 \begin{equation}
 \label{T-IIB-ii}
 \frac{2}{3}W^{\bar{3}}_5=W^{\bar{3}}_4
 \end{equation}
  in the UV-IR interpolating region/UV, implying a Klebanov-Strassler-like supersymmetry
  \cite{Butti et al [2004]}. Locally around $\theta_1\sim\frac{1}{N^{\frac{1}{5}}}, \theta_2\sim\frac{1}{N^{\frac{3}{10}}}$, the type IIA torsion classes of the delocalized SYZ type IIA mirror metric  were shown in \cite{NPB} to be:
 \begin{eqnarray}
 \label{T-IIA}
 T_{SU(3)}^{\rm IIA} &\in& W_2 \oplus W_3 \oplus W_4 \oplus W_5 \sim \gamma_2g_s^{-\frac{1}{4}} N^{\frac{3}{10}} \oplus g_s^{-\frac{1}{4}}N^{-\frac{1}{20}} \oplus g_s^{-\frac{1}{4}} N^{\frac{3}{10}} \oplus g_s^{-\frac{1}{4}} N^{\frac{3}{10}}\approx \gamma W_2\oplus W_4\oplus W_5\nonumber\\
 & & \stackrel{\scriptsize\rm fine\ tuning:\gamma\approx0}{\longrightarrow}\approx W_4\oplus W_5.
 \end{eqnarray}
   Further,
   \begin{equation}
   W_4\sim \Re e W_5
   \end{equation}
    indicative of supersymmetry after constructing the delocalized SYZ mirror.

The mirror type IIA metric after performing three T-dualities, first along $x$, then along $y$ and finally along $z$, utilizing the results of \cite{M. Becker et al [2004]} was worked out in \cite{MQGP}. The type IIA metric components  were worked out in \cite{MQGP}.

Apart from quantifying the departure from $SU(3)$ holonomy due to intrinsic contorsion supplied by the NS-NS three-form $H$, via the evaluation of the $SU(3)$ structure torsion classes, to our knowledge for the first time in the context of holographic thermal QCD {\bf at finite gauge coupling and for finite number of colors [in fact for $N_c=3$ in the IR]} in \cite{NPB}: \\
(i) the existence of approximate supersymmetry of the type IIB holographic dual of \cite{metrics} in the MQGP limit near the coordinate branch $\theta_1=\theta_2=0$ was demonstrated, which apart from the existence of a special Lagrangian three-cycle (as shown in \cite{transport-coefficients,NPB}) is essential for construction of the local SYZ type IIA mirror;\\
 (ii)   it was demonstrated that the large-$N$ suppression of the deviation of the type IIB resolved warped deformed conifold from being a complex manifold, is lost on being duality-chased to type IIA - it was also shown that one further fine tuning  $\gamma_2=0$ in $W_2^{\rm IIA}$ can ensure that the local type IIA mirror is complex;\\
(iii)  for the local type IIA $SU(3)$ mirror,  the possibility of surviving approximate supersymmetry was demonstrated which is essential from the point of view of the end result of application of the SYZ mirror prescription.

We can get a one-form type IIA potential from the triple T-dual (along $x, y, z$) of the type IIB $F_{1,3,5}$ in \cite{MQGP} and using which the following $D=11$ metric was obtained in \cite{MQGP} ($u\equiv\frac{r_h}{r}$):
\begin{eqnarray}
\label{Mtheory met}
& &\hskip -0.6in   ds^2_{11} = e^{-\frac{2\phi^{IIA}}{3}} \left[g_{tt}dt^2 + g_{\mathbb{R}^3}\left(dx^2 + dy^2 + dZ^2\right) +  g_{uu}du^2  +   ds^2_{IIA}({\theta_{1,2},\phi_{1,2},\psi})\right] \nonumber\\
& & \hskip -0.6in+ e^{\frac{4{\phi}^{IIA}}{3}}\Bigl(dx_{11} + A^{F_1}+A^{F_3}+A^{F_5}\Bigr)^2 \equiv\ {\rm Black}\ M3-{\rm Brane}+{\cal O}\left(\left[\frac{g_s M^2 \log N}{N}\right] \left(g_sN_f\right)\right).\nonumber\\
& &
\end{eqnarray}

Let us now briefly discuss $G_2$ structure. We will be following \cite{Grigorian,Bryant,karigiannis-2007}. If $V$ is a seven-dimensional real vector space, then a three-form $\varphi $ is said to be positive if it lies in the $GL\left( 7,
\mathbb{R}\right) $ orbit of $\varphi _{0}$, where $\varphi_0$ is a three-form on $\mathbb{R}^7$ which is preserved by $G_2$-subgroup of $GL(7,\mathbb{R})$.  The pair $\left( \varphi ,g\right)$ for a positive $3$-form $\varphi $ and corresponding metric $g$ constitute a $G_{2}$-structure.
The space of $p$-forms are known to decompose as following irreps of $G_{2}$ \cite{Grigorian}:
\begin{eqnarray}
\Lambda ^{1} &=&\Lambda _{7}^{1}  \label{l1decom} \nonumber\\
\Lambda ^{2} &=&\Lambda _{7}^{2}\oplus \Lambda _{14}^{2}  \label{l2decom} \nonumber\\
\Lambda ^{3} &=&\Lambda _{1}^{3}\oplus \Lambda _{7}^{3}\oplus \Lambda
_{27}^{3}  \label{l3decom} \nonumber\\
\Lambda ^{4} &=&\Lambda _{1}^{4}\oplus \Lambda _{7}^{4}\oplus \Lambda
_{27}^{4}  \label{l4decom} \nonumber\\
\Lambda ^{5} &=&\Lambda _{7}^{5}\oplus \Lambda _{14}^{5}  \label{l5decom} \nonumber\\
\Lambda ^{6} &=&\Lambda _{7}^{6}  \label{l6decom}
\end{eqnarray}
The subscripts denote the dimension of representation and components of same representation/dimensionality, are isomorphic to each other.
Let $M$ be a $7$-manifold with a $G_{2}$-structure $\left( \varphi ,g\right)
$.  Then the components of spaces of $2$-, $3$-, $4$-, and $5$-forms are given in \cite{Grigorian,Bryant}. The metric $g$ defines a reduction of the frame bundle F to a principal $SO\left( 7\right) $-sub-bundle  of oriented orthonormal frames. Now, $g$ also defines a Levi-Civita connection $\nabla $ on the tangent bundle
$TM$, and hence on $F$. However, the $G_{2}$-invariant $3$-form $\varphi $
reduces the orthonormal bundle further to a principal $G_{2}$-subbundle $Q$.
The Levi-Civita connection can be pulled back to $Q$. On $Q$,  $\nabla $ can be uniquely decomposed as
\begin{equation}
\nabla =\bar{\nabla}+\mathcal{T}  \label{tors}
\end{equation}
where $\bar{\nabla}$ is a $G_{2}$-compatible canonical connection, taking values in the sub-algebra $\mathfrak{g}_{2}\subset \mathfrak{so}%
\left( 7\right) $, while $\mathcal{T}$ is a $1$-form taking values in $%
\mathfrak{g}_{2}^{\perp }\subset \mathfrak{so}\left( 7\right) $; $\mathcal{T}$ is known as the intrinsic torsion of the
$G_{2}$-structure - the obstruction to the
Levi-Civita connection being $G_{2}$-compatible. Now $\mathfrak{so}%
\left( 7\right) $ splits under $G_{2}$ as
\begin{equation}
\mathfrak{so}\left( 7\right) \cong \Lambda ^{2}V\cong \Lambda _{7}^{2}\oplus
\Lambda _{14}^{2}.
\end{equation}
But $\Lambda _{14}^{2}\cong \mathfrak{g}_{2}$, so the orthogonal complement $\mathfrak{g%
}_{2}^{\perp }\cong \Lambda _{7}^{2}\cong V$. Hence $\mathcal{T}$ can be
represented by a tensor $T_{ab}$ which lies in $W\cong V\otimes V$. Now,
since $\varphi $ is $G_{2}$-invariant, it is $\bar{\nabla}$-parallel. So, the
torsion is determined by $\nabla \varphi $. Now, from the Lemma 2.24 of \cite{karigiannis-2007}:
\begin{equation}
\nabla \varphi \in \Lambda _{7}^{1}\otimes \Lambda _{7}^{3}\cong W.
\label{torsphiW}
\end{equation}%
Due to the isomorphism between the $\Lambda^{a=1,...,5}_7$s, $\nabla \varphi $ lies in the same space as $T_{AB}$ and thus
completely determines it. Equation (\ref{torsphiW}) is equivalent to:
\begin{equation}
\nabla _{A}\varphi _{BCD}=T_{A}^{\ \ E}\psi _{EBCD}  \label{fulltorsion}
\end{equation}%
where $T_{AB}$ is the full torsion tensor. Equation (\ref{fulltorsion}) can be inverted to yield:
\begin{equation}
T_{A}^{\ M}=\frac{1}{24}\left( \nabla _{A}\varphi _{BCD}\right) \psi ^{MBCD}.
\label{tamphipsi}
\end{equation}%
The tensor $T_A^{\ M}$, like the space W, possesses 49 components and hence fully defines $\nabla \varphi $. In general $T_{AB}$ cab be split into torsion components
as
\begin{equation}
T=T _{1}g+T _{7}\lrcorner \varphi +T _{14}+T _{27}
\label{torsioncomps}
\end{equation}
where $T _{1}$ is a function and gives the $\mathbf{1}$ component of $T$
. We also have $T _{7}$, which is a $1$-form and hence gives the $\mathbf{
7}$ component, and, $T _{14}\in \Lambda _{14}^{2}$ gives the $\mathbf{14}$
component. Further, $T _{27}$ is traceless symmetric, and gives the $\mathbf{27}$
component. Writing $T_i$ as $W_i$, we can split $W$ as
\begin{equation}
W=W_{1}\oplus W_{7}\oplus W_{14}\oplus W_{27}.  \label{Wsplit}
\end{equation}
From \cite{G2-Structure}, we see that a $G_2$ structure can be defined as:
\begin{equation}
\label{G_2_i}
\varphi_0 = \frac{1}{3!}{f}_{ABC}e^{ABC} = e^{-\phi^{IIA}}{f}_{abc}e^{abc} + e^{-\frac{2\phi^{IIA}}{3}}J\wedge e^{x_{10}},
\end{equation}
where $A,B,C=1,...,6,10; a,b,c,=1,...,6$ and ${f}_{ABC}$ are the structure constants of the imaginary octonions.
Using the same and  \cite{Chiossi+Salamon}:
\begin{eqnarray}
& & d\varphi_0 = 4 W_1 *_7\varphi_0 - 3 W_7\wedge\varphi_0 - *_7 W_{27}\nonumber\\
& & d*_7\varphi_0 = - 4 W_7\wedge *_7\varphi_0 - 2 *_7W_{14},
\end{eqnarray}
 the $G_2$-structure torsion classes were worked out around $\theta_1\sim\frac{1}{N^{\frac{1}{5}}}, \theta_2\sim\frac{1}{N^{\frac{3}{10}}}$ in \cite{NPB} to:
  \begin{equation}
  \label{G2}
  T_{G_2}\in W_2^{14} \oplus W_3^{27} \sim \frac{1}{\left(g_sN\right)^{\frac{1}{4}} }\oplus \frac{1}{\left(g_sN\right)^{\frac{1}{4}}}.
  \end{equation}
   Hence, the approach of the seven-fold, locally, to having a $G_2$ holonomy ($W_1^{G_2}=W_2^{G_2}=W_3^{G_2}=W_4^{G_2}=0$)  is accelerated in the MQGP limit.

As stated earlier, the global uplift to M-theory of the type $IIB$ background of \cite{metrics} is expected to involve a seven-fold of $G_2$ structure (not $G_2$-holonomy due to non-zero $G_4$). It is hence extremely important to be able to see this, at least locally. It is in this sense that the results of \cite{MQGP} are of great significance as one explicitly sees, for the first time, in the context of  holographic thermal QCD {\bf at finite gauge coupling}, though locally, the aforementioned $G_2$ structure having worked out the non-trivial $G_2$-structure torsion classes.

\section{SYZ Mirror of Ouyang Embedding}

To start off our study of  meson spectroscopy, we first need to understand how the flavor $D6$-branes are embedded in the mirror (constructed in \cite{MQGP}) of the resolved warped deformed conifold of \cite{metrics}, i.e., the delocalized SYZ mirror of the Ouyang embedding of the flavor
$D7$-branes in \cite{metrics}.

One can show that the delocalised type IIA mirror metric of the resolved warped deformed conifold metric as worked out in \cite{MQGP}, for fixed $\theta_1=\frac{\alpha_{\theta_1}}{N^{\frac{1}{5}}}$ in the $(\theta_2,T^3(x,y,z))$-subspace near $\theta_2=\frac{\alpha_{\theta_2}}{N^{\frac{3}{10}}}$ can be written as:
\begin{eqnarray}
\label{metric4x4_i}
\hskip -0.5in ds^2_{\rm IIA}(\theta_2,T^3(x,y,z)) & = & d\theta_2^2 N^{\frac{7}{10}}\left(\xi_{\theta_2\theta_2}\frac{\alpha_{\theta_1}^2}{\alpha_{\theta_2}^2}\sqrt{g_s} d\theta_2 + \xi_{\theta_2y}N^{-\frac{7}{20}}g_s^{\frac{1}{4}} dy - \xi_{\theta_2z}\frac{\log r M N_f}{\alpha_{\theta_2}} N^{-\frac{13}{20}} g_s^{\frac{7}{4}}dz\right) + ds^2(T^3(x,y,z))\nonumber\\
& & \hskip -0.5in \stackrel{N\gg1}{\longrightarrow} \xi_{\theta_2\theta_2} \frac{\alpha_{\theta_1}^2}{\alpha_{\theta_2}^2}\sqrt{g_s} d\theta_2^2  + ds^2(T^3(x,y,z)),
\end{eqnarray}
where the $T^3(x,y,z)$ metric is given by:
\begin{eqnarray}
\label{metric-T3-i}
& & g_{ij}(T^3(x,y,z)) = \nonumber\\
& & \left(
\begin{array}{ccc}
 \frac{3^{2/3} \left(\alpha_{\theta_1} ^2-\alpha_{\theta_2}^2 \sqrt[5]{\frac{1}{N}}\right)}{\alpha_{\theta_1} ^2} & \frac{2 \sqrt{2} \left(\alpha_{\theta_1} ^2 \alpha_{\theta_2} \sqrt{N}-2 \alpha_{\theta_2}^3
   N^{3/10}\right)}{3 \sqrt[6]{3} \alpha_{\theta_1} ^6} & \frac{2 \left(9 \sqrt{2} \sqrt[6]{3} \alpha_{\theta_1}  N^{4/5}-2\ 3^{2/3} N\right)}{27 \alpha_{\theta_1} ^2 \alpha_{\theta_2}^2} \\
 \frac{2 \sqrt{2} \left(\alpha_{\theta_1} ^2 \alpha_{\theta_2} \sqrt{N}-2 \alpha_{\theta_2}^3 N^{3/10}\right)}{3 \sqrt[6]{3} \alpha_{\theta_1} ^6} & 3^{2/3} & \frac{\sqrt{2}
   \left(\alpha_{\theta_2}^2-3 N^{3/5}\right)}{3 \sqrt[6]{3} \alpha_{\theta_2} N^{3/10}} \\
 \frac{2 \left(9 \sqrt{2} \sqrt[6]{3} \alpha_{\theta_1}  N^{4/5}-2\ 3^{2/3} N\right)}{27 \alpha_{\theta_1} ^2 \alpha_{\theta_2}^2} & \frac{\sqrt{2} \left(\alpha_{\theta_2}^2-3 N^{3/5}\right)}{3
   \sqrt[6]{3} \alpha_{\theta_2} N^{3/10}} & \frac{2 \left(\sqrt[5]{N} \alpha_{\theta_1} ^2+\alpha_{\theta_2}^2\right) N^{2/5}}{3 \sqrt[3]{3} \alpha_{\theta_1} ^2 \alpha_{\theta_2}^2}
\end{array}
\right).
\end{eqnarray}
Interestingly, one can diagonalize the local $T^3$ metric to:
\begin{eqnarray}
\label{metric-T3-ii}
 ds^2_{\rm IIA}(T^3(x,y,z))& = & \frac{2 d\tilde{x}^2 \left(9 \sqrt{2} \sqrt[6]{3} \alpha  N^{4/5}-2\ 3^{2/3} N\right)}{27 \alpha_{\theta_1}^22 \alpha_{\theta_2}^2}+\frac{2 d\tilde{y}^2 \left(2\ 3^{2/3}
   N-9 \sqrt{2} \sqrt[6]{3} \alpha  N^{4/5}\right)}{27 \alpha_{\theta_1}^22 \alpha_{\theta_2}^2}\nonumber\\
   & & +\frac{2 d\tilde{z}^2 \left(3^{2/3} \alpha_{\theta_1}^22 N^{3/5}+3^{2/3} \alpha_{\theta_2}^2
   N^{2/5}\right)}{27 \alpha_{\theta_1}^22 \alpha_{\theta_2}^2},
\end{eqnarray}
where:
\begin{eqnarray}
\label{metric-T3-iii}
& & d\tilde{x} = \frac{{dx} \left(3 \alpha_{\theta_1} ^2 \left(\frac{1}{N}\right)^{2/5}+4\right)}{4 \sqrt{2}}+\frac{{dz} \left(4-3 \alpha_{\theta_1} ^2 \left(\frac{1}{N}\right)^{2/5}\right)}{4
   \sqrt{2}}\nonumber\\
   & & +\frac{{dy} \sqrt{\frac{1}{N}} \left(2 \alpha_{\theta_2}^2 \sqrt[5]{\frac{1}{N}} \left(\left(54-3^{2/3}\right) \alpha_{\theta_1} ^6-54 \sqrt{6} \alpha_{\theta_1} ^3
   \alpha_{\theta_2}^2+66 \alpha_{\theta_2}^4\right)-\alpha_{\theta_1} ^2 \left(3^{2/3} \alpha_{\theta_1} ^6+72 \alpha_{\theta_2}^4\right)\right)}{16\ 3^{5/6} \alpha_{\theta_1} ^4 \alpha_{\theta_2}}\nonumber\\
   & & d\tilde{y} = \frac{{dx} \left(3 \alpha_{\theta_1} ^2 \left(\frac{1}{N}\right)^{2/5}-4\right)}{4 \sqrt{2}}+\frac{{dz} \left(3 \alpha_{\theta_1} ^2 \left(\frac{1}{N}\right)^{2/5}+4\right)}{4
   \sqrt{2}}\nonumber\\
   & & +\frac{{dy} \sqrt{\frac{1}{N}} \left(\alpha_{\theta_1} ^2 \left(-\left(3^{2/3} \alpha_{\theta_1} ^6+72 \alpha_{\theta_2}^4\right)\right)-2 \alpha_{\theta_2}^2
   \sqrt[5]{\frac{1}{N}} \left(\left(54+3^{2/3}\right) \alpha_{\theta_1} ^6+54 \sqrt{6} \alpha_{\theta_1} ^3 \alpha_{\theta_2}^2-66 \alpha_{\theta_2}^4\right)\right)}{16\ 3^{5/6} \alpha_{\theta_1} ^4
   \alpha_{\theta_2}}\nonumber\\
   & & d\tilde{z} = -\frac{9 \sqrt[6]{3} \sqrt{\alpha_{\theta_2}^2} {dx} \left(\frac{1}{N}\right)^{7/10} \left(\alpha_{\theta_2}^2 \sqrt[5]{\frac{1}{N}}-2 \alpha_{\theta_1} ^2\right)}{4
   \sqrt{2}}+\frac{{dy} \left(-3^{2/3} \alpha_{\theta_1} ^{12}-768 \alpha_{\theta_1} ^4 \alpha_{\theta_2}^2 N+1728 \sqrt[3]{3} \alpha_{\theta_2}^8\right)}{768 \alpha_{\theta_1} ^4
   \sqrt{\alpha_{\theta_2}^2} \alpha_{\theta_2} N}\nonumber\\
   & & -\frac{3 \sqrt[6]{3} \alpha_{\theta_2}^4 {dz} \sqrt{\frac{1}{N}} \left(3 \sqrt{3} \alpha_{\theta_1} ^3
   \sqrt[5]{\frac{1}{N}}+\sqrt{2} \alpha_{\theta_1} ^2-2 \sqrt{2} \alpha_{\theta_2}^2 \sqrt[5]{\frac{1}{N}}\right)}{2 \alpha_{\theta_1} ^4 \sqrt{\alpha_{\theta_2}^2}}.
\end{eqnarray}
Now, from the Buscher triple T-duality results for the NS-NS $B$ as given in (\ref{B}), one sees that for small $\phi_{1,2}$ (which ensures decoupling of $M_6(\theta_{1,2},\phi_{1,2},\psi,x^{10})$ from $M_5(\mathbb{R}^{1,3},r)$ in the M-theory uplift in the MQGP limit):
{\footnotesize
\begin{eqnarray}
\label{B-IIA-diag-nondiag}
& & B^{\rm IIA}\left(\theta_1=\frac{\alpha_{\theta_1}}{N^{\frac{1}{5}}},\theta_2\sim\frac{\alpha_{\theta_2}}{N^{\frac{3}{10}}}\right) = d\theta_2\wedge {dx} \left(-\frac{2 \sqrt{2} \sqrt[4]{\pi } \sqrt[4]{{g_s}} N^{3/4} \left(3 \sqrt{6} \alpha_{\theta_1}^3-2 \alpha_{\theta_1}^2 \sqrt[5]{N}+2 \alpha_{\theta_2}^2\right)}{27 \alpha_{\theta_1}^4
   \alpha_{\theta_2}}\right)\nonumber\\
   & & +d\theta_2\wedge {dz} \left(\frac{\sqrt[4]{\pi } \sqrt[4]{{g_s}} \left(5 \alpha_{\theta_2}^2 \sqrt[20]{\frac{1}{N}}-6 N^{11/20}\right)}{27 \sqrt{2} \alpha_{\theta_2}}\right) + d\theta_2\wedge {dy}\left(\frac{\sqrt[4]{\pi } \sqrt[4]{{g_s}} N^{3/20} \left(2 \alpha  \sqrt[10]{N}+\alpha_{\theta_2}\right)}{\sqrt{3} \alpha }\right)
  \nonumber\\
   & & = d\theta_2\wedge d\tilde{x} \left(-\frac{2 \sqrt[4]{\pi } \sqrt[4]{{g_s}} N^{3/4} \left(3 \sqrt{6} \alpha_{\theta_1}^3-2 \alpha_{\theta_1}^2 \sqrt[5]{N}+2 \alpha_{\theta_2}^2\right)}{27 \alpha_{\theta_1}^4 \alpha_{\theta_2}}\right)+d\theta_2\wedge d\tilde{y} \left(\frac{2 \sqrt[4]{\pi } \sqrt[4]{{g_s}} N^{3/4} \left(3 \sqrt{6} \alpha_{\theta_1}^3-2 \alpha_{\theta_1}^2 \sqrt[5]{N}+2 \alpha_{\theta_2}^2\right)}{27 \alpha_{\theta_1}^4 \alpha_{\theta_2}}\right)\nonumber\\
   & &  + d\theta_2\wedge d\tilde{z}\left(-\frac{\sqrt[4]{\pi } \alpha_{\theta_2} \sqrt[4]{{g_s}} N^{3/20} \left(2 \left(\sqrt[3]{3}-1\right) \alpha  \sqrt[10]{N}+\sqrt[3]{3}
   \alpha_{\theta_2}\right)}{3^{5/6} \alpha  \sqrt{\alpha_{\theta_2}^2}}\right)\nonumber\\
   & &
\end{eqnarray}}
There is an important message we must take in from (\ref{B-IIA-diag-nondiag}). As one realizes  from (\ref{B}) and therefore (\ref{B-IIA-diag-nondiag}), $B^{\rm IIA}$ is independent of $M$ even up to NLO in $N$ even though $B^{\rm IIB}$ is proportional to $M$. This will be important in obtaining the mesonic spectra in the subsequent sections and obtaining a good match with \cite{PDG} without having to invoke ${\cal O}\left(\frac{g_sM^2}{N}\right)$-corrections which the authors of \cite{Dasgupta_et_al_Mesons} had to use (and set to $0.5$ - and yet consider the same to 'small' - to get a reasonable match with \cite{PDG}).

The complete $10$ dimensional type IIA metric in large N limit is given as:
\begin{eqnarray}
\label{full IIA}
& & ds^{2}_{\rm IIA} \approx G^{\rm IIA}_{00} dx^{2}_{0}+G^{\rm IIA}_{11} dx^{2}_{1}+G^{\rm IIA}_{22} dx^{2}_{2}+G^{\rm IIA}_{33} dx^{2}_{3}+G^{\rm IIA}_{rr} dr^{2}+G^{\rm IIA}_{\theta_{1}\theta_{1}} d\theta_{1}^{2}+G^{\rm IIA}_{\theta_{1}\tilde{x}} d\theta_{1}d\tilde{x}+G^{\rm IIA}_{\theta_{1}\tilde{y}} d\theta_{1}d\tilde{y}\nonumber\\
 & &+G^{\rm IIA}_{\theta_{1}\tilde{z}} d\theta_{1}d\tilde{z}+G^{\rm IIA}_{\theta_{2}\theta_{2}} d\theta_{2}^{2}+ds^{2}(T^{3}(\tilde{x},\tilde{y},\tilde{z}))\nonumber\\
 & &
\end{eqnarray}
To obtain the pullback metric on the D6 branes, we choose the first branch of the Ouyang embedding where $(\theta_1,\tilde{x})=(0,0)$ and we consider the $\tilde{z}$ coordinate as a function of r, i.e. $\tilde{z}(r)$. We then use the equation of motion of the field to find the explicit functional dependence. The coordinates for D6 brane are $x^{\mu}={x^{(0,1,2,3)},r,\theta_2,\tilde{y}}$. The pull-back of the metric is given by:
\begin{eqnarray}
\label{pull-back}
& & G^{\rm IIA}_{6\mu \nu}dx^{\mu}dx^{\nu}= G^{\rm IIA}_{00} dx^{2}_{0}+G^{\rm IIA}_{11} dx^{2}_{1}+G^{\rm IIA}_{22} dx^{2}_{2}+G^{\rm IIA}_{33} dx^{2}_{3}+(G^{\rm IIA}_{rr}+G^{\rm IIA}_{\tilde{z}\tilde{z}}\tilde{z}\prime(r)^{2}) dr^{2}\nonumber\\
 & & G^{\rm IIA}_{\theta_{2}\theta_{2}}d\theta_{2}^{2}+G^{\rm IIA}_{\tilde{y}\tilde{y}}d\tilde{y}^{2}\nonumber\\
 & &
\end{eqnarray}
Near $\theta_1=\alpha_{\theta_1}N^{-1/5}$ and $\theta_2=\alpha_{\theta_2}N^{-3/10}$ the type IIA metric components upto NLO are given as following:
{\footnotesize
\begin{eqnarray}
\label{GIIA}
& & \hskip -0.8in  G^{\rm IIA}_{00}=-\frac{\left(r^4-{r_h}^4\right) \left(3 {g_s} M^2 \log (r) (-2 {g_s} {N_f} \log (\alpha_{\theta 1} \alpha_{\theta 2})+{g_s} {N_f} \log
   (N)-6 {g_s} {N_f}+{g_s} {N_f} \log (16)-8 \pi )-36 {g_s}^2 M^2 {N_f} \log ^2(r)+32 \pi ^2 N\right)}{64 \pi ^{5/2}
   \sqrt{{g_s}} N^{3/2} r^2}\nonumber\\
& & \hskip -0.8in  G^{\rm IIA}_{11}=\frac{r^2 \left(3 {g_s} M^2 \log (r) (-2 {g_s} {N_f} \log (\alpha_{\theta 1} \alpha_{\theta 2})+{g_s} {N_f} {\log N}-6 {g_s}
   {N_f}+{g_s} {N_f} \log (16)-8 \pi )-36 {g_s}^2 M^2 {N_f} \log ^2(r)+32 \pi ^2 N\right)}{64 \pi ^{5/2} \sqrt{{g_s}}
   N^{3/2}}\nonumber\\
& & \hskip -0.8in  G^{\rm IIA}_{22}=\frac{r^2 \left(3 {g_s} M^2 \log (r) (-2 {g_s} {N_f} \log (\alpha_{\theta 1} \alpha_{\theta 2})+{g_s} {N_f} {\log N}-6 {g_s}
   {N_f}+{g_s} {N_f} \log (16)-8 \pi )-36 {g_s}^2 M^2 {N_f} \log ^2(r)+32 \pi ^2 N\right)}{64 \pi ^{5/2} \sqrt{{g_s}}
   N^{3/2}}\nonumber\\
& & \hskip -0.8in  G^{\rm IIA}_{33}=\frac{r^2 \left(3 {g_s} M^2 \log (r) (-2 {g_s} {N_f} \log (\alpha_{\theta 1} \alpha_{\theta 2})+{g_s} {N_f} {\log N}-6 {g_s}
   {N_f}+{g_s} {N_f} \log (16)-8 \pi )-36 {g_s}^2 M^2 {N_f} \log ^2(r)+32 \pi ^2 N\right)}{64 \pi ^{5/2} \sqrt{{g_s}}
   N^{3/2}}\nonumber\\
& & \hskip -0.8in  G^{\rm IIA}_{rr}=\frac{\sqrt{{g_s}} r^2 \left(6 a^2+r^2\right) \left(3 {g_s} M^2 \log (r) (2 {g_s} {N_f} \log (\alpha_{\theta 1} \alpha_{\theta 2})-{g_s}
   {N_f} {\log N}+6 {g_s} {N_f}-2 {g_s} {N_f} \log (4)+8 \pi )+36 {g_s}^2 M^2 {N_f} \log ^2(r)+32 \pi ^2
   N\right)}{16 \pi ^{3/2} \sqrt{N} \left(9 a^2+r^2\right) \left(r^4-{r_h}^4\right)}\nonumber\\
 & & \hskip -0.8in G^{\rm IIA}_{\tilde{y}\tilde{y}}=-\frac{2 \left(9 \sqrt{2} \sqrt[6]{3} \alpha  N^{4/5}-2\ 3^{2/3} N\right)}{27 \alpha_{\theta_1}^22 \alpha_{\theta_2}^2}\nonumber\\
& & \hskip -0.8in  G^{\rm IIA}_{\theta_{2}\theta_{2}}=\frac{\sqrt{\pi } \sqrt{{g_s}} \sqrt{N} \left(\alpha_{\theta_1}^22 \sqrt[5]{N}+\alpha_{\theta_2}^2\right)}{\sqrt[3]{3} \alpha_{\theta_2}^2}.
\end{eqnarray}}
 One will assume that the embedding of the $D6$-brane will be given by $i:\Sigma^{1,6}\left(\mathbb{R}^{1,3},r,\theta_2\sim\frac{\alpha_{\theta_2}}{N^{\frac{3}{10}}},\tilde{y}\right)
 \hookrightarrow M^{1,9}$, effected by: $\tilde{z} = \tilde{z}(r)$. The pull back of the B-field along the directions of the D6-branes is given by:
 \begin{eqnarray}
 \label{pull-back flux}
 & &  \delta\left(\theta_2 - \frac{\alpha_{\theta_2}}{N^{\frac{3}{10}}}\right) i^*B= \delta\left(\theta_2 - \frac{\alpha_{\theta_2}}{N^{\frac{3}{10}}}\right)\left[ -B_{\theta_{2}\tilde{z}}\tilde{z}^\prime(r)dr\wedge d\theta_{2}+B_{\theta_2\tilde{y}}d\theta_2\wedge d\tilde{y}+B_{\theta_{2}\tilde{x}}d\theta_{2}\wedge d\tilde{x}\right],\nonumber\\
 & &
 \end{eqnarray}
 where $B_{\theta_2\tilde{x}}, B_{\theta_2\tilde{y}}, B_{\theta_2\tilde{z}}$ can be read off from (\ref{B-IIA}).
 Now, one can show that:
\begin{eqnarray}
\label{DBI-det_i}
& & {\rm det}\left(i^*(g+B)\right) = \Sigma_0(r;g_s,N_f,N,M) + \Sigma_1(r;g_s,N_f,M,N)\left(\tilde{z}^\prime(r)\right)^2,
\end{eqnarray}
where the embedding functions $\Sigma_{0,1}(r;g_s,N_f,M,N)$ are given in (\ref{DBI-det_ii}).

Thus, the Euler-Lagrange equation of motion yields:
\begin{equation}
\label{EL-DBI-i}
\frac{d}{dr}\left(\frac{\tilde{z}^\prime(r)}{\sqrt{\Sigma_0(r;g_s,N_f,N,M) + \Sigma_1 (r;g_s,N_f,N,M) (\tilde{z}^\prime)^2}}\right) = 0.
\end{equation}
Like \cite{Dasgupta_et_al_Mesons}, $\tilde{z}=$constant, is a solution of (\ref{EL-DBI-i}). Alternatively,
(\ref{EL-DBI-i}) is equivalent to:\\ $\frac{\tilde{z}^\prime(r)}{\sqrt{\Sigma_0(r;g_s,N_f,N,M) + \Sigma_1 (r;g_s,N_f,N,M) (\tilde{z}^\prime)^2}} = K$. Hence, the Euler-Lagrange equation for the $\tilde{z}(r)$ from the DBI action can be written in the following form:
\begin{eqnarray}
\label{embedding eq}
& &z'(r)^2-\frac{24461180928 \pi ^{19/2} \alpha_{\theta_1}^{16} \alpha_{\theta_2}^8 {C_1} {g_s}^4 K N^{49/5}}{{C_2}^2-24461180928 \pi ^{19/2} \alpha_{\theta_1}^{16}
   \alpha_{\theta_2}^8 {C_3} {g_s}^4 K N^{49/5}}=0\nonumber\\
   & &
\end{eqnarray}
where $K$ is an arbitary constant while $C_1(r;g_s,N_f,N), C_2(r;g_s,N_f,N)$ and $C_3(r;g_s,N_f,N)$ up to NLO-in-$N$ after large$N$ expansion have the following forms:
\begin{eqnarray}
\label{coefficients}
& & C_1(r;g_s,N_f,N) \nonumber\\
& & = \frac{4194304 \pi ^{17/2} \sqrt[5]{N} r^6 \left(6 a^2+r^2\right) \left(27 \sqrt[3]{3} \alpha_{\theta_1}^6-12 \sqrt{6} \alpha_{\theta_1}^3 \alpha_{\theta_2}^2+4 \alpha_{\theta_1}^2
   \alpha_{\theta_2}^2 \sqrt[5]{N}-8 \alpha_{\theta_2}^4\right)}{\alpha_{\theta_1}^6 \alpha_{\theta_2}^4 {g_s} \left(9 a^2+r^2\right)}\nonumber\\
   & & C_2(r;g_s,N_f,N) \nonumber\\
   & & = \frac{8388608 \pi ^8 \alpha_{\theta_1}^2 \sqrt{{g_s}} N^{26/5} r^4 \left(r^4-{r_h}^4\right) \left(81 \alpha_{\theta_1}^6-36 \sqrt{2} \sqrt[6]{3} \alpha_{\theta_1}^3 \alpha_{\theta_2}^2+4
   3^{2/3} \alpha_{\theta_1}^2 \alpha_{\theta_2}^2 \sqrt[5]{N}-4 3^{2/3} \alpha_{\theta_2}^4\right)}{27 \alpha_{\theta_2}^2}\nonumber\\
   & & C_3(r;g_s,N_f,N)\nonumber\\
   & &  = \frac{4194304 \pi ^8 N^{3/10} r^4 \left(r^4-{r_h}^4\right) \left(81 \alpha_{\theta_1}^6-36 \sqrt{2} \sqrt[6]{3} \alpha_{\theta_1}^3 \alpha_{\theta_2}^2+4 3^{2/3} \alpha_{\theta_1}^2
   \alpha_{\theta_2}^2 \sqrt[5]{N}-4 3^{2/3} \alpha_{\theta_2}^4\right)}{27 \alpha_{\theta_1}^6 \alpha_{\theta_2}^6 {g_s}^{3/2}}.\nonumber\\
   & &
\end{eqnarray}
Substituting the values of $C_1(r;g_s,N_f,N), C_2(r;g_s,N_f,N)$ and $C_3(r;g_s,N_f,N)$ in the differential equation presented above and keeping terms only up to NLO-in-$N$ after taking a large $N$ expansion the differential equation acquires the following form:
\begin{eqnarray}
\label{diff_eqn}
& & {\tilde{z}}'(r)^2 -\frac{59049 3^{2/3} \pi ^2 \alpha_{\theta_1}^4 \alpha_{\theta_2}^6 {g_s}^2 K \left(\frac{1}{N}\right)^{3/5} \left(6 a^2+r^2\right)}{2 r^2 \left(9 a^2+r^2\right)
   \left(r^4-{r_h}^4\right)^2}\nonumber\\
   & & -\frac{531441 \pi ^2 \alpha_{\theta_1}^5 \alpha_{\theta_2}^4 {g_s}^2 K \left(\frac{1}{N}\right)^{4/5} \left(6 a^2+r^2\right) \left(4
   \sqrt{2} \sqrt[6]{3} \alpha_{\theta_2}^2-9 \alpha_{\theta_1}^3\right)}{8 r^2 \left(9 a^2+r^2\right) \left(r^4-{r_h}^4\right)^2}=0\nonumber\\
   & &
\end{eqnarray}
Analogous to \cite{Dasgupta_et_al_Mesons}, from (\ref{EL-DBI-i}), one sees that $\tilde{z}=$constant, is a valid solution and by choosing $\tilde{z} = \pm{\cal C}\frac{\pi}{2}$, one can choose
the $D6/\overline{D6}$-branes to be at  ``antipodal" points. Using a very similar computation for a thermal background with no black-hole ($r_h=0$), one can show that this constant embedding of
$D6$-branes, is still valid.

\section{Vector Meson Spectroscopy in a Black-Hole Background for All Temperatures}

Equipped with the embedding of the flavor $D6$-branes in the delocalized SYZ mirror of resolved warped deformed conifold of \cite{metrics} from section {\bf 3}, we now proceed to obtaining the spectra as the Kaluza-Klein modes of the massless sector of open strings in type IIA at finite gauge coupling. In this and the next section, we do not worry about the issues like the black hole gravity dual is not considered at low temperatures wherein one must consider a thermal background. Happily, in section {\bf 6}, we will via an explicity computation verify that the mesonic spectra obtained in sections {\bf 4} (for [pseudo-]vector mesons) and {\bf 5} (for [pseudo-]scalar mesons), are nearly isospectral with one obtained by working with a thermal background without a black-hole valid at only low temperatures.

We evaluate the masses of the (pseudo-)vector and (pesudo-)scalar mesons separately - the former by considering gauge fluctuations of a background gauge field along the world volume of the embedded flavor $D6$-branes and the latter (without turning on a background gauge field) by looking at fluctuations of the embedding transverse to the world volume of the embedded $D6$-branes.

As done in \cite{Dasgupta_et_al_Mesons}, let us redefine $(r,\tilde{z})$ in terms of new variables $(Y,Z)$:
\begin{eqnarray}
\label{YZ}
& & r = r_h e^{\sqrt{Y^2+Z^2}}\nonumber\\
& & \tilde{z} = {\cal C} \arctan \frac{Z}{Y},
\end{eqnarray}
and the constant embedding of $D6(\overline{D6})$-branes can correspond to $\tilde{z} = \frac{\pi}{2}$ for ${\cal C}=1$ for $D6$-branes and $\tilde{z} = -\frac{\pi}{2}$ for ${\cal C}=-1$ for $\overline{D6}$-branes, both corresponding to $Y=0$. Now, consider turning on a gauge field fluctuation $\tilde{F}\frac{\sigma^3}{2}$ about a small background gauge field $F_0\frac{\sigma^3}{2}$ and the backround $i^*(g+B)$. This implies:
\begin{eqnarray}
\label{DBI-i}
& & {\rm Str}\left.\sqrt{{\rm det}_{\mathbb{R}^{1,3},Z,\theta_2,\tilde{y}}\left(i^*(G+B) + (F_0 + \tilde{F})\frac{\sigma^3}{2}\right)}\right|_{Y=0}\delta\left(\theta_2 - \frac{\alpha_{\theta_2}}{N^{\frac{3}{10}}}\right)\nonumber\\
& & = \sqrt{{\rm det}_{\theta_2,\tilde{y}}\left(i^*(g+B)\right)}\left. {\rm Str}\sqrt{{\rm det}_{\mathbb{R}^{1,3},Z}\left(i^*(g+B) + (F_0 + \tilde{F})\frac{\sigma^3}{2}\right)}\right|_{Y=0}\delta\left(\theta_2 - \frac{\alpha_{\theta_2}}{N^{\frac{3}{10}}}\right)
\nonumber\\
& & = \left.\sqrt{{\rm det}_{\theta_2,\tilde{y}}\left(i^*(g+B)\right)}\sqrt{{\rm det}_{\mathbb{R}^{1,3},Z}(i^*g)} {\rm Str}\left({\bf 1}_2 - \frac{1}{2}\left[(i^*g)^{-1}\left((F_0 + \tilde{F})\frac{\sigma^3}{2}\right)\right]^2 + ....\right)\right|_{Y=0}\delta\left(\theta_2 - \frac{\alpha_{\theta_2}}{N^{\frac{3}{10}}}\right).
\nonumber\\
\end{eqnarray}
Concentrating on the terms quadratic in $\tilde{F}$:
{\footnotesize
\begin{eqnarray}
\label{DBI-ii}
& & \hskip -0.4in S_{D6} = \nonumber\\
 & & \hskip -0.4in -\frac{1}{2}\int d^4x dZ d\theta_2d\tilde{y}\left.\sqrt{{\rm det}_{\theta_2,\tilde{y}}\left(i^*(G+B)\right)}\sqrt{{\rm det}_{\mathbb{R}^{1,3},Z}(i^*G)}\left[2\sqrt{h}G^{ZZ}\tilde{G}^{\mu\nu}\tilde{F}_{\mu Z}\tilde{F}_{\nu Z} + h \tilde{G}^{\mu\mu_1}\tilde{G}^{\nu\nu_1}\tilde{F}_{\mu_1\nu}\tilde{F}_{\nu_1\mu}\right]\right|_{Y=0}\delta\left(\theta_2 - \frac{\alpha_{\theta_2}}{N^{\frac{3}{10}}}\right),\nonumber\\
& &
\end{eqnarray}}
where $\tilde{G}^{\mu\nu}$ are the unwarped $\mathbb{R}^{1,3}$ metric components.
Substituting:
\begin{eqnarray}
\label{AZ+Amu}
& & A_\mu(x^\nu,Z) = \sum_{n=1} B_\mu^{(n)}(x^\nu)\alpha_n^{\left\{\mu\right\}}(Z),\ \text{no summation w.r.t.}\ \mu,\nonumber\\
& & A_Z(x^\nu,Z) = \sum_{n=1} \phi^{(n)}(x^\nu)\beta_n(Z),
\end{eqnarray}
one obtains:
\begin{eqnarray}
\label{Ftildesq-i}
& & \int d^4x dZ \left({\cal V}_2(Z)F^{(n)}_{\mu\nu}F^{\mu\nu}_{(n)}\alpha_m^{\left\{\mu\right\}}(Z)\alpha_n^{\left\{\mu\right\}}(Z) + {\cal V}_1(Z)B^{(m)}_\mu B^{(n)}_\nu\dot{\alpha}_m^{\left\{\mu\right\}}\dot{\alpha}_n^{\left\{\mu\right\}}\right),
\end{eqnarray}
where ${\cal V}_{1,2}$ are given in (\ref{v1v2}).

  Now, $F_{\mu\nu}(x^\rho,|Z|) = \sum_n\partial_{[\mu}B_{\nu]}^{(n)}\alpha_n(Z)\equiv F_{\mu\nu}^{(n)}\alpha_n(Z)$. The EOM satisfied by $B_\mu(x^\nu)^{(n)}$ is: $\partial_\mu\tilde{F}^{\mu\nu}_{(n)} + \partial_\mu\log\sqrt{G_{\mathbb{R}^{1,3},|Z|}}\tilde{F}^{\mu\nu}_{(n)} = \partial_\mu\tilde{F}^{\mu\nu}_{(n)} = {\cal M}_{(n)}^2B^\nu_{(n)}$. After integrating by parts once, and utilizing the EOM for $B^{(n)}_\mu$, one rewrites (\ref{Ftildesq-i}) as:
  \begin{eqnarray}
  \label{Ftildesq-ii}
& & \int d^4x dZ\  \left(-2 {\cal V}_2(Z) {\cal M}_{(m)}^2\alpha_n^{B_\mu}\alpha_m^{B_\mu} + {\cal V}_1(Z)\dot{\alpha}_n^{B_\mu}\dot{\alpha}_m^{B_\mu}\right)B^{\mu (n)}B_{\mu}^{(m)},
\end{eqnarray}
which yields the following equations of motion:
\begin{eqnarray}
\label{eoms_alpha_n_Bmu}
& & \alpha^{\left\{0\right\}}_{m}: \frac{d}{dZ\ }\left({\cal V}_1(Z) \tilde{G}^{00}(Z)\dot{\alpha}_{m}^{\left\{0\right\}}\right) + 2 {\cal V}_2(Z)\tilde{G}^{00}{\cal M}_{(m)}^2\alpha^{\left\{0\right\}}_m = 0,\nonumber\\
& & \alpha^{\left\{i\right\}}_{m}: \frac{d}{dZ\ }\left({\cal V}_1(Z) \dot{\alpha}_{m}^{\left\{i\right\}}\right) + 2 {\cal V}_2(Z){\cal M}_{(m)}^2\alpha^{\left\{i\right\}}_m = 0.
\end{eqnarray}

Writing $a = r_h\left(0.6 + 4\frac{g_s M^2}{N}\left(1 + \log r_h\right)\right), m = \tilde{m}\frac{r_h}{\sqrt{4\pi g_s N}}$, one hence obtains the following EOMs:
\begin{eqnarray}
\label{EOMs-i}
& & \hskip -0.8in  \alpha_n^{\left\{i\right\}}\  ^{\prime\prime}(Z)\nonumber\\
& & \hskip -0.8in + \alpha_n^{\left\{i\right\}}\ ^{\prime}(Z) \Biggl(\frac{-{g_s} {N_f} \left(e^{4 |Z|} (-2 {\log N}+6 |Z|+3)-2 {\log N}+6 |Z|-3\right)-6 {g_s} {N_f} \left(e^{4
   |Z|}+1\right) \log ({r_h})+8 \pi  \left(e^{4 |Z|}+1\right)}{\left(e^{4 |Z|}-1\right) ({g_s} {N_f} ({\log N}-3 |Z|)-3 {g_s}
   {N_f} \log ({r_h})+4 \pi )}\nonumber\\
   & & \hskip -0.8in -\frac{1}{N^2 ({g_s} {N_f} ({\log N}-3. |Z|)-3. {g_s} {N_f} \log
   ({r_h})+12.5664)^2}\Biggl\{1.5 e^{-2 |Z|} \left(4. {g_s} M^2 \log ({r_h})+4. {g_s} M^2+0.6 N\right)^2
  \nonumber\\
 & & \hskip -0.8in \times   \Biggl[{g_s}^2 {N_f}^2 \left(2. {\log N}^2-12. {\log N} |Z|-6. {\log N}+18. Z^2+18. |Z|+9.\right)+18. {g_s}^2 {N_f}^2
   \log ^2({r_h})\nonumber\\
   & & \hskip -0.8in +{g_s} {N_f} \log ({r_h}) ({g_s} {N_f} (-12. {\log N}+36. |Z|+18.)-150.796)+{g_s} {N_f}
   (50.2655 {\log N}-150.796 |Z|-75.3982)\nonumber\\
   & &  \hskip -0.8in+315.827\Biggr]\Biggr\}\Biggr) + \alpha_n^{\left\{i\right\}}(Z) \frac{\tilde{m}^2  \left(e^{2 |Z|}-\frac{3. \left(4. {g_s} M^2 \log ({r_h})+4. {g_s} M^2+0.6
   N\right)^2}{N^2}\right)}{e^{4 |Z|}-1}= 0,\nonumber\\
   & &\hskip -0.8in
\end{eqnarray}
and
\begin{eqnarray}
\label{EOMs-ii}
& & \hskip -0.8in
{\alpha_n^{\left\{0\right\}}}\ ^{\prime\prime}(Z)\nonumber\\
& & \hskip -0.8in + \frac{\alpha_n^{\left\{0\right\}}\ ^{\prime}(Z)}{2 ({g_s} {N_f} {\log N}-3 {g_s} {N_f} \log ({r_h})-3 {g_s}
   {N_f} |Z|+4 \pi )^2}\nonumber\\
   & &\hskip -0.8in \times\Biggl\{ \Biggl(\frac{1}{N^2}\Biggl\{e^{-2 |Z|} \left(4. {g_s} M^2 \log ({r_h})+4. {g_s} M^2+0.6 N\right)^2 (9 {g_s}
   {N_f} (-{g_s} {N_f} {\log N}+3 {g_s} {N_f} \log ({r_h})+3 {g_s} {N_f} |Z|-4 \pi )\nonumber\\
   & & \hskip -1in+(2 {g_s} {N_f} \log
   (N)-6 {g_s} {N_f} \log ({r_h})-6 {g_s} {N_f} |Z|-3 {g_s} {N_f}+8 \pi ) (-3 {g_s} {N_f} {\log N}+9 {g_s}
   {N_f} \log ({r_h})+9 {g_s} {N_f} |Z|+9 {g_s} {N_f}-12 \pi ))\Biggr\}\nonumber\\
   & & \hskip -1in+2 (2 {g_s} {N_f} {\log N}-6 {g_s}
   {N_f} \log ({r_h})-6 {g_s} {N_f} |Z|-3 {g_s} {N_f}+8 \pi ) ({g_s} {N_f} {\log N}-3 {g_s} {N_f} \log
   ({r_h})-3 {g_s} {N_f} |Z|+4 \pi )\Biggr)\Biggr\} \nonumber\\
   & & \hskip -1in+ {\alpha_n^{\left\{0\right\}}}(Z)\frac{\tilde{m}^2  \left({r_h}^2 e^{2 |Z|}-\frac{3. {r_h}^2 \left(4. {g_s} M^2 \log
   ({r_h})+4. {g_s} M^2+0.6 N\right)^2}{N^2}\right)}{{r_h}^2 \left(e^{4 |Z|}-1\right)}=0\nonumber\\
   & &
\end{eqnarray}

We will now proceed to obtaining the (pseudo-)vector meson spectrum by three routes. The first will cater exclusively to an IR computation where we solve the $\alpha^{\left\{i\right\}}_n(Z)$ and $\alpha^{\left\{0\right\}}_n(Z)$ EOMs near the horizon. Imposing Neumann boundary condition at the horizon results in quantization of the (pseudo-)vector meson masses  and via $N_f$- and $M$-dependent contributions, we extract the temperature dependence of the (pseudo-)vector meson spectrum. We will see that up to LO in N, in the IR, there is a near isospectrality in the (pseudo-)vector meson spectrum obtained by solving the $\alpha^{\left\{i\right\}}_n(Z)$ and $\alpha^{\left\{0\right\}}_n(Z)$ EOMs. The second route will be to convert the $\alpha^{\left\{i\right\}}_n(Z)$ and $\alpha^{\left\{0\right\}}_n(Z)$ EOMs into Schr"{o}dinger-like EOMs and to solve the same in the IR and UV separately and obtain (pseudo-)vector mass quantization by imposing Neumann boundary conditions at the horizon (IR)/asymptotic boundary (UV). It turns out the former yields a result, which up to LO in $N$, is of the same order as the IR results of route one. The UV computations satisfy Neumann and/or Dirichlet boundary conditions without any mass quantization condition. The third route catering to the IR-UV interpolating region and what gives us our main results that are directly compared with PDG results, is obtaining the (pseudo-)vector meson masses via WKB quantization condition. We also show that an IR WKB quantization (pseudo-)vector meson spectroscopy is nearly isospectral with the results of route one.

\subsection{Vector Meson Spectrum from Solution of  EOMs near $r=r_h$}

The $\alpha_n^i(Z)$ EOM, near the horizon, i.e., $Z=0 (Y=0)$, is of the form:
\begin{equation}
\label{alpha_n^i_EOM-i}
\alpha_n^i\ ^{\prime\prime}(Z) + \left(\frac{1}{|Z|} + \alpha_1\right)\alpha_n^i\ ^\prime(Z) + \left(\frac{\beta_2}{|Z|} + \alpha_2\right)\alpha_n^i(Z) = 0,
\end{equation}
whose solution is given by:
\begin{eqnarray}
\label{alpha_n^i_EOM-ii}
& & \alpha_n^{\left\{i\right\}}(Z) = c_1 e^{\frac{1}{2} |Z| \left(-\sqrt{\alpha_1^2-4 \alpha_2}-\alpha_1\right)} U\left(-\frac{-\alpha_1+2 \beta_2-\sqrt{\alpha_1^2-4 \alpha_2}}{2
   \sqrt{\alpha_1^2-4 \alpha_2}},1,\sqrt{\alpha_1^2-4 \alpha_2} |Z|\right)\nonumber\\
   & & +c_2 e^{\frac{1}{2} |Z| \left(-\sqrt{\alpha_1^2-4 \alpha_2}-\alpha_1\right)}
   L_{\frac{-\sqrt{\alpha_1^2-4 \alpha_2}-\alpha_1+2 \beta_2}{2 \sqrt{\alpha_1^2-4 \alpha_2}}}\left(|Z| \sqrt{\alpha_1^2-4 \alpha_2}\right).
\end{eqnarray}
One sets $c_2=0$ as satisfying the Neumann boundary condition for the associate Laguerre function will not be feasible. From (\ref{eoms_alpha_n_Bmu}), one notes that the differential operator is even under $Z\rightarrow-Z$ - relevant to parity and charge conjugation \cite{Sakai-Sugimoto-1} -  and therefore one can think of solutions that are even or odd under $Z\rightarrow-Z$. As one knows from \cite{Sakai-Sugimoto-1} that $\alpha^{\left\{i\right\}}_{2n}(-Z) = - \alpha^{\left\{i\right\}}_{2n}(Z)$ and $\alpha^{\left\{i\right\}}_{2n+1}(-Z) = \alpha^{\left\{i\right\}}_{2n+1}(Z)$, (\ref{alpha_n^i_EOM-ii}) ($c_2=0$) must be understood as:
\begin{eqnarray}
\label{vector-i_P_C_wavefunction}
& &\hskip -0.3in \alpha^{\left\{i\right\}}_n(Z) = \left(\delta_{n,2\mathbb{Z}^+}{\rm Sign}(Z) + \delta_{n,(2\mathbb{Z}^+\cup\left\{0\right\})+1}\right)e^{\frac{1}{2} |Z| \left(-\sqrt{\alpha_1^2-4 \alpha_2(n)}-\alpha_1\right)} \nonumber\\
 & & U\left(-\frac{-\alpha_1+2 \beta_2(n)-\sqrt{\alpha_1^2-4 \alpha_2(n)}}{2
   \sqrt{\alpha_1^2-4 \alpha_2(n)}},1,\sqrt{\alpha_1^2-4 \alpha_2(n)} |Z|\right).\nonumber\\
\end{eqnarray}

Setting $c_2=0$, one sees:
\begin{eqnarray}
\label{alpha_n^i_EOM-iii}
& & \alpha_n^{\left\{i\right\}}\ ^\prime(Z) = -\frac{1}{2} e^{-\frac{1}{2} |Z| \left(\sqrt{\alpha_1^2-4 \alpha_2}+\alpha_1\right)} \Biggl[\left(\sqrt{\alpha_1^2-4 \alpha_2}+\alpha_1\right)
   U\left(\frac{\alpha_1-2 \beta_2+\sqrt{\alpha_1^2-4 \alpha_2}}{2 \sqrt{\alpha_1^2-4 \alpha_2}},1,\sqrt{\alpha_1^2-4 \alpha_2}
   |Z|\right)\nonumber\\
   & & +\left(\sqrt{\alpha_1^2-4 \alpha_2}+\alpha_1-2 \beta_2\right) U\left(\frac{\alpha_1-2 \beta_2+3 \sqrt{\alpha_1^2-4 \alpha_2}}{2
   \sqrt{\alpha_1^2-4 \alpha_2}},2,\sqrt{\alpha_1^2-4 \alpha_2} |Z|\right)\Biggr]\nonumber\\
   & & = -\frac{1}{|Z| \Gamma \left(\frac{\alpha_1-2 \beta_2+\sqrt{\alpha_1^2-4 \alpha_2}}{2 \sqrt{\alpha_1^2-4 \alpha_2}}\right)}\nonumber\\
   & & +\frac{1}{2 \Gamma \left(\frac{\alpha_1-2 \beta_2+\sqrt{\alpha_1^2-4 \alpha_2}}{2 \sqrt{\alpha_1^2-4
   \alpha_2}}\right)}\Biggl\{\beta_2 \log
   \left(\alpha_1^2-4 \alpha_2\right)+\left(\sqrt{\alpha_1^2-4 \alpha_2}+\alpha_1\right) \psi ^{(0)}\left(\frac{\alpha_1-2 \beta_2+\sqrt{\alpha_1^2-4
   \alpha_2}}{2 \sqrt{\alpha_1^2-4 \alpha_2}}\right)\nonumber\\
   & & -\left(\sqrt{\alpha_1^2-4 \alpha_2}+\alpha_1-2 \beta_2\right) \psi ^{(0)}\left(\frac{\alpha_1-2
   \beta_2+3 \sqrt{\alpha_1^2-4 \alpha_2}}{2 \sqrt{\alpha_1^2-4 \alpha_2}}\right)\nonumber\\
   & & +2 \sqrt{\alpha_1^2-4 \alpha_2}+2 \alpha_1+2 \beta_2 {\log |Z|}-2
   \beta_2+4 \gamma  \beta_2\Biggr\}+ {\cal O}\left(|Z|\right)
\end{eqnarray}
One therefore sees that one can impose the Neumann/Dirichlet boundary condition $\alpha_n^i\ ^\prime({r = r_h})=0$ provided the following condition is imposed:
\begin{equation}
\label{alpha_n^i_EOM-iv}
\frac{\sqrt{\alpha_1^2-4 \alpha_2}+\alpha_1-2 \beta_2}{2 \sqrt{\alpha_1^2-4 \alpha_2}} = - n\in\mathbb{Z}^-.
\end{equation}
One can show that (\ref{alpha_n^i_EOM-iv}) in the context of the EOM (\ref{EOMs-i}), for $a = 0.6 r_h$ (\cite{EPJC-2,Sil_Yadav_Misra-EPJC-17}):
\begin{eqnarray}
\label{alphas+betas}
& & \alpha_1 = -1.08 -\frac{9. {g_s} M^2 (4.8 \log ({r_h})+4.8)}{{\log N} N}+\frac{1.5 M^2 (4.8 \log ({r_h})+4.8) (-18. {g_s} {N_f} \log
   ({r_h})+9 {g_s} {N_f}+75.398)}{N {N_f} \log ^2(N)}\nonumber\\
   & & +\frac{3. {g_s} M^2 (4.8 \log ({r_h})+4.8)}{N}+\frac{(-4.14
   {g_s} {N_f}-3.016) \log ({r_h})}{{g_s} {N_f} \log ^2(N)}+\frac{0.24}{{\log N}},\nonumber\\
& & \alpha_2 = \frac{{g_s} M^2 \tilde{m}^2 (7.2 \log ({r_h})+7.2)}{N}+0.54 \tilde{m}^2,\nonumber\\
& & \beta_2 = \frac{{g_s} M^2 \tilde{m}^2 (-3.6 \log ({r_h})-3.6)}{N}-0.02 \tilde{m}^2.\nonumber\\
\end{eqnarray}
Up to LO in $N$, NLO in $\log N$ (and assuming large $|\log r_h|$) LO in $\log r_h$, would yield the following meson spectrum:
\begin{eqnarray}
\label{meson-spectroscopy-i}
& & \tilde{m}_n^{\alpha^{\left\{i\right\}}_n} = 0.5 \sqrt{-10800. n^2-10800. n+10800. \sqrt{(n+0.36) (n+0.5) (n+0.5)
   (n+0.64)}-2592.}\nonumber\\
   & & +\frac{0.25 \left(\frac{23.04 (n+0.5)^2}{\sqrt{(n+0.36) (n+0.5) (n+0.5) (n+0.64)}}-24.\right)}{{\log N} \sqrt{-10800. n^2-10800.
   n+10800. \sqrt{(n+0.36) (n+0.5) (n+0.5) (n+0.64)}-2592.}}\nonumber\\
   & & + \frac{1}{{g_s} (\log N)^2 \sqrt{-10800. n^2-10800. n+10800. \sqrt{(n+0.36) (n+0.5) (n+0.5) (n+0.64)}-2592.}
   {N_f}}\nonumber\\
   & & \Biggl\{0.25 \log ({r_h}) \Biggl(\frac{1}{((n+0.36) (n+0.5) (n+0.5) (n+0.64))^{3/2}}\nonumber\\
   & & \times\Biggl\{n^6 (-397.44 {g_s} {N_f}-289.529)+n^5 (-1192.32 {g_s} {N_f}-868.588)+n^4 (-1482.61
   {g_s} {N_f}-1080.06)\nonumber\\
   & & +n^3 (-978.02 {g_s} {N_f}-712.473)+n^2 (-360.915 {g_s} {N_f}-262.921)+n (-70.6251 {g_s}
   {N_f}-51.4493)\nonumber\\
   & & -5.72314 {g_s} {N_f}-4.16922\Biggr\}+414. {g_s}
   {N_f}+301.593\Biggr)\Biggr\}+ {\cal O}\left(\frac{1}{(\log N)^3},\frac{1}{N}\right)\nonumber\\
   \end{eqnarray}
Disregarding $n=0$ (as it yields a ${\cal O}\left(\frac{1}{N^2}\right)$-suppressed though imaginary value) one sees:
\begin{eqnarray}
\label{meson-spectroscopy-ii}
& & \tilde{m}_{n=1}^{\alpha^{\left\{i\right\}}_0} = \frac{0.18 (414 {g_s} {N_f}+0.089 (-4487.65 {g_s} {N_f}-3269.19)+301.6) \log ({r_h})}{{g_s} {N_f} \log
   ^2(N)}-\frac{0.15}{{\log N}}+0.69,\nonumber\\
& & \tilde{m}_{n=2}^{\alpha^{\left\{i\right\}}_n} = \frac{1}{{g_s} {N_f} \log ^2(N)}\Biggl\{0.173 \log ({r_h}) (0.004 (-5.723 {g_s} {N_f}+16 (-1482.61 {g_s} {N_f}-1080.06)\nonumber\\
& & +32 (-1192.32 {g_s}
   {N_f}-868.588)+8 (-978.02 {g_s} {N_f}-712.473)+4 (-360.915 {g_s} {N_f}-262.92)\nonumber\\
   &&+2 (-70.625 {g_s} {N_f}-51.449)+64
   (-397.44 {gsNf}-289.529)-4.169)+414. {g_s} {N_f}+904.77)\Biggr\}\nonumber\\
 & &   -\frac{0.16}{{\log N}}+0.721.
\end{eqnarray}
Given that one is solving the EOM near the horizon, i.e., the IR, one expects the masses to be small, something verified by (\ref{meson-spectroscopy-ii}).

Now, the EOM (\ref{EOMs-ii}), near $r=r_h$,  can be written as:
\begin{eqnarray}
\label{alpha_n^0_EOM-ii}
& & \alpha_n^{\left\{0\right\}}\ ^{\prime\prime}(Z) + \alpha_1\alpha_n^{\left\{0\right\}}\ ^{\prime}(Z) + \left(\frac{\beta_2}{|Z|} + \alpha_2\right)\alpha_n^{\left\{0\right\}}(Z) = 0,
\end{eqnarray}
whose solution is given by:
\begin{eqnarray}
\label{alpha_n^0_EOM-iii}
& & \alpha_n^{\left\{0\right\}}(Z) = c_1 |Z| e^{-\frac{1}{2} |Z| \left(\sqrt{\alpha_1^2-4 \alpha_2(n)}+\alpha_1\right)}
   U\left(1-\frac{\beta_2}{\sqrt{\alpha_1^2-4 \alpha_2(n)}},2,\sqrt{\alpha_1^2-4 \alpha_2(n)} |Z|\right)\nonumber\\
& & + c_2 |Z| e^{-\frac{1}{2} |Z| \left(\sqrt{\alpha_1^2-4 \alpha_2(n)}+\alpha_1\right)} \, _1F_1\left(1-\frac{\beta_2(n)}{\sqrt{\alpha_1^2-4
   \alpha_2(n)}};2;\sqrt{\alpha_1^2-4 \alpha_2(n)} |Z|\right).
\end{eqnarray}
As $\frac{d}{dZ\ }\left[c_2 |Z| e^{-\frac{1}{2} |Z| \left(\sqrt{\alpha_1^2-4 \alpha_2}+\alpha_1\right)} \, _1F_1\left(1-\frac{\beta_2}{\sqrt{\alpha_1^2-4
   \alpha_2}};2;\sqrt{\alpha_1^2-4 \alpha_2} |Z|\right)\right]$ vanishes at $|Z|\rightarrow0^+$ only for $c_2=0$, one sets $c_2=0$ at the very outset.  Similar to (\ref{vector-i_P_C_wavefunction}):
   \begin{eqnarray}
\label{vector-ii_P_C_wavefunction}
& &\hskip -0.3in \alpha^{\left\{0\right\}}_n(Z) = \left(\delta_{n,2\mathbb{Z}^+}{\rm Sign}(Z) + \delta_{n,(2\mathbb{Z}^+\cup\left\{0\right\})+1}\right)e^{-\frac{1}{2} |Z| \left(\sqrt{\alpha_1^2-4 \alpha_2(n)}+\alpha_1\right)}
   U\left(1-\frac{\beta_2(n)}{\sqrt{\alpha_1^2-4 \alpha_2(n)}},2,\sqrt{\alpha_1^2-4 \alpha_2(n)} |Z|\right).\nonumber\\
\end{eqnarray}

   Now:
   \begin{eqnarray}
   \label{alpha_n^0_EOM-iv}
   & & \left.\alpha_n^{\left\{0\right\}}\ ^{\prime}(Z)\right|_{c_2=0} = -\frac{1}{2} c_1 e^{-\frac{1}{2} |Z| \left(\sqrt{\alpha_1^2-4 \alpha_2}+\alpha_1\right)}\nonumber\\
   & & \times \Biggl[\left(|Z| \sqrt{\alpha_1^2-4 \alpha_2}+\alpha_1 |Z|-2\right)
   U\left(1-\frac{\beta_2}{\sqrt{\alpha_1^2-4 \alpha_2}},2,\sqrt{\alpha_1^2-4 \alpha_2} |Z|\right)\nonumber\\
   & & +2 |Z| \left(\sqrt{\alpha_1^2-4 \alpha_2}-\beta_2\right)
   U\left(2-\frac{\beta_2}{\sqrt{\alpha_1^2-4 \alpha_2}},3,\sqrt{\alpha_1^2-4 \alpha_2} |Z|\right)\Biggr]\nonumber\\
   & & = \frac{c_1 \left(\beta_2 \log \left(\alpha_1^2-4 \alpha_2\right)+2 \beta_2 \psi ^{(0)}\left(1-\frac{\beta_2}{\sqrt{\alpha_1^2-4
   \alpha_2}}\right)+\sqrt{\alpha_1^2-4 \alpha_2}+\alpha_1+2 \beta_2 {\log |Z|}+4 \gamma  \beta_2\right)}{2 \beta_2 \Gamma
   \left(-\frac{\beta_2}{\sqrt{\alpha_1^2-4 \alpha_2}}\right)} + {\cal O}\left(|Z|\right).\nonumber\\
   & &
   \end{eqnarray}
   Hence, one can successfully impose Neumann/Dirichlet boundary condition at the horizon: $\alpha_n^{\left\{0\right\}}\ ^{\prime}({r = r_h})=0$ by demanding:
   \begin{equation}
   \label{Neumann_f0}
   \frac{\beta_2}{\sqrt{\alpha_1^2-4 \alpha_2}} = n \in \mathbb{Z}^+.
   \end{equation}
   In the context of (\ref{EOMs-ii}):
   \begin{eqnarray}
   \label{alphas+betas}
    \alpha_1 & = & \frac{-\frac{3.016}{{g_s} {N_f}}+0.72 \log ({r_h})-4.86}{{\log N}^2}+\frac{43.2 {g_s} M^2 {N_f} \log ({r_h})+43.2
   {g_s} M^2 {N_f}}{{\log N} N {N_f}}\nonumber\\
   & & +\frac{{g_s} M^2 (-14.4 \log ({r_h})-14.4)}{N}+\frac{0.24}{{\log N}}+0.92,\nonumber\\
    \alpha_2 & = & \frac{{g_s} M^2 \tilde{m}^2 (7.2 \log ({r_h})+7.2)}{N}+0.54 \tilde{m}^2,\nonumber\\
   \beta_2  & = & \frac{{g_s} M^2 \tilde{m}^2 (-3.6 \log ({r_h})-3.6)}{N}-0.02 \tilde{m}^2.
   \end{eqnarray}
   One sees that (\ref{Neumann_f0}) can be satisfied for only a single value of $ \tilde{m}_{n}^{\alpha^{\left\{0\right\}}_n}$ - which we declare to be the ground state (largely due to the proximity of its value to (\ref{meson-spectroscopy-i})) - which satisfies the following condition:
   \begin{eqnarray}
   \label{condition-single-mtilde}
   && \frac{1}{{g_s}^2
   {\log N}^4 {N_f}^2}\Biggl\{\Biggl({g_s}^2 {\log N} M^2 {N_f} (43.2-14.4 {\log N})\nonumber\\
   & & +{g_s} {N_f} \log ({r_h}) \left({g_s} {\log N} M^2
   (43.2-14.4 {\log N})+0.72 N\right)\nonumber\\
   & & +N (0.92 {g_s} ({\log N}-2.17166) ({\log N}+2.43253) {N_f}-3.016)\Biggr)^2\Biggr\}\nonumber\\
   & & +\tilde{m}^2 N \left(-28.8 {g_s} M^2 \log ({r_h})-28.8 {g_s} M^2-2.16 N\right)=0.\nonumber\\
   & &
   \end{eqnarray}
The solution when expanded in powers of $N$ and $\log N$
   \begin{eqnarray}
   \label{meson-spectroscopy-iii}
  & &   \tilde{m}_{n=0}^{\alpha^{\left\{0\right\}}_n} = \frac{0.163}{{\log N}} + 0.626 + \frac{-\frac{2.052}{{g_s} {N_f}}+0.49 \log ({r_h})-3.307}{{\log N}^2}
   \nonumber\\
   & & +\frac{{g_s} M^2 (28.305 \log
   ({r_h})+28.305)}{N{\log N} }+\frac{{g_s} M^2 (-13.971 \log ({r_h})-13.971)}{N} + ...
   \end{eqnarray}
  Now, from (\ref{meson-spectroscopy-ii}) and (\ref{meson-spectroscopy-iii}), disregarding
  ${\cal O}\left(\frac{\log r_h}{(\log N)^2}\right)$ terms, one sees that
\begin{equation}
\label{match-ground-states}
 \tilde{m}_{n=1}^{\alpha^{\left\{i\right\}}} = \tilde{m}_{n=0}^{\alpha^{\left\{0\right\}}},\ {\rm for}\ N=105.
\end{equation}
Hence, from (\ref{meson-spectroscopy-ii}) and (\ref{meson-spectroscopy-iii}), one sees an IR isospectrality in the spectra of $\alpha_{n=1}^i$ and $\alpha_{n=0}^0$ mesons.
This equation (\ref{meson-spectroscopy-iii}) beautifully captures the conformal ($N\rightarrow\infty$), the non-conformal
($N_f,M$-dependent) contributions as well as the temperature dependence via $\log r_h$ of vector mesons, and explicitly captures. Also, from both, (\ref{meson-spectroscopy-ii}) and (\ref{meson-spectroscopy-iii}), we see that the temperature dependence entering via $\log r_h$ does so at
${\cal O}\left(\frac{1}{N}\right)$.

\subsection{Vector Meson Spectrum from Conversion of $\alpha^{\left\{i\right\}}_n(Z)$'s EOM to Schr\"{o}dinger-like Equations}

The $\alpha^{\left\{i\right\}}_n$ EOM (\ref{EOMs-i}), written as $\alpha^{\left\{i\right\}}_n\ ^{\prime\prime}(Z) + A(Z)\alpha^{\left\{i\right\}}_n\ ^{\prime}(Z) + B(Z)\alpha^{\left\{i\right\}}_n(Z) = 0$, with a field redefinition:  $\psi_n^i(Z) = \sqrt{{\cal C}_1(Z)}\alpha_n^i(Z)$, is converted to:
\begin{equation}
\label{WKB-i-1}
\psi_n^{\left\{i\right\}}\ ^{\prime\prime}(Z) + V(\alpha^{\left\{i\right\}}_n)\psi_n^{\left\{i\right\}}(Z) = 0,
\end{equation}
where:
$V = \frac{{\cal C}_1^{\prime\prime}}{2{\cal C}_1} - \frac{1}{4}\left(\frac{{\cal C}_1^\prime}{{\cal C}_1}\right)^2 + B$. This potential for $\alpha^{\left\{i\right\}}_n(Z)$ can be easily worked out but due to the cumbersome nature of the expression so obtained, we will not be giving its analytical expression.

For $\alpha^{\left\{i\right\}}_n$ vector mesons,
\begin{eqnarray}
\label{C1_WKB}
& & {\cal C}_1 = -\frac{1}{2 {r_h}^2}\Biggl\{e^{-4 |Z|} \left(e^{4 |Z|}-1\right) \Biggl({g_s} {N_f} {\log N} \left(3 a^2+2 {r_h}^2 e^{2 |Z|}\right)-3 {g_s}
   {N_f} \log ({r_h}) \left(3 a^2+2 {r_h}^2 e^{2 |Z|}\right)\nonumber\\
   & & -9 a^2 {g_s} {N_f} |Z|-9 a^2 {g_s} {N_f}+12 \pi
   a^2-6 {g_s} {N_f} {r_h}^2 e^{2 |Z|} |Z|+8 \pi  {r_h}^2 e^{2 |Z|}\Biggr)\Biggr\}.
\end{eqnarray}

\subsubsection{IR}

The potential $V(\alpha^{\left\{i\right\}}_n)$, performing first a large-$N$ and then a small-$|Z|$ expansion, for $a=r_h
\left(0.6 + 4 \frac{g_s M^2}{N}\left(1 + \log r_h\right)\right)$ \cite{EPJC-2}, is given by:
\begin{eqnarray}
\label{V-WKB}
& & \hskip -0.4in V(\alpha^{\left\{i\right\}}_n) = \nonumber\\
& & \hskip -0.4in \frac{1}{\left(e^{4
   |Z|}-1.\right)^2}\Biggl\{e^{-2 |Z|} \left(e^{6 |Z|} \left(6.-1.08 \tilde{m}^2\right)+e^{4 |Z|} \left(2.16- \tilde{m}^2\right)+
   \tilde{m}^2 e^{8 |Z|}+\left(1.08 \tilde{m}^2-1\right) e^{2 |Z|}- e^{10 |Z|}-2.16\right)\Biggr\}
   \nonumber\\
   & & \hskip -0.4in +\frac{e^{-2 |Z|} \left({g_s}^3 {N_f}^3 \left(4.86-3. e^{2 |Z|}-3.24 e^{4 |Z|}-1.62 e^{8 |Z|}+3. e^{10
   |Z|}\right)\right)}{{g_s}^3 {\log N} {N_f}^3 \left(e^{4
   |Z|}-1.\right)^2} + {\cal O}\left(\frac{1}{{(\log N)^2}},\frac{g_sM^2}{N}\right).
\end{eqnarray}
In the IR, (\ref{V-WKB}) yields:
\begin{eqnarray}
\label{V-WKB-alphai-IR}
& & V(\alpha^{\left\{i\right\}}_n; IR) = \frac{-0.02 \tilde{m}^2-\frac{0.12}{{\log N}}+0.54}{|Z|}+0.54 \tilde{m}^2+\frac{4.86}{{\log N}}+\frac{0.25}{Z^2}-3.49333 + {\cal O}\left(|Z|,\frac{1}{{(\log N)^2}},\frac{g_sM^2}{N}\right).\nonumber\\
& &
\end{eqnarray}
The solution to (\ref{V-WKB}) is given in terms of Whittaker functions:
\begin{eqnarray}
\label{solution-Schrodinger}
& & \psi_n^{\left\{i\right\}}(Z) = c_1 M_{\frac{{\log N} \left(0.27-0.01 \tilde{m}^2\right)-0.06}{\sqrt{{\log N}} \sqrt{-0.54 {\log N}
   \tilde{m}^2+3.49333 {\log N}-4.86}},0}\left(\frac{2 \sqrt{-0.54 {\log N} \tilde{m}^2+3.49333
   {\log N}-4.86} |Z|}{\sqrt{{\log N}}}\right)\nonumber\\
   & & +c_2 W_{\frac{{\log N} \left(0.27-0.01
   \tilde{m}^2\right)-0.06}{\sqrt{{\log N}} \sqrt{-0.54 {\log N} \tilde{m}^2+3.49333
   {\log N}-4.86}},0}\left(\frac{2 \sqrt{-0.54 {\log N} \tilde{m}^2+3.49333 {\log N}-4.86} |Z|}{\sqrt{{\log N}}}\right).\nonumber\\
   & &
\end{eqnarray}
One can show that:
\begin{eqnarray}
\label{inf_der_horizon}
& & \left.\frac{d}{dZ\ }\left(\frac{M_{\frac{{\log N} \left(0.27-0.01 \tilde{m}^2\right)-0.06}{\sqrt{{\log N}} \sqrt{-0.54 {\log N}
   \tilde{m}^2+3.49333 {\log N}-4.86}},0}\left(\frac{2 \sqrt{-0.54 {\log N} \tilde{m}^2+3.49333
   {\log N}-4.86} |Z|}{\sqrt{{\log N}}}\right)}{\sqrt{C_1}}\right)\right|_{{r = r_h}}=0,\nonumber\\
   & &
\end{eqnarray}
implies
\begin{equation}
\label{alpha-Sch-IR}
\tilde{m} = \left(2.543-\frac{1.769}{{\log N}}\right) + {\cal O}\left(\left(\frac{1}{{\log N}}\right)^{3/2}\right)
\end{equation}
One can also show that
\begin{eqnarray}
\label{Neumann_horizon}
& & \left.\frac{d}{dZ\ }\left(\frac{W_{\frac{{\log N} \left(0.27-0.01 \tilde{m}^2\right)-0.06}{\sqrt{{\log N}} \sqrt{-0.54 {\log N}
   \tilde{m}^2+3.49333 {\log N}-4.86}},0}\left(\frac{2 \sqrt{-0.54 {\log N} \tilde{m}^2+3.49333 {\log N}-4.86}
   |Z|}{\sqrt{{\log N}}}\right)}{\sqrt{{C_1}}}\right)\right|_{{r = r_h}}=0\nonumber\\
   & &
\end{eqnarray}
implies:
\begin{equation}
\label{Neumann-Schrodinger}
\tilde{m}_n^{\alpha_n^i} = 0.5 \sqrt{-10800. n^2+10800. \sqrt{(n+0.376679) (n+0.623321) \left(n^2+n+0.25\right)}-10800. n-2592}.
   \end{equation}
 The $n=0$ result of (\ref{Neumann-Schrodinger}) - 2.479 - is close to the LO result in (\ref{alpha-Sch-IR}).  Once again, from considerations of parity and charge conjugation, similar to
 (\ref{vector-i_P_C_wavefunction}) and (\ref{vector-ii_P_C_wavefunction}),
{\footnotesize \begin{eqnarray}
 \label{alphai-wavefunction-Sch-IR}
 & & \hskip -0.5in \alpha^{\left\{i\right\}}_{n=0}(Z) ={\rm Sign}(Z) \frac{ {M\ {or}\ W}_{\frac{{\log N} \left(0.27-0.01 \tilde{m}^2\right)-0.06}{\sqrt{{\log N}} \sqrt{-0.54 {\log N}
   \tilde{m}^2+3.49333 {\log N}-4.86}},0}\left(\frac{2 \sqrt{-0.54 {\log N} \tilde{m}^2+3.49333
   {\log N}-4.86} |Z|}{\sqrt{{\log N}}}\right)}{\sqrt{{\cal C}_1(Z)}}.\nonumber\\
   & &
 \end{eqnarray}}

\subsubsection{UV}

Neglecting $N_f, M$-dependent terms in the potential in the UV (as both become very small), one obtains:
\begin{eqnarray}
\label{V_alphai_UV}
& & V(\alpha^{\left\{i\right\}}_n;UV)  \nonumber\\
& & = \frac{e^{-2 |Z|} \left(e^{6 |Z|} \left(6.-1.08 \tilde{m}^2\right)+e^{4 |Z|} \left(2.16-1\tilde{m}^2\right)+\tilde{m}^2
   e^{8 |Z|}+\left(1.08 \tilde{m}^2-1\right) e^{2 |Z|}- e^{10 |Z|}-2.16\right)}{\left(e^{4 |Z|}-1\right)^2}\nonumber\\
   & & = -1 + \left(2.16 + \tilde{m}^2\right)e^{-2 |Z|} + {\cal O}\left(e^{-4 |Z|}\right).
\end{eqnarray}
The solution to the Schr\"{o}dinger-like equation is:
\begin{eqnarray}
\label{solution-Sch-UV}
& & \psi_n^{\left\{i\right\}}(|Z|\in{\rm UV}) =\left(\delta_{n,2\mathbb{Z}^+}{\rm Sign}(Z) + \delta_{n,(2\mathbb{Z}^+\cup\left\{0\right\})+1}\right)\nonumber\\
& & \times \left[c_1 I_{1}\left(0.2i e^{-|Z|} \sqrt{25. \tilde{m}^2(n)+54.}\right)+c_2 K_{1}\left(0.2i e^{-|Z|}
   \sqrt{25 \tilde{m}^2(n)+54}\right)\right].\nonumber\\
   & &
\end{eqnarray}
One can show that the Neumann boundary condition:
\begin{equation}
\label{dI}
\lim_{Z\rightarrow\infty}\frac{d}{dZ\ }\left(\frac{c_1 I_{1}\left(0.2 i e^{-|Z|} \sqrt{25. \tilde{m}^2+54.}\right)}{\sqrt{{C_1}}}\right)=0
\end{equation}
as well as the Dirichlet boundary conditions are identically satisfied. This hence does not yield
values for $\tilde{m}$. Similarly, one can show that the Neumann boundary condition:
\begin{equation}
\label{dK}
\lim_{Z\rightarrow\infty}\frac{d}{dZ\ }\left(\frac{c_2 K_{1}\left(0.2 i e^{-|Z|} \sqrt{25. \tilde{m}^2+54.}\right)}{\sqrt{{C_1}}}\right)=0,
\end{equation}
but not Dirichlet boundary condition, is identically satisfied, therefore not providing values for $\tilde{m}$.

\subsection{Vector Meson Spectrum from Conversion of $\alpha^{\left\{0\right\}}_n(Z)$'s EOM to Schr\"{o}dinger-like Equations}

\subsubsection{IR}

One can show that:
\begin{eqnarray}
\label{V-alpha0-IR}
& & V(\alpha^{\left\{0\right\}}_n;IR) = -1 + |Z| \left(\frac{3.24}{{\log N}}-0.526667 \tilde{m}^2\right)-\frac{0.02 \tilde{m}^2}{|Z|}+0.54 \tilde{m}^2-\frac{3.24 Z^2}{\log
   (N)}+\frac{1.38}{{\log N}} \nonumber\\
   & &  + {\cal O}\left(\frac{1}{(\log N)^2},\frac{g_s M^2}{N},Z^3\right).
\end{eqnarray}
The solution to:
\begin{equation}
\psi^0_n\ ^{\prime\prime} (Z) + \left(\frac{a_1}{|Z|} + b_1\right)\psi^0_n(Z) = 0,
\end{equation}
is given by:
\begin{eqnarray}
\label{solution-alpha0-IR-i}
& & \psi^0_n(Z) = c_2 |Z| e^{-\sqrt{-{b_1}} |Z|}\ _1F_1 \left(\frac{- {a_1}}{2 \sqrt{-{b_1}}}+1;2;2 \sqrt{-{b_1}} |Z|\right)+c_1 |Z| e^{-\sqrt{-{b_1}} |Z|} U\left(\frac{- {a_1}}{2 \sqrt{-{b_1}}}+1,2,2 \sqrt{-{b_1}}
   |Z|\right).\nonumber\\
   & &
\end{eqnarray}
One can show that near ${r = r_h}$:
\begin{eqnarray}
\label{dU}
& & \frac{d}{dZ\ }\left(\frac{c_1 |Z| e^{-\sqrt{-{b_1}}|Z|} U\left(\frac{- {a_1}}{2 \sqrt{-{b_1}}}+1,2,2 \sqrt{-{b_1}} |Z|\right)}{{\sqrt{C_1}}(Z)}\right)\nonumber\\
   & & = \frac{1}{2 {a_1} {C_1}(0)^{3/2} \left({a_1} \sqrt{-{b_1}}+2
   {b_1}\right) \Gamma \left(-\frac{{a_1}}{2 \sqrt{-{b_1}}}\right)}\nonumber\\
   & & \times\Biggl\{c_1 \Biggl(2 {a_1}^2 \sqrt{-{b_1}} {C_1}(0) \log |Z|+4 \gamma  {a_1}^2 \sqrt{-{b_1}} {C_1}(0)+{a_1}
   \sqrt{-{b_1}} {C_1}'(0)+4 {a_1} {b_1} {C_1}(0) \log |Z|\nonumber\\
   & & -2 {a_1} {b_1} {C_1}(0)+8 \gamma  {a_1} {b_1}
   {C_1}(0)+2 {a_1} {C_1}(0) \left({a_1} \sqrt{-{b_1}}+2 {b_1}\right) \log \left(2 \sqrt{-{b_1}}\right)
   \nonumber\\
   & & +2 {a_1}
   {C_1}(0) \left({a_1} \sqrt{-{b_1}}+2 {b_1}\right) \psi ^{(0)}\left(1-\frac{{a_1}}{2 \sqrt{-{b_1}}}\right)+2 {b_1}
   {C_1}'(0)+4 \sqrt{-{b_1}} {b_1} {C_1}(0)\Biggr)\Biggr\}+ {\cal O}(Z).\nonumber\\
& & \end{eqnarray}
Hence by  requiring:
\begin{equation}
\label{alpha0-IR-quantization}
\frac{a_1}{2\sqrt{-b_1}} = n \in\mathbb{Z}^+\cup\left\{0\right\},
\end{equation}
one can impose Neumann boundary condition at the horizon, ${r = r_h}$.
With:
\begin{eqnarray}
\label{a1+b1}
& & a_1 = -0.02 \tilde{m}^2\nonumber\\
& & b_1 =  -1 + 0.54 \tilde{m}^2+\frac{1.38}{{\log N}},
\end{eqnarray}
This yields:
\begin{eqnarray}
\label{m-alpha0-IR}
& & m^{\alpha^{\left\{0\right\}}_n}_n = 0.5 \sqrt{10800. \sqrt{n^4+0.001 n^2}-10800 n^2}-\frac{2.56 n^2}{{\log N} \sqrt{n^4+0.001 n^2} \sqrt{10800. \sqrt{n^4+0.001
   n^2}-10800 n^2}} \nonumber\\
   & & + {\cal O}\left(\frac{1}{(\log N)^2}\right)\nonumber\\
   & & = \left\{1.36059-\frac{0.938489}{{\log N}},1.36077-\frac{0.93885}{{\log N}},1.3608-\frac{0.938917}{{\log N}},1.36081-\frac{0.938941}{{\log N}},...\right\}.
\end{eqnarray}

\subsubsection{UV}

In the UV disregarding the $M$ and $N_f$ (as there is no net $D7$-brane and $D5$-brane charge in the UV in \cite{metrics} and therefore in their mirror in \cite{MQGP}):
\begin{eqnarray}
\label{V-UV-alpha-0}
& & V(\alpha^{\left\{0\right\}};UV) = \frac{e^{-2 |Z|} \left(3. e^{2 |Z|}-1.62\right)}{{\log N}}+\frac{\tilde{m}^2 e^{2 |Z|}}{e^{4 |Z|}-1.}-\frac{1.08 \tilde{m}^2}{e^{4 |Z|}-1.}-1\nonumber\\
& & = -1 + e^{-2 |Z|} \left(\tilde{m}^2-\frac{1.62}{{\log N}}\right)+\frac{3}{{\log N}} +
{\cal O}(e^{-4|Z|}).
\end{eqnarray}
The solution to:
\begin{equation}
\psi^{\left\{0\right\}}_n "(Z) + \left(A + B e^{-|Z|}\right)\psi^{\left\{0\right\}}_n (Z) = 0
\end{equation}
is given by:
\begin{equation}
\psi^{\left\{0\right\}}_n(Z) = c_1 J_{-i \sqrt{A}}\left(\sqrt{B} \sqrt{e^{-2 |Z|}}\right)+c_2 J_{i \sqrt{A}}\left(\sqrt{B} \sqrt{e^{-2 |Z|}}\right).
\end{equation}
One can show that $\psi^0_n(Z)$ does not satisfy the Dirichlet boundary condition in the UV but
the Neumann boundary condition:
\begin{equation}
\lim_{Z\rightarrow\infty}\frac{d}{dZ\ }\left(\frac{\psi^{\left\{0\right\}}_n(Z)}{\sqrt{{\cal C}_1(Z)}}\right)=0,
\end{equation}
is identically satisfied in the UV and hence one does not obtain any quantization condition on the masses $\tilde{m}$.

\subsection{$\alpha^{\left\{i\right\}}_n(Z)$ Meson Spectroscopy from WKB Quantization}

The potential  in the Schr\"{o}dinger-like EOM having converted the $\alpha^{\left\{i\right\}}_n(Z)$-EOM to the same, is given by (\ref{WKB-i-1}). To keep the calculations tractable, we first perform a large-$N$ expansion of the potential  and work up to LO in $N$, then expand $\sqrt{V(\alpha^{\left\{i\right\}}_n)}$ up to NLO in $\log N$.

Performing first a large-$N$ expansion, one obtains the following:
{
\begin{eqnarray}
\label{V_f_large_LogN}
& & \sqrt{V^{\alpha^{\left\{i\right\}}_n}(\tilde{m},N)} \nonumber\\
& & = \sqrt{\frac{e^{-2 |Z|} \left(e^{6 |Z|} \left(6.-1.08 \tilde{m}^2\right)+e^{4 |Z|} \left(2.16-
   \tilde{m}^2\right)+\tilde{m}^2 e^{8 |Z|}+\left(1.08 \tilde{m}^2-1.\right) e^{2 |Z|}-e^{10
   |Z|}-2.16\right)}{\left(e^{4 |Z|}-1.\right)^2}}\nonumber\\
   & & -\frac{0.75 \left(e^{4 |Z|}-1\right) \left(2. e^{2 |Z|}-1.08 e^{4 |Z|}+2. e^{6
   |Z|}-3.24 \right)}{{\log N} \left(-\tilde{m}^2 e^{8 |Z|}+\left(1-1.08
   \tilde{m}^2\right) e^{2 |Z|}+\left(\tilde{m}^2-2.16\right) e^{4 |Z|}+\left(1.08 \tilde{m}^2-6.\right) e^{6 |Z|}+e^{10
   |Z|}+2.16\right)}\nonumber\\
   & & \times\sqrt{\frac{e^{-2 |Z|} \left(e^{6 |Z|} \left(6.-1.08 \tilde{m}^2\right)+e^{4 |Z|} \left(2.16-1.
   \tilde{m}^2\right)+\tilde{m}^2 e^{8 |Z|}+\left(1.08 \tilde{m}^2-1.\right) e^{2 |Z|}-e^{10
   |Z|}-2.16\right)}{\left(e^{4 |Z|}-1\right)^2}} \nonumber\\
   & & + {\cal O}\left(\left(\frac{1}{{\log N}}\right)^2\right).
\end{eqnarray}}

\paragraph{(a) Large-$\tilde{m}$ expansion : UV regime, i.e., $r>0.6\sqrt{3}r_h$ or $|Z|>0.04$}

One notes from (\ref{V_f_large_LogN}) that $\sqrt{V}\in\mathbb{R}$ in the UV for large $\tilde{m}$:
\begin{eqnarray}
\label{range-Exp|Z|}
& & \hskip -0.7in \sqrt{0.5 \tilde{m}^2-0.1 \sqrt{25. \tilde{m}^4-108. \tilde{m}^2}}<e^{|Z|}<\sqrt{0.5 \tilde{m}^2+0.1 \sqrt{25.
   \tilde{m}^4-108. \tilde{m}^2}},
\end{eqnarray}
or
\begin{eqnarray}
\label{range-|Z|}
& & |Z|\in\left[\log\left(1.039 + \frac{0.561}{\tilde{m}^2} + {\cal O}\left(\frac{1}{\tilde{m}^3}\right)\right), \log\left(\tilde{m} - \frac{0.54}{\tilde{m}} + {\cal O}\left(\frac{1}{\tilde{m}^3}\right)\right)\right].
\end{eqnarray}
    Thus, after performing a large-$N$ expansion, followed by a large-$\tilde{m}$ expansion and then a large-$|Z|$ expansion, one obtains:
\begin{eqnarray}
\label{large-mtilde}
& & \sqrt{V^{\alpha^{\left\{i\right\}}_n}(\tilde{m},N)} = \left(e^{-|Z|} - 0.54 e^{-3|Z|}\right)\tilde{m} + \frac{1}{\tilde{m}}\left(-{0.5}e^{-|Z|} - {0.27}e^{-|Z|} + {2.03}e^{-3|Z|}\right) \nonumber\\
& &  + \frac{1}{\tilde{m}N}\left(1.5 e^{-|Z|} + 2.47 e^{-3|Z|}\right)  + {\cal O}\left(\frac{1}{\tilde{m}^2},\frac{1}{(\log N)^2},e^{-5|Z|}\right).
\end{eqnarray}
Finally, the WKB quantization condition:
\begin{eqnarray}
\label{WKB-alphai-1}
& & \int_{\log\left(1.039 + \frac{0.561}{\tilde{m}^2}\right)}^{\log\left(\tilde{m} - \frac{0.54}{\tilde{m}}\right)}\sqrt{V} = \left(n + \frac{1}{2}\right)\pi
\end{eqnarray}
up to ${\cal O}\left(\frac{1}{\log N}\right)$ obtains:
\begin{table}[h]
\begin{center}
\begin{tabular}{|c|c|c|c|c|}\hline
& (Pseudo-)Vector Meson Name & $J^{PC}$ & $m_{n>0}$ & PDG Mass (MeV) \\
&&&(units of $\frac{r_h}{\sqrt{4\pi g_s N}}$) & \\ \hline
$B^{(1)}_{\mu}$& $\rho[770]$ & $1^{++}$ & 7.649 - $\frac{1.759}{\log N}$ &775.49 \\ \hline
$B^{(2)}_{\mu}$ &$a_1[1260]$ & $1^{--}$ & 11.60 - $\frac{1.792}{\log N}$ & 1230 \\ \hline
$B^{(3)}_{\mu}$ &$\rho[1450]$ &$1^{++}$ & 15.535 - $\frac{1.81}{\log N}$ & 1465\\  \hline
$B^{(4)}_{\mu}$ & $a_1[1640]$ & $1^{--}$ & 19.462 - $\frac{1.821}{\log N}$ & 1647 \\ \hline
\end{tabular}
\end{center}
\caption{(Pseudo-)Vector Meson masses from WKB Quantization applied to $V(\alpha^{\left\{i\right\}}_n)$}
\end{table}

\paragraph{(b) Small-$\tilde{m}$ Expansion}

 We expand $\sqrt{V(\alpha^{\left\{i\right\}}_n)}$ up to ${\cal O}\left(\frac{1}{\log N},\tilde{m}^4\right)$. One can show that $\sqrt{V}\in\mathbb{R}$ for
$Z\in[0.01,0.47]$; given that $Z=0.0385$ corresponding to $r={\cal R}_{D5/\overline{D5}}$ - the $D5-\overline{D5}$ separation - we put the lower limit by hand as 0.01. One can show from the WKB quantization condition:
\begin{equation}
\label{WKB-quantization-alphain}
\int_{0.01}^{0.47}dZ\sqrt{V\left(\alpha^{\left\{i\right\}}_n;Z\right)} = \left(n + \frac{1}{2}\right)\pi,
\end{equation}
the following IR vector meson spectrum is generated:
{\scriptsize
\begin{eqnarray}
\label{mn_IR}
& & m_n (IR) = 0.5 \sqrt{\frac{3.036-0.1136 {\log N}}{0.068 {\log N}+56.946}+2. \sqrt{\frac{(0.854513 n-0.0765252) \log ^2(N)+(715.605 n-67.1225)
   {\log N}-134.138}{(0.068 {\log N}+56.946)^2}}}.\nonumber\\
& &
\end{eqnarray}}
Happily, the ground state is non-zero and  for $N=6000$, is $0.81$ - not that far off from the value  $0.694 - \frac{0.155}{\log N}$ in (\ref{meson-spectroscopy-ii}) obtained by solving the $\alpha^{\left\{i\right\}}_n(Z)$ equation of motion near $r=r_h$ or $Z=0$ - for $N=6000$ the same yields $0.677$.

\subsection{$\alpha^{\left\{0\right\}}_n$ Spectroscopy from WKB Quantization}

Writing $m=\tilde{m}\frac{r_h}{\sqrt{4 \pi g_s N}s},\ a = r_h\left(0.6 + 4 \frac{g_sM^2}{N}\left(1 + \log r_h\right)\right)$, one can obtain the Schr\"{o}dinger-like potential for $\alpha^{\left\{0\right\}}_n(Z)$ - due to its cumbersome form, we will not be giving the explicit form of its analytical expression.

After retaining terms up LO in $N$ in the potential, the square root of the Schr\"{o}dinger-like potential for $\alpha^{\left\{0\right\}}_n(Z)$ after a large-$(\log)N$ expansion yields:
\begin{eqnarray}
\label{sqrtValpha0n}
& & \sqrt{V^{\alpha^{\left\{0\right\}}_n}(|Z|,N,\tilde{m})} = \sqrt{\frac{\tilde{m}^2 e^{2 |Z|}}{e^{4 |Z|}-1}-\frac{1.08 \tilde{m}^2}{e^{4 |Z|}-1.}+0. e^{-2 |Z|}-1.}+\frac{e^{-2 |Z|} \left(1.5 e^{2
   |Z|}-0.81\right)}{{\log N} \sqrt{\frac{\tilde{m}^2 e^{2 |Z|}}{e^{4 |Z|}-1.}-\frac{1.08 \tilde{m}^2}{e^{4 |Z|}-1}-1.}} \nonumber\\
   & & + {\cal O}\left(\left(\frac{1}{{\log N}}\right)^2\right).
\end{eqnarray}

\subsubsection{Large-$\tilde{m}$ Expansion}

One can show that $V(\alpha^{\left\{0\right\}}_n)\in\mathbb{R}$ provided:
\begin{eqnarray}
\label{range_e^|Z|_alpha0}
& & \hskip -0.5in 0.5 \log \left(0.1 \left(5 \tilde{m}^2-\sqrt{25. \tilde{m}^4-108\tilde{m}^2+100}\right)\right)<|Z|<0.5 \log \left(0.1 \left(5
   \tilde{m}^2+\sqrt{25 \tilde{m}^4-108 \tilde{m}^2+100}\right)\right),\nonumber\\
   & &
\end{eqnarray}
or
\begin{eqnarray}
\label{range_|Z|_alpha0}
& & |Z|\in\left[0.0385 + {\cal O}\left(\frac{1}{\tilde{m}^2}\right),\log \tilde{m} - \frac{0.54}{\tilde{m}^2} + {\cal O}\left(\frac{1}{\tilde{m}^3}\right)\right]\nonumber\\
& & \approx \left[0.0385,\log \tilde{m}\right],
\end{eqnarray}
which will be the turning points for the WKB quantization condition implementation.

One obtains the following large-$\tilde{m}$ expansion from (\ref{sqrtValpha0n}):
\begin{eqnarray}
\label{sqrtVlargemtilde}
& &  \sqrt{V^{\alpha^{\left\{0\right\}}_n}(|Z|,N,\tilde{m})} = \tilde{m}\left(e^{-|Z|} - 0.54 e^{-3|Z|}\right) + \frac{1}{\tilde{m}}\left(-\frac{e^{-|Z|}}{2} - 0.27 e^{-|Z|} + 0.03 e^{-3|Z|}\right)\nonumber\\
& &  + \frac{1}{\tilde{m}\log N}\left(1.5 e^{|Z|}- 0.531 e^{-3|Z|}\right)  + {\cal O}\left(\frac{1}{(\log N)^2},\frac{1}{\tilde{m}^2},e^{-5|Z|}\right).
\end{eqnarray}
The WKB quantization condition:
\begin{equation}
\label{WKB-alpha0_largem}
\int_{0.0385}^{\log\tilde{m}} dZ\sqrt{V^{\alpha^{\left\{0\right\}}_n}(Z)} = \left(n + \frac{1}{2}\right)\pi
\end{equation}
yields a cubic of the form: $a + b \tilde{m} + \frac{c}{\tilde{m}} + \frac{d}{\tilde{m}^2} = g$ where:
\begin{eqnarray}
\label{abcdg}
& & a = -1.5 + \frac{1.5}{\log N}\nonumber\\
& & b = 0.802\nonumber\\
& & c = 0.269 - \frac{1.717}{\log N}\nonumber\\
& & d = 0.27\nonumber\\
& & g = \left(n + \frac{1}{2}\right)\pi.
\end{eqnarray}
The only real root up to ${\cal O}\left(\frac{1}{\log N}\right)$ yields the following vector meson spectrum (disregarding $n=0$ as it does not satisfy the large-$\tilde{m}$ assumption):
\begin{table}[h]
\begin{center}
\begin{tabular}{|c|c|c|c|c|}\hline
& (Pseudo-)Vector Meson Name & $J^{PC}$ & $m_{n>0}$ & PDG Mass (MeV) \\
&&&(units of $\frac{r_h}{\sqrt{4\pi g_s N}}$) & \\ \hline
$B^{(1)}_{\mu}$& $\rho[770]$ & $1^{++}$ & 7.698 - $\frac{1.604}{\log N}$ &775.49 \\ \hline
$B^{(2)}_{\mu}$ &$a_1[1260]$ & $1^{--}$ & 11.634 - $\frac{1.692}{\log N}$ & 1230 \\ \hline
$B^{(3)}_{\mu}$ &$\rho[1450]$ &$1^{++}$ & 15.56 - $\frac{1.736}{\log N}$ & 1465\\  \hline
$B^{(4)}_{\mu}$ & $a_1[1640]$ & $1^{--}$ & 19.483 - $\frac{1.762}{\log N}$ & 1647 \\ \hline
\end{tabular}
\end{center}
\caption{(Pseudo-)Vector Meson masses from WKB Quantization applied to $V(\alpha^{\left\{0\right\}}_n)$}
\end{table}

One hence notes a near isospectrality between the (pseudo-)vector meson spectra from Tables 1 and 2, and as will be seen in Table 4, upon comparison with PDG, it is the results of Table 2 that are slightly more closer to the PDG values than those of Table 1.

The WKB quantization does not work for $\alpha^{\left\{0\right\}}_n(Z)$ for small $\tilde{m}$ as it can be easily shown that in the large-$N$ limit there are no turning points of the potential.

\section{Scalar Meson Spectroscopy using a Black-Hole Background for All Temperatures}

Unlike vector meson spectroscopy, the scalar meson spectrum will be obtained by considering  fluctuation of the $D6$-brane world volume along $Y$ by switching off any $D6$-brane world-volume fluxes as in \cite{Dasgupta_et_al_Mesons}. Now, $Y\neq0$ and the $D6$-brane metric (\ref{pull-back}), using (\ref{YZ}) and the embedding:
\begin{equation}
\label{Y-embedding}
Y = Y(x^\mu,Z),
\end{equation}
 is therefore:
\begin{eqnarray}
\label{metric_IIA_pull_back}
 G^{\rm IIA}_{6\mu \nu}dx^{\mu}dx^{\nu} &=& G^{\rm IIA}_{\mu\nu}\left(1 + {\cal C}_1(x^\kappa,Z) G_{IIA}^{\rho\lambda}\partial_\rho{Y} \partial_\lambda{Y} \right)dx^\mu dx^\nu + {\cal C}_2(x^\kappa,Z,\dot{Y}) dZ^2 + {\cal C}_3(x^\kappa,Z,\dot{Y})dx^\mu dZ \partial_\mu Y
\nonumber\\
 & & G^{\rm IIA}_{\theta_{2}\theta_{2}}d\theta_{2}^{2}+G^{\rm IIA}_{\tilde{y}\tilde{y}}d\tilde{y}^{2},
\end{eqnarray}
where:
\begin{eqnarray}
\label{Cs}
& & {\cal C}_1(x^\kappa,Z) = {\cal A} Y^2 + {\cal B} Z^2,\nonumber\\
& & {\cal C}_2(x^\kappa,Z,\dot{Y}) = \left({\cal A}Y^2 + {\cal B}Z^2\right)\dot{Y}^2
 + \left({\cal A}Y^2 + {\cal B}Z^2\right) + 2 Y Z\left({\cal A} - {\cal B}\right)\dot{Y},\nonumber\\
 & & {\cal C}_3(x^\kappa,Z,\dot{Y}) = 2 \left({\cal A}Y^2 + {\cal B}Z^2\right)\dot{Y} + 2 Y Z \left({\cal A} - {\cal B}\right),
\end{eqnarray}
wherein:
\begin{eqnarray}
\label{AB}
& & {\cal A} = \frac{G^{\rm IIA}_{rr}r_h^2 e^{2\sqrt{Y^2 + Z^2}}}{\left(Y^2 + Z^2\right)},\nonumber\\
& & {\cal B} = \frac{G^{\rm IIA}_{\tilde{z}\tilde{z}}}{\left(Y^2 + Z^2\right)^2}.
\end{eqnarray}
  $B^{\rm IIA}_{NS-NS}$\cite{MQGP} in diagonal basis $(\theta_2,\tilde{x},\tilde{y},\tilde{z})$ is given by:
\begin{eqnarray}
\label{B-IIA}
B^{IIA} &=& B_{\theta_2\tilde{y}}d\theta_2\wedge d\tilde{y} + B_{\theta_2\tilde{z}}d\theta_2\wedge d\tilde{z}+ B_{\theta_2\tilde{x}}d\theta_{2}\wedge d\tilde{x}.
\end{eqnarray}
Thus, its pull-back on $D_6$ is given by:
\begin{eqnarray}
\label{i*B}
i^* B^{\rm IIA} &=& B_{\theta_2\tilde{y}}d\theta_2\wedge d\tilde{y} + {\cal C}_4(x^\kappa,Z,\dot{Y})dZ\wedge d\theta_2 + {\cal C}_5(x^\kappa,Z)\partial_\mu Y dx^\mu\wedge d\theta_2
\end{eqnarray}
where:
\begin{eqnarray}
\label{BIIA-Cs}
& &   {\cal C}_4(x^\kappa,Z,\dot{Y}) = \left(\frac{B^{\rm IIA}_{\theta_2\tilde{z}}}{Y^2 + Z^2}\right)
\left(\dot{Y} Z - Y\right),\nonumber\\
& & {\cal C}_5(x^\kappa,Z) = \left(\frac{B^{\rm IIA}_{\theta_2\tilde{z}}}{Y^2 + Z^2}\right)Z.
\end{eqnarray}
Now, $B_{\theta_2\tilde{z}}$ and $B_{\theta_2\tilde{y}}$ are as given in (\ref{B-IIA-diag-nondiag}).

Therefore:
\begin{equation}
\label{i*(g+B)_7x7-i}
i^*(G + B)^{\rm IIA} = \left(\begin{array}{cc} \mathbb{A}_{4\times4} & \mathbb{B}_{4\times3}\\
\mathbb{C}_{3\times4} & \mathbb{D}_{3\times3}\end{array} \right),
\end{equation}
where:
\begin{eqnarray}
& & \mathbb{A} = \left(\begin{array}{cccc} G^{\rm IIA}_{00}{\cal T} & 0 & 0 & 0 \\
0 & G^{\rm IIA}_{x^1x^1}{\cal T} & 0 & 0 \\
0 & 0 & G^{\rm IIA}_{x^2x^2}{\cal T}& 0 \\
0 & 0 & 0 & G^{\rm IIA}_{x^3x^3}{\cal T}\end{array}\right),\nonumber\\
& & {\cal T}=\left(1 + {\cal C}_1 G_{IIA}^{\rho\lambda}\partial_\rho{Y} \partial_\lambda{Y} \right),\nonumber\\
& & \mathbb{B}_{4\times3} = \left(\begin{array}{ccc} i^* G^{\rm IIA}_{x^0 Z} & i^* B^{\rm IIA}_{x^0\theta_2} &
0 \\
 i^* G^{\rm IIA}_{x^1 Z} & i^* B^{\rm IIA}_{x^1\theta_2} &
0 \\
 i^* G^{\rm IIA}_{x^2 Z} & i^* B^{\rm IIA}_{x^2\theta_2} &
0 \\
 i^* G^{\rm IIA}_{x^3 Z} & i^* B^{\rm IIA}_{x^3\theta_2} &
0
\end{array}\right),\nonumber\\
& & \mathbb{D}_{3\times3} = \left(\begin{array}{ccc} i^* G^{\rm IIA}_{Z Z} & i^* B^{\rm IIA}_{Z\theta_2} & 0
\nonumber\\
- i^* B^{\rm IIA}_{Z\theta_2}& i^* G^{\rm IIA}_{\theta_2 \theta_2} & i^* B^{\rm IIA}_{\theta_2\tilde{y}}\\
0 & -i^* B^{\rm IIA}_{\theta_2\tilde{y}} & i^* G^{\rm IIA}_{\tilde y \tilde y}
\end{array}\right).
\end{eqnarray}
Now, ${\rm det}\left(i^*(G + B)^{\rm IIA} \right) = {\rm det}\mathbb{A}{\rm det}\left(\mathbb{D} - \mathbb{C}\mathbb{A}^{-1}\mathbb{B}\right)$, and retaining terms in the following up to\\
${\cal O}\left(Y^2,\dot{Y}^2,\partial_\mu Y\partial_\nu Y\right)$ (indicated by a tilde  below), one obtains:
\begin{eqnarray}
\label{DBI-quad-Ys}
& & \sqrt{{\rm det}\mathbb{A}} \sim \sqrt{-G^{\rm IIA}}_{\mathbb{R}^{1,3}}\left(1 + \frac{{\cal C}_1(Y=0)}{2}G^{\mu\nu}_{\rm IIA}\partial_\mu Y\partial_\nu Y\right),\nonumber\\
& & {\rm det}\left(\mathbb{D} - \mathbb{C}\mathbb{A}^{-1}\mathbb{B}\right) \sim \dot{Y}^2\Omega_1 + Y^2\Omega_2 + \Omega_3 G_{\rm IIA}^{\mu\nu}\partial_\mu Y\partial_\nu Y + \Omega_4,
\nonumber\\
& & {\rm implying}:\nonumber\\
& & \sqrt{{\rm det}\left(\mathbb{D} - \mathbb{C}\mathbb{A}^{-1}\mathbb{B}\right)}\sim
\sqrt{\Omega_4}\left(1 + \frac{\Omega_1}{\Omega_4}\frac{\dot{Y}^2}{2} + \frac{\Omega_2}{\Omega_4}\frac{\dot{Y}^2}{2} + \frac{\Omega_3}{\Omega_4}\frac{G_{\rm IIA}^{\mu\nu}}{2}\partial_\mu Y\partial_\nu Y\right);\ \cdot\equiv\frac{d}{dZ}.
\end{eqnarray}
Finally, one thus obtains the following DBI action for $N_f$ $D6$-branes (setting the tension to unity):
\begin{eqnarray}
\label{DBI-scalar}
& & \hskip -0.9in S_{D6}\nonumber\\
& &  \hskip -0.9in = N_f\left.\int d^4x dZ d\theta_2 d\tilde{y} \delta\left(\theta_2 - \frac{\alpha_{\theta_2}}{N^{\frac{3}{10}}}\right)e^{-\phi^{\rm IIA}}\sqrt{\Omega_4}\sqrt{-G^{\rm IIA}}_{\mathbb{R}^{1,3}}\left[1
+ \frac{{\cal C}_1  + \frac{\Omega_3}{\Omega_4}}{2}G_{\rm IIA}^{\mu\nu}\partial_\mu Y\partial_\nu Y  + \frac{\Omega_1}{\Omega_4}\frac{\dot{Y}^2}{2} + \frac{\Omega_2}{\Omega_4}\frac{\dot{Y}^2}{2}\right]\right|_{\theta_1=\frac{\alpha_{\theta_1}}{N^{\frac{1}{5}}},\tilde{x}=0} \nonumber\\
& & \hskip -0.9in = N_f\left.\int d^4x dZ d\theta_2 d\tilde{y}\delta\left(\theta_2 - \frac{\alpha_{\theta_2}}{N^{\frac{3}{10}}}\right)\left[{\cal S}_1(Z)G_{\rm IIA}^{\mu\nu}\partial_\mu Y\partial_\nu Y  + {\cal S}_2(Z)\dot{Y}^2 + {\cal S}_3(Z)Y^2 \right]\right|_{\theta_1=\frac{\alpha_{\theta_1}}{N^{\frac{1}{5}}},\tilde{x}=0},
\end{eqnarray}
where ${\cal S}_{1,2,3}$ are defined in (\ref{S123}).

Now, similar to \cite{Dasgupta_et_al_Mesons}, we make the KK ansatz:
\begin{equation}
\label{KK-Y}
Y(x^\mu,Z) = \sum_{n=1} {\cal Y}^{(n)}(x^\mu){\cal Z}_n(Z),
\end{equation}
  together with the following identifications, normalization and EOM:
\begin{eqnarray}
\label{mass-term-identification+EOM}
& & \int dZ\left({\cal S}_2(Z)\dot{\cal Z}_m(Z)\dot{\cal Z}_n(Z) + {\cal S}_3(Z){\cal Z}_m(Z){\cal Z}_n(Z)\right) = \frac{m_n^2}{2}\delta_{mn},\nonumber\\
& & \int dZ {\cal S}_1(Z) {\cal Z}_m(Z) {\cal Z}_n(Z) = \frac{1}{2}\delta_{mn};\nonumber\\
& & -\partial_Z\left({\cal S}_2(Z)\partial_Z{\cal Z}_n\right) + {\cal S}_3(Z){\cal Z}_n(Z)
 = {\cal S}_1(Z)m_n^2{\cal Z}_n.
\end{eqnarray}
Making a field redefinition: ${\cal Z}_n(Z) = |Z| {\cal G}_n(Z)$, one obtains the following EOM for
${\cal G}(Z)$:
{\footnotesize
\begin{eqnarray}
\label{G_EOM}
& &\hskip -0.8in{\cal G}_n''(Z) + {\cal G}_n'(Z)\frac{1}{(2 {g_s} {N_f} {\log N}-6 {g_s} {N_f} (\log ({r_h})+|Z|)+8 \pi )^2}\nonumber\\
& & \hskip -0.8in\times\Biggl\{ \Biggl(\frac{2 (2 {g_s} {N_f}
   {\log N}-6 {g_s} {N_f} (\log ({r_h})+|Z|)+8 \pi ) \left(4 {g_s} {N_f} e^{4 |Z|} {\log N}-12 {g_s} {N_f} e^{4 |Z|} (\log
   ({r_h})+|Z|)-3 {g_s} {N_f} e^{4 |Z|}+3 {g_s} {N_f}+16 \pi  e^{4 |Z|}\right)}{e^{4 |Z|}-1}
 \nonumber\\
& & \hskip -0.8in   -3 e^{-2 |Z|} \left(\frac{4 {g_s} M^2 \log
   ({r_h})}{N}+\frac{4 {g_s} M^2}{N}+0.6\right)^2\nonumber\\
   & & \hskip -0.8in\times \Biggl[-24 {g_s}^2 {N_f}^2 {\log N} (\log ({r_h})+|Z|)+4 {g_s}^2 {N_f}^2
   \log ^2(N)-12 {g_s}^2 {N_f}^2 {\log N}+36 {g_s}^2 {N_f}^2 (\log ({r_h})+|Z|)^2+36 {g_s}^2 {N_f}^2 (\log
   ({r_h})+|Z|)\nonumber\\
   & & \hskip -0.8in+18 {g_s}^2 {N_f}^2+32 \pi  {g_s} {N_f} {\log N}-96 \pi  {g_s} {N_f} (\log ({r_h})+|Z|)-48 \pi  {g_s}
   {N_f}+64 \pi ^2\Biggr]\Biggr)\Biggr\}\nonumber\\
   & &  \hskip -0.8in+ {\cal G}_n(Z)\frac{\tilde{m}^2  \left(\alpha_{\theta_1}^22 \sqrt[5]{N}-\alpha_{\theta_2}^2\right) \left(e^{2 |Z|}-\frac{3. \left(4. {g_s} M^2 \log ({r_h})+4.
   {g_s} M^2+0.6 N\right)^2}{N^2}\right)}{\alpha_{\theta_1}^22 \sqrt[5]{N} \left(e^{4 |Z|}-1\right)} = 0.\nonumber\\
   & &
\end{eqnarray}}

Analogous to obtaining the (pesudo-)vector meson spectrum in Section {\bf 4}, we will now proceed to obtaining the (pseudo-)scalar meson spectrum by three routes. The first will cater exclusively to an IR computation where we solve the ${\cal G}_n(Z)$  EOM near the horizon. Imposing Neumann boundary condition at the horizon results in quantization of the (pseudo-)scalar meson masses.  The second route will be to convert the ${\cal G}_n(Z)$ EOM into Schr\"{o}dinger-like EOM and to solve the same in the IR and UV separately and obtain (pseudo-)scalar mass quantization by imposing Neumann boundary conditions at the horizon (IR)/asymptotic boundary (UV). It turns out the former yields a result, which up to LO in $N$, is of the same order as the IR results of route one. The UV computations satisfy Neumann and/or Dirichlet boundary conditions without any mass quantization condition. The third route catering to the IR-UV interpolating region and what gives us our main results that are directly compared with PDG results, is obtaining the (pseudo-)scalar meson masses via WKB quantization condition.

\subsection{Scalar Meson Spectrum from Solution to EOM near ${r = r_h}$}

Analogous to (\ref{alpha_n^i_EOM-i}) - (\ref{alpha_n^i_EOM-iv}), one can rewrite (\ref{G_EOM}) and solve the same near ${r = r_h}$, impose Neumann boundary condition at ${r = r_h}$ with the following identifications:
\begin{eqnarray}
\label{alphas+betas}
& & \alpha_1 = 0.92+\frac{0.24}{{\log N}},\nonumber\\
& & \alpha_2 = \frac{0.02 \alpha_{\theta_2}^2 \tilde{m}^2}{\alpha_{\theta_1}^22 \sqrt[5]{N}}+\frac{{g_s} M^2 \tilde{m}^2 (-3.6 \log ({r_h})-3.6)}{N}-0.02
   \tilde{m}^2,\nonumber\\
& & \beta_2 = -\frac{0.54 \alpha_{\theta_2}^2 \tilde{m}^2}{\alpha_{\theta_1}^22 \sqrt[5]{N}}+\frac{{g_s} M^2 \tilde{m}^2 (7.2 \log ({r_h})+7.2)}{N}+0.54
   \tilde{m}^2.
\end{eqnarray}
The analog of (\ref{alpha_n^i_EOM-iv}) for scalar mesons up to ${\cal O}\left(\frac{1}{\log N}\right)$:
yields:
\begin{eqnarray}
\label{mn-Dirichlet}
& & \tilde{m}_n = 0.5 \sqrt{0.548697 n^2+0.548697 \sqrt{\left(n^2+n+0.25\right) \left(n^2+n+166.926\right)}+0.548697 n+3.54458}\nonumber\\
& & +\frac{0.25 \left(\frac{22.9689
   (n+0.5)^2}{\sqrt{\left(n^2+n+0.25\right) \left(n^2+n+166.926\right)}}+0.888889\right)}{{\log N} \sqrt{0.548697 n^2+0.548697
   \sqrt{\left(n^2+n+0.25\right) \left(n^2+n+166.926\right)}+0.548697 n+3.54458}} \nonumber \\ & & + {\cal O}\left(\frac{1}{(\log N)^2}\right).
\end{eqnarray}
The lightest scalar meson masses are:
\begin{table}[h]
\begin{center}
\begin{tabular}{|c|c|}\hline
$m_{n=1}$ & 1.331 - $\frac{0.167}{\log N}$\\ \hline
$m_{n=2}$ & 1.958 - $\frac{0.226}{\log N}$ \\ \hline
\end{tabular}
\end{center}
\caption{The lightest Sector Meson masses}
\end{table}

Our result implies that $\frac{m_{n=1}^2}{m_{n=0}^2} = 2.16$ if one disregards the ${\cal O}\left(\frac{1}{N}\right)$ corrections.
On comparison with the PDG table for scalar meson masses, if one assumes that the lightest scalar mesons are : $f0[980]/a0[980], f0[1370]$ then their mass-squared ratio is 1.95 - not too far from our result.

\subsection{Scalar Mass Spectrum from Solution of the Schr\"{o}dinger-Like Equation}

\subsubsection{IR}

In the IR, one can show that the potential in a Schr\"{o}dinger-like potential, simplifies to:
\begin{eqnarray}
\label{V_IR_scalar}
& & V(IR) = \frac{\frac{-0.36 {g_s}^2 {N_f}^2 \log ({r_h})+2.43 {g_s}^2 {N_f}^2+1.50796 {g_s} {N_f}}{{g_s}^2 {\log N}^2
   {N_f}^2}-\frac{0.12}{{\log N}}-0.02 \tilde{m}^2-0.46}{|Z|}\nonumber\\
   & & +\frac{13.86 {g_s}^4 {N_f}^4 \log ({r_h})+{g_s}^2
   {N_f}^2 \left(-2.97 {g_s}^2 {N_f}^2-58.0566 {g_s} {N_f}\right)}{{g_s}^4 {\log N}^2 {N_f}^4}+0.54
   \tilde{m}^2+\frac{0.25}{Z^2}-3.413.\nonumber\\
& & \end{eqnarray}
The solution to the Schr\"{o}dinger-like equation: $\Phi^{\prime\prime}_n(Z) + V(IR)(Z)\Phi_n(Z) = 0$, where $\Phi_n(Z) = \sqrt{{\cal C}_1}{\cal G}_n(Z)$, and $V(IR)(Z) = \frac{c_1}{Z^2}+\frac{a_1}{|Z|}+b_1$ with:
\begin{eqnarray}
\label{c1a1b1}
& & c_1 = 0.25,\nonumber\\,
& & a_1 = \frac{-0.36 {g_s}^2 {N_f}^2 \log ({r_h})+2.43 {g_s}^2 {N_f}^2+1.50796 {g_s} {N_f}}{{g_s}^2 {N_f}^2 \log
   ^2(N)}-0.02 \tilde{m}^2-\frac{0.12}{\log (N)}-0.46,\nonumber\\
   & & b_1 = 0.54 \tilde{m}^2 - 3.413,
\end{eqnarray}
is given by:
\begin{eqnarray}
\label{solution_Sch_IR}
& & \Phi_n(Z) = c_1 M_{\frac{{g_s} \left(\left(-0.01 \tilde{m}^2(n)-0.23\right) {\log N}^2-0.06 {\log N}+1.215\right) {N_f}-0.18 {g_s}
   \log ({r_h}) {N_f}+0.753982}{{g_s} {\log N}^2 \sqrt{3.41333-0.54 \tilde{m}^2(n)} {N_f}},0}\left(2 \sqrt{3.41333-0.54
   \tilde{m}^2(n)} |Z|\right)\nonumber\\
   & & +c_2 W_{\frac{{g_s} \left(\left(-0.01 \tilde{m}^2(n)-0.23\right) {\log N}^2-0.06 {\log N}+1.215\right)
   {N_f}-0.18 {g_s} \log ({r_h}) {N_f}+0.753982}{{g_s} {\log N}^2 \sqrt{3.41333-0.54 \tilde{m}^2(n)}
   {N_f}},0}\left(2 \sqrt{3.41333-0.54 \tilde{m}^2(n)} |Z|\right).
\end{eqnarray}
Now,
\begin{eqnarray}
\label{C1_scalar}
& & {\cal C}_1(Z) = -\frac{1}{2 {r_h}^2}\Biggl\{e^{-2 |Z|} \left(e^{4 |Z|}-1\right)\nonumber\\
& & \times \Biggl(2 {g_s} {N_f} \log (N) \left(3 {r_h}^2 \left(\frac{4 {g_s} M^2 \log
   ({r_h})}{N}+\frac{4 {g_s} M^2}{N}+0.6\right)^2+2 {r_h}^2 e^{2 |Z|}\right)\nonumber\\
   & & -6 {g_s} {N_f} (\log ({r_h})+|Z|) \left(3
   {r_h}^2 \left(\frac{4 {g_s} M^2 \log ({r_h})}{N}+\frac{4 {g_s} M^2}{N}+0.6\right)^2+2 {r_h}^2 e^{2 |Z|}\right)\nonumber\\
   & & +2
   \left({r_h}^2 (12 \pi -9 {g_s} {N_f}) \left(\frac{4 {g_s} M^2 \log ({r_h})}{N}+\frac{4 {g_s} M^2}{N}+0.6\right)^2+8
   \pi  {r_h}^2 e^{2 |Z|}\right)\Biggr)\Biggr\}.
\end{eqnarray}
One can show that the Neumann boundary condition is identically satisfied by ${\cal Z}_n(Z) = |Z| {\cal G}_n(Z) = |Z| \frac{\Phi_n(Z)}{\sqrt{{\cal C}_1(Z)}}$ and is hence uninteresting, and will therefore not be implemented. We will implement Neumann boundary condition on ${\cal G}_n(Z) =  \frac{\Phi_n(Z)}{\sqrt{{\cal C}_1(Z)}}$. One sees that:
\begin{eqnarray}
\label{DMoversqrtC1}
& & \left.\frac{d}{dZ}\left(\frac{M_{\frac{{g_s} \left(\left(-0.01 \tilde{m}^2-0.23\right) {\log N}^2-0.06 {\log N}+1.215\right) {N_f}-0.18 {g_s}
   \log ({r_h}) {N_f}+0.753982}{{g_s} {\log N}^2 \sqrt{3.41333-0.54 \tilde{m}^2} {N_f}},0}\left(2 \sqrt{3.41333-0.54
   \tilde{m}^2} |Z|\right)}{\sqrt{{\cal C}_1(Z)}}\right)\right|_{|Z|\sim0}\nonumber\\
   & & = \frac{1}{N^2}\left(0.54 \tilde{m}^2-3.41333\right)^{\frac{1}{4}}\left(\omega_1(g_s, M, N_f)+ \omega_2(g_s, M, N_f; \log r_h)\tilde{m}^2\right) + {\cal O}(Z).
\end{eqnarray}
Hence, either:
\begin{equation}
\label{mtilde-i-Sch-scalar-IR}
0.54 \tilde{m}^2-3.41333 = 0,\ {\rm implying}\  \tilde{m} = 2.514,
\end{equation}
or
\begin{eqnarray}
\label{mtilde-ii-Sch-scalar-IR}
& & \omega_1(g_s, M, N_f)+ \omega_2(g_s, M, N_f; \log r_h)\tilde{m}^2 = 0,\ {\rm implying}\nonumber\\
& & \tilde{m} = 3.07694 +\frac{\frac{95.6605}{{g_s} {N_f}}-22.8373 \log ({r_h})+3.34479}{\log ^2(N)}-\frac{7.61242}{\log (N)}+ {\cal O}\left(\frac{1}{(\log N)^2}\right).
\end{eqnarray}

One can show that:
\begin{eqnarray}
\label{DWoversqrtC1}
& & \left.\frac{W_{-\frac{i {a_1}}{2 \sqrt{{b_1}}},0}\left(2 i \sqrt{{b_1}} |Z|\right)}{\sqrt{{\cal C}_1(Z)}}\right|_{|Z|\sim0}
\nonumber\\
& & = \frac{\frac{2 i \sqrt{{b_1}} \left(\psi ^{(0)}\left(\frac{i
   {a_1}}{2 \sqrt{{b_1}}}-\frac{1}{2}\right)+\log \left(2 i
   \sqrt{{b_1}}\right)+\log |Z|+2 \gamma \right)}{\Gamma
   \left(\frac{i {a_1}}{2
   \sqrt{{b_1}}}-\frac{1}{2}\right)}+\frac{{a_1} \left(\psi
   ^{(0)}\left(\frac{i {a_1}}{2
   \sqrt{{b_1}}}+\frac{1}{2}\right)+\log \left(2 i
   \sqrt{{b_1}}\right)+\log |Z|+2 \gamma \right)}{\Gamma
   \left(\frac{i {a_1}}{2
   \sqrt{{b_1}}}+\frac{1}{2}\right)}}{\sqrt{2} \sqrt{i
   \sqrt{{b_1}}} \sqrt{{\cal C}_1(0)} \sqrt{|Z|}}\nonumber\\
   & & +\frac{\sqrt{|Z|}
   \left(\frac{2 {b_1} {C1}'(0) \log |Z|}{\Gamma \left(\frac{i
   {a_1}}{2 \sqrt{{b_1}}}-\frac{1}{2}\right)}-\frac{2 {b_1}
   {C1}'(0) \log |Z|}{\Gamma \left(\frac{i {a_1}}{2
   \sqrt{{b_1}}}+\frac{1}{2}\right)}-\frac{i {a_1}
   \sqrt{{b_1}} {C1}'(0) \log |Z|}{\Gamma \left(\frac{i
   {a_1}}{2 \sqrt{{b_1}}}+\frac{1}{2}\right)}\right)}{2 \sqrt{2}
   \left(i \sqrt{{b_1}}\right)^{3/2}
   {\cal C}_1(0)^{3/2}}+O\left(Z^{3/2}\right).
\end{eqnarray}
One therefore notices that one can satisfy the Neumann boundary condition at ${r = r_h}$ if:
\begin{equation}
\label{Neumann_scalar_W_IR}
\frac{1}{2} -\frac{i {a_1}}{2 \sqrt{{b_1}}} = - n\in\mathbb{Z}^-\cup\left\{0\right\},
\end{equation}
which yields the following quantization condition on $\tilde{m}$:
{\footnotesize\begin{eqnarray}
\label{mtilde_W}
& & \hskip -0.8in m_n = 0.5 \sqrt{-10800. n^2+10800. \sqrt{\left(n^2+n+0.25\right) \left(n^2+n+0.271719\right)}-10800. n-2792}\nonumber\\
& & \hskip -0.8in + \frac{6. n^2-6. \sqrt{\left(n^2+n+0.25\right) \left(n^2+n+0.271719\right)}+6. n+1.5}{{\log N} \sqrt{\left(n^2+ n+0.25\right)
   \left(n^2+n+0.271719\right)} \sqrt{-10800. n^2+10800. \sqrt{\left(n^2+n+0.25\right) \left(n^2+n+0.271719\right)}-10800.
   n-2792.}}\nonumber\\
   & & \hskip -0.8in+ {\cal O}\left(\frac{1}{(\log N)^2}\right).
\end{eqnarray}}s
One sees that the $n=0$ value $2.39 - \frac{0.51}{\log N}$ is close to $2.51$ of (\ref{mtilde-i-Sch-scalar-IR}), and not too far from
$3.08 - \frac{7.61}{\log N}$ of (\ref{mtilde-ii-Sch-scalar-IR}).

\subsubsection{UV}

In the UV, we will assume $M$ and $N_f$ to be quite small and hence approximate the potential by:
{\begin{eqnarray}
\label{V_scalar_UV}
& & \hskip -0.8in V(UV) =  = \frac{e^{-2 |Z|} \left(e^{6 |Z|} \left(8-1.08 \tilde{m}^2\right)+e^{4 |Z|} \left(- \tilde{m}^2\right)+1.08 \tilde{m}^2 e^{2
   |Z|}+\left(\tilde{m}^2+1.08\right) e^{8 |Z|}-4 e^{10 |Z|}-1.08\right)}{\left(e^{4 |Z|}-1\right)^2}
   \nonumber\\
   & & \hskip -0.8in +\frac{\alpha_{\theta_2}^2 \tilde{m}^2
   \left(e^{2 |Z|}+1.08 e^{4 |Z|}- e^{6 |Z|}-1.08\right)}{\alpha_{\theta_1}^2 \sqrt[5]{N} \left(e^{4 |Z|}-1\right)^2}.
\end{eqnarray}}
The solution to the Schr\"{o}dinger-like equation $\Phi^{\prime\prime}(Z) + V(Z)\Phi(Z) = 0$ is given by:
\begin{eqnarray}
\label{solution_UV_scalar}
& & \Phi(Z) = c_1 J_{-i \sqrt{{x_2}}}\left(e^{|Z|} \sqrt{{x_1}}\right)+c_2 J_{i \sqrt{{x_2}}}\left(e^{|Z|} \sqrt{{x_1}}\right),
\end{eqnarray}
where:
\begin{eqnarray}
\label{x1-x2}
& & x_1 = 1.08 + \tilde{m}^2,\nonumber\\
& & x_2 = 8 - 1.08 \tilde{m}^2.
\end{eqnarray}
One can show that in the UV:
\begin{eqnarray}
\label{Neumann_UV_scalar}
& & \frac{d}{dZ\ }\left(\frac{J_{\pm i \sqrt{{x_2}}}\left(e^{|Z|}\sqrt{{x_1}}\right)}{\sqrt{{\cal C}_1(Z)}}\right)\sim \cos\left(e^{|Z|}\sqrt{x_1}\right)\times {\cal O}\left(e^{-\frac{3|Z|}{2}}\right),
\end{eqnarray}
which tells us that the Neumann boundary condition is identically satisfied and one does not obtain any mass quantization condition in the UV from this approach.

\subsection{Scalar Mass Spectrum via WKB Quantization Condition}

For the purpose of simplification of this calculation, we will be disregarding $N_f$ and $M$ because one can show that the WKB quantization condition integral fails to converge for
IR-valued $\tilde{m}$ and in the UV, $M$ and $N_f$ are very small. One will hence work in the large-$\tilde{m}$/UV limit. From (\ref{V_scalar_UV}), one thus obtains:
{\begin{eqnarray}
\label{Sqrt[V]_scalar}
 & & \sqrt{V}  =  \tilde{m} \sqrt{\frac{1.08-e^{2 |Z|}-1.08 e^{4 |Z|}+e^{6 |Z|}}{1.-2e^{4 |Z|}+e^{8 |Z|}}}\nonumber\\
 & & +\frac{0.5 \left(8e^{6 |Z|}+1.08 e^{8 |Z|}-4e^{10
   |Z|}-1.08\right) \sqrt{\frac{1.08-e^{2 |Z|}-1.08 e^{4 |Z|}+e^{6 |Z|}}{1.-2e^{4 |Z|}+e^{8 |Z|}}}}{\tilde{m} \left(1.08 e^{2 |Z|}-e^{4
   |Z|}-1.08 e^{6 |Z|}+e^{8 |Z|}\right)}\nonumber\\
   & & + \frac{0.5 \alpha_{\theta_2}^2 \tilde{m} \left(e^{2 |Z|}+1.08 e^{4 |Z|}-e^{6 |Z|}-1.08\right)}{\alpha_{\theta_1}^2 \sqrt[5]{N} \left(e^{4 |Z|}-1.\right)^2
   \sqrt{\frac{1.08-e^{2 |Z|}-1.08 e^{4 |Z|}+e^{6 |Z|}}{1.-2e^{4 |Z|}+e^{8 |Z|}}}} + {\cal O}\left(\frac{1}{\tilde{m}N^{\frac{1}{5}}},\frac{1}{\tilde{m}^2}\right).
\end{eqnarray}}
One sees that $\sqrt{V}\in\mathbb{R}$ for $|Z|\in\left[0.0385, 0.5\log\left(0.54 + \frac{1.013\times 10^9}{\tilde{m}^2}\right)\right]$ which we will approximate as $|Z|\in\left[0.0385, 0.5\log(1.013\times 10^9)\approx10.368\right]$. The WKB quantization condition:
\begin{equation}
\label{WKB-scalar}
\int_{0.0385}^{10.368}d Z\sqrt{V(Z)} = \left(n + \frac{1}{2}\right)\pi,
\end{equation}
yields the following scalar meson spectrum:
\begin{eqnarray}
\label{scalar-meson-WKB-i}
& & \hskip -0.3in \tilde{m}_n = \frac{\alpha_{\theta_1}^2 (157.08 n+78.5398)+0.5 \sqrt{\alpha_{\theta_1}^4 \left(98696n^2+98696 n+2.08747\times 10^9\right)-1.04372\times 10^9 \alpha_{\theta_1}^2
   \alpha_{\theta_2}^2 \sqrt[5]{N}}}{82\alpha_{\theta_1}^2-41\alpha_{\theta_2}^2 \sqrt[5]{N}}.
   \nonumber\\
   & &
\end{eqnarray}
One can argue that $Y(x^{\mu},Z)$ is even under parity: $(x^{1,2,3},Z)\rightarrow (-x^{1,2,3},-Z)$. The idea is the following. The type IIB setup of \cite{metrics} includes
$D3$-branes with world-volume coordinates $x^{0,1,2,3}$ and $D7$-branes with world-volume coordinates $(x^{0,1,2,3},r,\tilde{x},\theta_1,\tilde{z})$\footnote{There are also $D5$-branes with world volume coordinates $(x^{0,1,2,3},\theta_1,\tilde{x})$ and $\overline{D5}$-branes with
world volume coordinates $(x^{0,1,2,3},\theta_1,\tilde{x})$ which, relative to the $D5$-branes are at the antipodal point of the resolved $S^2_a(\theta_2,\phi_2)$; their bound state however is equivalent to producing a net $D3$-brane charge provided a certain topological condition is satisfied (See \cite{EPJC-2} and references therein).}, which after three T-dualities along $\tilde{x},\tilde{y},\tilde{z}$ yield two sets of $D6$-branes, one set with world-volume coordinates $(x^{0,1,2,3},\tilde{x},\tilde{y},\tilde{z})$ (obtained from a triple T-dual of the $D3$-branes) and the other set with world-volume coordinates $(x^{0,1,2,3},r,\theta_2,\tilde{y})$ (obtained from a triple T-dual of the $D7$-branes). One hence sees that the two sets of $D6$-branes are separated in $r$
or $Y$. In the type IIB setup of \cite{metrics}, the flavor $D7$-branes never touch the $D3$-branes which in the SYZ or triple T-dual picture implies that the two sets of $D6$-branes never touch each other. This, like \cite{Sakai-Sugimoto-1}, \cite{Dasgupta_et_al_Mesons}, implies one can construct a $C_5\sim Y dx^0\wedge dx^1\wedge dx^2\wedge dx^3\wedge dZ$  which vanishes precisely when the two sets of $D6$-branes touch. From this $C_5$, one can construct a Chern-Simons action: $\int_{{w.v. of }D6}F_2\wedge C_5$ where $F_2=dA_1$ correponds to a gauge field on the $D6$-brane world volume. If one demands the Cherns-Simons action be invariant under parity - which includes $Z\rightarrow-Z$ - given that $F_2$ is even, one sees that $Y$ is even under parity. Similarly, under charge conjugation - which includes $Z\rightarrow - Z$ - and noting that $F_2$ is charge-conjugation odd implies that $Y$ is charge-conjugation even. From (\ref{KK-Y}), under 5D parity, ${\cal Z}_n(-Z) = (-)^{n+1}{\cal Z}_n(Z),
{\cal Y}^{(n)}(-x^{\mu}) = (-)^{n+1}{\cal Y}^{(n)}(x^{\mu}),  n\in\mathbb{Z}^+$ \cite{Sakai-Sugimoto-1}.

We assume that the three lightest scalar mesons from the PDG  are $f0[980]/a0[980], f0[1370]$ and $f0[1450]$. We could choose $\alpha_{\theta_1}$ and $\alpha_{\theta_2}$ to match $\frac{m_{n=3}}{m_{n=1}}$ with PDG exactly (this is not normalizing our $\frac{m_{n=3}}{m_{n=1}}$ result to match PDG values)! This is effected by imposing the following condition on
$\alpha_{\theta_1}, \alpha_{\theta_2}$:
\begin{eqnarray}
\label{match_Dasgupta_et_al}
\frac{m_{n=3}}{m_{n=1}} = \frac{549.7787 \alpha_{\theta_1}^2+0.5 \sqrt{2.08767\times 10^9 \alpha_{\theta_1}^4-1.04372\times 10^9 \alpha_{\theta_1}^2 \alpha_{\theta_2}^2 \sqrt[5]{N}}}{235.6194 \alpha
   ^2+0.5 \sqrt{2.08747\times 10^9 \alpha_{\theta_1}^4-1.04372\times 10^9 \alpha_{\theta_1}^2 \alpha_{\theta_2}^2 \sqrt[5]{N}}} = \frac{1350}{980},
\end{eqnarray}
which is satisfied by $\alpha_{\theta_1} = 0.70765N^{\frac{1}{10}}\alpha_{\theta_2}$.  Having done so, the ratio $\frac{m_{n=5}}{m_{n=1}}$ is not too far off of the PDG value - see Table 8 in Section {\bf 7}! The (pseudo-)scalar meson ($0^{--}$)$0^{++}$ masses are listed in Table 4 below. (The entries against $0^{--}$ are blanks as there are, as of now, no known candidates with this $J, P, C$ assignment.)
\begin{table}[h]
\begin{center}
\begin{tabular}{|c|c|c|c|c|}\hline
& (Pseudo-)Scalar Meson Name & $J^{PC}$ & $m_{n>0}$ & PDG Mean Mass (MeV) \\
&&&(units of $\frac{r_h}{\sqrt{4\pi g_s N}}$) & \\ \hline
${\cal Y}^{(1)}$& $f0[980]/a0[980$ & $0^{++}$ & 9207.44 & 980 \\ \hline
${\cal Y}^{(2)}$ & -- & $0^{--}$ & 10861.9 & -- \\ \hline
${\cal Y}^{(3)}$ &f0[1370] &$0^{++}$ & 12683.7 & 1350\\  \hline
${\cal Y}^{(4)}$ & -- & $0^{--}$ & 14640.8 & -- \\ \hline
${\cal Y}^{(5)}$ & f0[1450] & $0^{++}$ & 16704 & 1474 \\ \hline
\end{tabular}
\end{center}
\caption{(Pseudo-)Scalar Meson masses from WKB Quantization}
\end{table}

\section{Meson Spectroscopy in a Thermal Background and Near Isospectrality with Black-Hole Background}

In this section we show an interesting near isospectrality of the (pseudo-) vector meson spectrum (in {\bf 6.1} and its sub-sub-sections) and (pseudo-)scalar meson spectrum (in {\bf 6.2} and its sub-sub-sections) obtained using a thermal background which is valid for low temperatures, with the corresponding results of sections {\bf 4} and {\bf 5} obtained using a black-hole background (expected to be valid/stable at high temperatures) for all temperatures.

As the techniques are similar   to and in fact simpler than the ones used in sections {\bf 4} and
{\bf 5}, we will only be presenting the main results to substantiate our claim.

\subsection{Vector Meson Spectroscopy in a Thermal Background}

\subsubsection{Solving the EOM near an IR cut-off $r=r_0$}

Writing $r = r_0 e^{\sqrt{Y^2+Z^2}}$ - $r_0$ being an IR cut-off \footnote{This is not actually a parameter put in by hand. In the spirit of a top-down approach, one can show that a Hawking-Page transition occurs at a temperature at an $r_0$ given in terms of  $r_h$ and an ${\cal O}(1)$ constant of proportionality relating the modulus of the Ouyang embedding parameter corresponding to the holomorphic embedding of type IIB flavor branes to $r_0$  - see \cite{EPJC-2}.} - and defining $m = \tilde{m}\frac{r_0}{\sqrt{4\pi g_s N}}$, setting $r_h=0$ and introducing a bare resolution parameter $a = \gamma r_0$ (to ensure that ${\cal R}_{D5/\overline{D5}}=\sqrt{3}a\neq0$), one can show that the $\alpha_N(Z)$ EOM - there is no need to attach a superscript to $\alpha_N$ anymore as $r_h=0$ - near the horizon can be written in the form:
\begin{equation}
\label{alphaEOM-rh=0}
\alpha_n^{\prime\prime}(Z) + (a_1 + b_1 |Z|)\alpha_n^\prime(Z) + (a_2 + b_2 |Z|)\alpha_n(Z) = 0,
\end{equation}
where up to ${\cal O}\left(\frac{1}{\log N}\right)$:
\begin{eqnarray}
\label{a1-b1-c1}
& & a_1 = 2 - 3\gamma^2 - \frac{3}{\log N} + \frac{9\gamma^2}{\log N},\nonumber\\
& & b_1 = 6\gamma^2 - 18 \frac{\gamma^2}{\log N},\nonumber\\
& & a_2 = \left(1 - 3\gamma^2\right)\tilde{m}^2,
\end{eqnarray}
whose solution is given by:
\begin{eqnarray}
\label{solution_EOM_alpha-rh=0}
& & \alpha_n(Z) = e^{-{a_1} |Z|+\frac{{b_2} |Z|}{{b_1}}-\frac{{b_1} Z^2}{2}} \Biggl(c_2 \left(\frac{{b_1}^3-{a_2} {b_1}^2+{a_1} {b_2}
   {b_1}-{b_2}^2}{2 {b_1}^3};\frac{1}{2};\frac{\left(|Z| {b_1}^2+{a_1} {b_1}-2 {b_2}\right)^2}{2 {b_1}^3}\right)\nonumber\\
   & & +c_1
   H_{\frac{-{a_1} {b_1} {b_2}+{a_2} {b_1}^2-{b_1}^3+{b_2}^2}{{b_1}^3}}\left(\frac{{a_1} {b_1}+{b_1}^2 |Z|-2
   {b_2}}{\sqrt{2} {b_1}^{3/2}}\right)\Biggr).
\end{eqnarray}
By imposing Neumann boundary condition at $r=r_0$ by assuming $c_2=0$, numerically, e.g., for $N=6000, \gamma=0.6$ (similar to $a(r_h\neq0) = \left(0.6 + {\cal O}\left(\frac{g_sM^2}{N}\right)\right)r_h $), one sees one obtains as a root - corresponding to the lightest vector meson - $\tilde{m}\approx 1.04$ - of the same order as the LO value in (\ref{meson-spectroscopy-ii}). One gets only the root $\tilde{m}=0$ if $c_1=0$.

\subsubsection{Schr\"{o}dinger-like EOM}

One can rewrite the EOM as a Schr\"{o}dinger-like equation in terms of $\psi_n(Z) = \sqrt{{\cal C}_1(Z)}\alpha_n(Z)$ where $\psi_n(Z)$ satisfies: $\psi^{\prime\prime}_n(Z) + V(Z)\psi_n(Z) = 0$,
where:
\begin{eqnarray}
\label{C1-alpha-rh=0}
&& {\cal C}_1(Z) = \frac{1}{2} \Biggl[3 {\gamma}^2 ({g_s} {N_f} (\log (N)-3 |Z|-3)+4 \pi )-3 {g_s} {N_f} \left(3 {\gamma}^2+2 e^{2 |Z|}\right)
   \log ({r_0})\nonumber\\
   & & +2 e^{2 |Z|} ({g_s} {N_f} (\log (N)-3 |Z|)+4 \pi )\Biggr],
\end{eqnarray}
and
\begin{equation}
\label{V_vectormesons_rh=0}
V(Z) = -1 + e^{-2|Z|}\tilde{m}^2 - 3 e^{-4|Z|}\gamma^2\tilde{m}^2 + \frac{3 - \frac{9}{2}e^{-2|Z|}\gamma^2}{\log N} + {\cal O}\left(\frac{1}{(\log N)^2}\right).
\end{equation}
Near $r=r_0$ - the IR - the EOM can written as $\psi_n^{\prime\prime}(Z) + (a + b|Z|)\psi_n(Z)$,
where:
\begin{eqnarray}
\label{abc}
& & a = - 1 + \frac{3}{\log N} - \frac{9 \gamma^2}{2\log N} + \tilde{m}^2(1 - 3 \gamma^2),\nonumber\\
& & b = \frac{9\gamma^2}{\log N} + \tilde{m}^2(12\gamma^2 - 2),
\end{eqnarray}
 the solution to which is given by:
\begin{equation}
\label{solution_psi-rh=0}
{\psi}(Z) = c_1 {Ai}\left(-\frac{a+b |Z|}{(-b)^{2/3}}\right)+c_2 {Bi}\left(-\frac{a+b |Z|}{(-b)^{2/3}}\right).
\end{equation}
In the large $|\log r_0|$-limit and setting $\gamma=0.6$, one can show that one obtains the value:
$\tilde{m} = 0.36$.

In the UV, $V(Z) = -1 + e^{-2|Z|}\tilde{m}^2 + {\cal O}\left(e^{-4|Z|}\right)$, and the solution to the EOM is given by $J_1\left(e^{-|Z|}\tilde{m}\right)$ and $Y_1\left(e^{-|Z|}\tilde{m}\right)$ which satisfy the
Neumann/Dirichlet boundary condition, identically, in the UV and do not provide $\tilde{m}$
quantization.

\subsubsection{WKB Quantization Condition}

In the UV, one can show that:
\begin{eqnarray}
\label{sqrtV-alpha-rh=0}
& & \sqrt{V(\alpha_n)} = \frac{3-\frac{9}{2} {\gamma}^2 e^{-2 |Z|}}{2 \log (N) \sqrt{-3 {\gamma}^2 {\tilde{m}}^2 e^{-4 |Z|}+{\tilde{m}}^2 e^{-2
   |Z|}-1}}+\sqrt{-3 {\gamma}^2 {\tilde{m}}^2 e^{-4 |Z|}+{\tilde{m}}^2 e^{-2 |Z|}-1}\nonumber\\
   & & +{\cal O}\left(\left(\frac{1}{{\log N}}\right)^2\right).
\end{eqnarray}
One can see that $\sqrt{V}\in\mathbb{R}$ for $|Z|\in\left[\log\left(\sqrt{3}\gamma + {\cal O}\left(\frac{1}{\tilde{m}^2}\right)\right),\log\left(\tilde{m} - \frac{3}{2\tilde{m}} + {\cal O}\left(\frac{1}{\tilde{m}^2}\right)\right)\right]$. One can then show that:
\begin{equation}
\label{WKB_alpha_rh=0}
\int_{\log\left(\sqrt{3}\gamma\right)}^{\log\left(\tilde{m} - \frac{3}{2\tilde{m}} \right)}\sqrt{V} = \left(n + \frac{1}{2}\right)\pi,
\end{equation}
yields:
\begin{eqnarray}
\label{mn-alpha-rh=0}
& & m_n^{\alpha_n} = \frac{3}{20} \sqrt{3} \left(\sqrt{2} \sqrt{{\gamma}^2 \left(2 (2 \pi  n+\pi )^2+12
   \pi  (2 n+1)+13\right)}+2 {\gamma} (2 \pi  n+\pi +3)\right)\nonumber\\
& & \frac{\sqrt{3} \left(\frac{{\gamma}^2 (7-36 \pi  (2 n+1))}{\sqrt{2} \sqrt{{\gamma}^2 \left(2 (2 \pi  n+\pi )^2+12 \pi  (2
   n+1)+13\right)}}-18 {\gamma}\right)}{20 \log (N)}+ {\cal O}\left(\left(\frac{1}{\log N}\right)^2\right).
   \end{eqnarray}
Hence, disregarding $n=0$, the following spectrum, {\it nearly isospectral with Table 1 gotten using a black-hole background}, is obtained:
\begin{table}[h]
\begin{center}
\begin{tabular}{|c|c|c|c|c|}\hline
& (Pseudo-)Vector Meson Name & $J^{PC}$ & $m_{n>0}$ & PDG Mass (MeV) \\
&&&(units of $\frac{r_0}{\sqrt{4\pi g_s N}}$) & \\ \hline
$B^{(1)}_{\mu}$& $\rho[770]$ & $1^{++}$ & 7.716 - $\frac{1.636}{\log N}$ &775.49 \\ \hline
$B^{(2)}_{\mu}$ &$a_1[1260]$ & $1^{--}$ & 11.644 - $\frac{1.714}{\log N}$ & 1230 \\ \hline
$B^{(3)}_{\mu}$ &$\rho[1450]$ &$1^{++}$ & 15.567 - $\frac{1.753}{\log N}$ & 1465\\  \hline
$B^{(4)}_{\mu}$ & $a_1[1640]$ & $1^{--}$ & 19.488 - $\frac{1.776}{\log N}$ & 1647 \\ \hline
\end{tabular}
\end{center}
\caption{(Pseudo-)Vector Meson masses from WKB Quantization applied to $V(\alpha^{\left\{0\right\}}_n)$}
\end{table}

\subsection{Scalar Meson Spectroscopy in a Thermal Background}

\subsubsection{Solving the EOM near an IR cut-off $r = r_0$}

The EOM for ${\cal G}_n(Z)$, near the horizon, is again of the form (\ref{alphaEOM-rh=0}) wherein:
\begin{eqnarray}
\label{a1_b1_a2_b2-scalar}
& & a_1 = 4 - 3\gamma^2 +\frac{(9\gamma^2 - 3)}{\log N} + \frac{(27\gamma^2 - 9)\log r_0}{(\log N)^2},\nonumber\\
& & b_1 = 6\gamma^2 - \frac{18\gamma^2}{\log N} - \frac{54\gamma^2\log r_0}{(\log N)^2},
\nonumber\\
& & a_2 = \tilde{m}^2(1 - 3 \gamma^2),\nonumber\\
& & b_2 = \tilde{m}^2(12\gamma^2 - 2).
\end{eqnarray}
Quite interestingly, this IR computation is able to resolve $f0[980](m_{f0[980]}=990 MeV)$ and $a0[980](m_{a0[980]}=980 MeV)$ because, for
$\gamma=0.6$, numerically one can show that the two smallest roots of the equation obtained by imposing Neumann boundary condition on ${|Z|\cal G}_n(r=r_0)$ by setting $c_2=0$ are:
$1.83$ and $1.94$ - the second in particular not far off of the results of Table 1 gotten using a black-hole gravity dual - and $\frac{1.94}{1.83} = 1.06$ and $\frac{m_{f0[980]}}{m_{a0[980]}}=1.01$ - very close indeed! A black-hole computation could not do so.

\subsubsection{Schr\"{o}dinger-like EOM}

With:
\begin{eqnarray}
\label{C1-scalar}
& & {\cal C}_1(Z) = e^{2 |Z|} \Biggl(3 {\gamma}^2 ({g_s} {N_f} (\log (N)-3 |Z|-3)+4 \pi )-3 {g_s} {N_f} \left(3
   {\gamma}^2+2 e^{2 |Z|}\right) \log ({r_0})\nonumber\\
   & & +2 e^{2 |Z|} ({g_s} {N_f} (\log (N)-3 |Z|)+4 \pi
   )\Biggr),
\end{eqnarray}
and the potential in the Schr\"{o}dinger-like EOM (analogous to (\ref{V_vectormesons_rh=0})) given by:
\begin{eqnarray}
\label{V-scalar-rh=0}
& & V({\cal G}_n) = -4 - 3 e^{-4|Z|}\gamma^2\tilde{m}^2 + e^{-2|Z|}(3\gamma^2 + \tilde{m}^2)
\nonumber\\
& & + \frac{6 - \frac{27}{2}e^{-2|Z|}\gamma^2}{\log N} + {\cal O}\left(\frac{1}{(\log N)^2}\right),
\end{eqnarray}
the $a, b$, analogous to (\ref{abc}), are given by:
\begin{eqnarray}
\label{ab-rh=0}
& & a = -4 + 3\gamma^2 + \tilde{m}^2(1 - 3 \gamma^2) + \frac{(6 - 27 \gamma^2)}{\log N},\nonumber\\
& & b = - 6 \gamma^2 + \tilde{m}^2(12\gamma^2 - 2) + \frac{27 \gamma^2}{\log N}.
\end{eqnarray}
One gets a solution analogous to (\ref{solution_psi-rh=0}); only for $c_2=0$ we can  show numerically that for $\gamma=0.6$, $\tilde{m}\approx 0.85$ - not too far off of the smallest root in {\bf 6.2.1} and about the same order as the results of {\bf 5.2.1} - one can approximately satisfy the Neumann boundary condition at $r=r_0$.

\subsubsection{WKB Quantization Condition}

Once again, as was assumed for the black-hole background computation, given that scalars are more typically more massive than vector mesons implying that we address the UV-IR interpolating/UV region in which $M, N_f$ are very small, one sees that:
\begin{eqnarray}
\label{sqrtV_scalar_rh=0}
& & \sqrt{V({\cal G}_n)} = \sqrt{-4 - 3 e^{-2|Z|}\gamma^2\tilde{m}^2 + e^{-2|Z|}\left(3\gamma^2 + \tilde{m}^2\right)} + {\cal O}\left(\frac{1}{N^{\frac{1}{5}}}\right).
\end{eqnarray}
One sees that $|Z|\in\left[\log\left(\sqrt{3}\gamma\right),\log\left(\frac{\tilde{m}}{2} - \frac{9\gamma^2}{4\tilde{m}^2}\right)\right]$, $\sqrt{V}\in\mathbb{R}$ and the WKB quantization condition:
\begin{equation}
\label{QKB_scalar_rh=0}
\int_{\log\left(\sqrt{3}\gamma\right)}^{\log\left(\frac{\tilde{m}}{2} - \frac{9\gamma^2}{4\tilde{m^2}}\right)}\sqrt{V\left({\cal G}_n(Z)\right)} = \left(n + \frac{1}{2}\right)\pi,
\end{equation}
yields:
\begin{equation}
\label{mn-scalar-rh=0}
m_n = \frac{1}{20} \sqrt{3} {\gamma} \left(6 (2 \pi  n+\pi +6)+\sqrt{2} \sqrt{18 (2 \pi  n+\pi )^2+216 \pi  (2
   n+1)+349}\right).
\end{equation}
Hence, disregarding $n=0$, the following spectrum is generated:
\begin{table}[h]
\begin{center}
\begin{tabular}{|c|c|c|c|c|}\hline
& (Pseudo-)Scalar Meson Name & $J^{PC}$ & $m_{n>0}$ & PDG Mean Mass (MeV) \\
&&&(units of $\frac{r_0}{\sqrt{4\pi g_s N}}$) & \\ \hline
${\cal Y}^{(1)}$& $f0[980]/a0[980]$ & $0^{++}$ & 15.745  & 980 \\ \hline
${\cal Y}^{(2)}$ & -- & $0^{--}$ & 22.359 & -- \\ \hline
${\cal Y}^{(3)}$ &f0[1370] &$0^{++}$ & 28.934 & 1350\\  \hline
\end{tabular}
\end{center}
\caption{(Pseudo-)Scalar Meson masses from WKB Quantization}
\end{table}
For a low-temperature thermal gravity dual, we do not trust values $n>3$ and hence have not quoted the same.

\section{Summary, New Insights into Thermal QCD and Future Directions}

A top-down finite-gauge-coupling finite-number-of-colors holographic thermal QCD calculation pertaining to meson spectroscopy \footnote{For glueball spectroscopy, see \cite{Sil_Yadav_Misra-EPJC-17}.}, has thus far, been missing in the literature. This paper fills in this gap. We should keep in mind that even though lattice QCD is a good tool to deal with IR Physics but it is hard to include fundamental fermions in the same. However, incorporation of fermions is easily taken care of in the top-down type IIB construct of \cite{metrics} and its type IIA mirror in \cite{MQGP}. In this paper, we have calculated (pseudo-)vector and (pseudo-)scalar meson spectra from the delocalized type IIA SYZ mirror (constructed in \cite{MQGP}) of the UV-complete top-down type IIB holographic dual of large-$N$ thermal QCD (constructed in \cite{metrics}), at finite coupling and with finite number of colors (part of the `MQGP' limit), and compared our results with \cite{Sakai-Sugimoto-1}, \cite{Dasgupta_et_al_Mesons} and \cite{PDG}. We first do a computation with a black hole background assuming the same to be valid for all temperatures, low and high (similar in spirit to the computations in \cite{BH_all_T}).   We then repeat the computation in a thermal background with no black hole which is valid for low temperatures. What we learn about QCD is that the mirror of \cite{metrics} when considered in the 'MQGP limit' - involving finite gauge/string coupling and finite $N_c = M$ (at the end of a Seiberg duality cascade) and not just a large t'Hooft coupling - can, almost without any fine tuning - generate the low-lying vector and scalar meson spectra from the massless string sector. An analytical finite-gauge-coupling computation in perturbative (thermal) QCD is very hard if not undoable. This however, is easily done as a classical supergravity computation in our setup.

\begin{itemize}
\item
{\bf Summary of New Results Obtained (Points 1. - 6.) and the New Insights Gained into Thermal QCD (Point 7.)}
\begin{enumerate}
\item
In tables 1 and 2, even if we  drop the ${\cal O}\left(\frac{1}{\log N}\right)$ terms in  the  vector meson masses (BH/thermal background) obtained by a WKB quantization condition, and assume $n=1,2,3,4$ to correspond respectively to $\rho[770], a_1[1260], \rho[1450], a_1[1640]$, then the following table compares mass ratios from our results at LO in $N$ (obtained from a WKB quantization condition) with those from \cite{Sakai-Sugimoto-1}, \cite{Dasgupta_et_al_Mesons} (up to first order in $\delta=\frac{g_sM^2}{N}<1$) and \cite{PDG}:

\begin{table}[h]
\begin{center}
\begin{tabular}{|c|c|c|c|c|c|c|}\hline
                      & ratio ($\alpha_{n}^{\left\{i\right\}}$)&ratio ($\alpha_{n}^{\left\{0\right\}}$)&ratio  &Sakai-Sugimoto & Best value & Exp. value PDG\\
                      &(BH)&(BH)&(Thermal)&{\scriptsize (as given in \cite{Dasgupta_et_al_Mesons})}& in \cite{Dasgupta_et_al_Mesons}:$\delta=0.5$ &{\scriptsize (as given in \cite{Dasgupta_et_al_Mesons})}\\ \hline
$\frac{m_{a_1[1260]}^{2}}{m_{\rho[770]}^{2}}$ & 2.30& 2.28 &2.28 & 2.32 &2.31& 2.52\\ \hline
$\frac{m_{\rho[1450]}^{2}}{m_{\rho[770]}^{2}}$ & 4.12 & 4.09 & 4.07& 4.22 &4.09& 3.57\\ \hline
$\frac{m_{a_1[1640]}^{2}}{m_{\rho[770]}^{2}}$ & 6.47 & 6.41& 6.38& 6.62 &5.93& 4.51\\ \hline
\end{tabular}\begin{center}
\end{center}
\end{center}
\caption{Comparison of  Mesons masses ratio }
\end{table}

The authors of \cite{Dasgupta_et_al_Mesons} obtain a variety of values by adjusting the values of  and working up to first order in $\delta$, as well as a constant appearing in a `squashing factor' in the metric. Their best values for (pseudo-)vector meson mass ratios are quoted in Table 5 column 5. But they need to do a lot of fine tuning, incorporate contributions to the results from the ${\cal O}(\delta)$ terms and choose $\delta=0.5$, which in fact can not justify disregarding terms of higher powers of $\delta$ as $\delta=0.5$ is not very small to warrant the same. Our results, specially coming from the WKB quantization condition applied to $V(\alpha^{\left\{0\right\}}_n)$ for the BH gravitational dual or $V(\alpha_n)$ for the thermal gravitational background, working even up to LO in $N$ without having to explicitly numerically compute the ${\cal O}(\delta)$ ($\delta\sim0.001$ for our calculations and thereby justifying dropping higher powers of $\delta$) contribution, display the following features:
\begin{itemize}
\item
our $m_{a_1[1260]}^{2}/m_{\rho[770]}^{2}$ is close to \cite{Sakai-Sugimoto-1} and\cite{Dasgupta_et_al_Mesons}, and not too far off of the PDG value

\item
our $m_{\rho[1450]}^{2}/m_{\rho[770]}^{2}$ is the same as (for BH background)/very close to (for thermal background) \cite{Dasgupta_et_al_Mesons} (but without any fine tuning and already at LO in $N$) - within $\approx$ 15$\%$ of the PDG value

\item
our $m_{a_1[1640]}^{2}/m_{\rho[770]}^{2}$ is closer to the PDG value than \cite{Sakai-Sugimoto-1}
\end{itemize}

\item
there is a near isospectrality between the lightest (pseudo-)vector meson masses calculated by a BH and thermal backgrounds

\item
The thermal background, to the order permissible by our analytical/numerical computations, does not provide a temperature dependence of $\tilde{m}$ at low temperatures - in agreement with one's expectations \cite{BH_all_T}. Encouraged by the aforementioned isospectrality, by solving the EOMs for the gauge field fluctuations along the $D6$ world volume in a BH gravitational dual, close to the horizon, we are able to capture the explicit temperature dependence of the lowest lying vector meson mass with the temperature dependence appearing at ${\cal O}\left(\frac{1}{(\log N)^2}, \frac{g_s M^2}{N}\right)$. The temperature-dependent meson mass $\tilde{m}$ will have the following form:
\begin{equation}
\label{rh_T_meson_mass}
\hskip -0.4in \tilde{m}_{\rm lightest} = \alpha + \frac{\beta}{\log N} + \frac{\left(\delta_1 + \delta_2 \log r_h\right)}{\left(\log N\right)^2} + \frac{g_sM^2\left(\kappa_1 + \kappa_2\log r_h\right)}{N} + {\cal O}\left(\frac{g_sM^2}{N}\frac{\log r_h}{\log N}\right),
\end{equation}
$\delta_2>0$.
The temperature, assuming the resolution to be larger than the deformation in the resolved warped deformed conifold in the type IIB background of \cite{metrics} in the MQGP limit, and utilizing the IR-valued warp factor $h(r,\theta_1\sim N^{-\frac{1}{5}},\theta_2\sim N^{-\frac{3}{10}})$, is \cite{EPJC-2}:
\begin{eqnarray}
\label{T}
& & \hskip -0.4in T = \frac{\partial_rG^{\cal M}_{00}}{4\pi\sqrt{G^{\cal M}_{00}G^{\cal M}_{rr}}}\nonumber\\
& & \hskip -0.4in = {r_h} \left[\frac{1}{2 \pi ^{3/2} \sqrt{{g_s} N}}-\frac{3 {g_s}^{\frac{3}{2}} M^2 {N_f} \log ({r_h}) \left(-\log
   {N}+12 \log ({r_h})+\frac{8 \pi}{g_s N_f} +6-\log (16)\right)}{64 \pi ^{7/2} N^{3/2}} \right]\nonumber\\
   & & \hskip -0.4in+ a^2 \left(\frac{3}{4 \pi ^{3/2} \sqrt{{g_s}} \sqrt{N} {r_h}}-\frac{9 {g_s}^{3/2} M^2 {N_f} \log ({r_h})
   \left(\frac{8 \pi }{{g_s} {N_f}}-\log (N)+12 \log ({r_h})+6-2 \log (4)\right)}{128 \pi ^{7/2} N^{3/2}
   {r_h}}\right).\nonumber\\
   & &
\end{eqnarray}
Using (\ref{T}) and the arguments of \cite{EPJC-2}, one can  invert (\ref{T}) and express $r_h$ in terms of $T$ \cite{Gale+Misra} in the MQGP limit. Assuming $\log r_h$ in (\ref{rh_T_meson_mass}) to be in fact
$\log\left(\frac{r_h}{\Lambda}\right), \Lambda>r_h$ being the scale at which confinement occurs, one sees that, as per expectations, the vector meson masses decrease with temperature
\cite{BH_all_T} with the same being large-$N$ suppressed \cite{Herzog_Tc} (and references therein).

\item
On comparing scalar meson mass ratios obtained from (\ref{scalar-meson-WKB-i}) using a black hole gravitational dual WKB quantization and PDG values, we obtain Table 8:
\begin{table}[h]
\begin{center}
\begin{tabular}{|c|c|}\hline
Our results & PDG values \\ \hline\hline
$\frac{m_{n=3}}{m_{n=1}}$ & $\frac{m_{{f0}[1370]}}{m_{{f0}[980]}}$\\ \hline
1.38 & 1.38 \\ \hline\hline
$\frac{m_{n=5}}{m_{n=1}}$ & $\frac{m_{{f0}[1450]}}{m_{{f0}[980]}}$\\ \hline
1.81 & 1.50 \\ \hline\hline
\end{tabular}
\end{center}
\caption{The lightest Scalar Meson mass ratios}
\end{table}

The agreement with the PDG values for the lightest three scalar meson candidates (if assumed to be $f0[980], f0[1370], f0[1450]$) is quite nice. We do not expect the agreement for more massive scalar mesons. This is for the following reason. As discussed in \cite{Imoto:2010ef,Dasgupta_et_al_Mesons}\footnote{One of us (AM) thanks K. Dasgupta for emphasizing this point to him.}  massive open string excitations can contribute to the  meson (specially scalar) masses (as scalar mesons are typically heavier than (pseudo-)vector mesons). We do not attempt to estimate the same as open string quantization in a curved RR background is a hard problem, and in this paper, like \cite{Dasgupta_et_al_Mesons}, we have assumed that the mesons arise from the massive KK modes of the massless open string sector. The difference between our results and the PDG results for the mass ratios of vector and scalar mesons, for heavier mesons, could be attributable to the contributions coming to meson masses from the massive  open string sector (which we have not calculated) in addition to the ones coming from the massless open string sector (which we have calculated in this paper).

\item
Using a thermal background, though on one hand, it appears to be possible to  in fact resolve $a0[980]$ (average mass of $980 MeV$) and $f0[980]$ (average mass $990 MeV$), on the other hand assuming $f0[980]/a0[980], f0[1370], f0[1450]$ to be the lightest scalar mesons,  a WKB quantization condition yields a mass ratio of the first two as 1.83, the mass ratio of $f0[1370]$ and $a0[980]$ being 1.38; as $f0[1370]$ is expected to have mass range of $1200 - 1500 MeV$ \cite{PDG}, with $1500 MeV$ the ratio - as per PDG values - increases to 1.53.

The thermal background is not able to correctly account for $f0[1470]$. The black hole background, as is evident from Table 8, does a good job though.

\item
The $0^{--}$ pseudo-scalar mesons in Table 4 do not, thus far, have corresponding entries in the PDG tables. This is one point of difference between our results and PDG tables.

\item
There are {\bf three main insights} we gain into thermal QCD by evaluating mesonic (vector/scalar) spectra and comparing with PDG values. \\
(a) The first is the realization that from a type IIA perspective, meson spectroscopy in the mirror of top-down holographic type IIB duals of large-$N$ thermal QCD at {\it finite coupling and number of colors} \footnote{In the IR, as explained in {\bf 2.1}, $N_c=M$ which can be ${\cal O}(1)$ in the MQGP limit of \cite{MQGP} - taken to be three in this paper - and not $N$.} (closer to a realistic QCD calculation) which are UV complete - we know of only \cite{metrics} that falls in this category for which the mirror was worked out in \cite{MQGP} - will give results closer to PDG values rather than a single T-dual of the same. Even though obtaining the mirror requires a lot of work, but once obtained, one can obtain very good agreement with PDG tables already at ${\cal O}\left(\left(\frac{g_sM^2}{N}\right)^0\right)$ (for vector mesons, {\it without any fine tuning}). This is a major lesson we learn from our computation. There are two major reasons for the same.
\begin{itemize}
\item
 As noted in section {\bf 3}, the mirror of \cite{metrics} picks up sub-dominant terms in $N$ of ${\cal O}(N^0)$ in $B^{\rm IIA}$ which are therefore bigger than the ${\cal O}(\frac{g_sM^2}{N})$ contributions, and were missed in \cite{Dasgupta_et_al_Mesons}. This is the reason why the authors of \cite{Dasgupta_et_al_Mesons} had to set $\frac{g_sM^2}{N}=0.5$ - not a small enough fraction to warrant disregarding ${\cal O}\left(\left(\frac{g_sM^2}{N}\right)^2\right)$ contributions which they did - to obtain a reasonable match with \cite{PDG}.

\item
 In the context of \cite{metrics}, this is expected to be related to the following \footnote{One of us (AM) thanks K. Dasgupta for discussion on this point.}. The brane construct of \cite{metrics} involves $N\ D3$-branes, $M\ D5$-branes wrapping the vanishing $S^2$, $M\ \overline{D5}$-branes wrapping the same $S^2$ but placed at the antipodal points of the resolved $S^2(a)$ relative to the $D3, D5$-branes, $N_f$ flavor $D7$-branes wrapping an $S^3$ and radially extending all the way into the IR starting from the UV and $N_f\ \overline{D7}$-branes wrapping the same $S^3$ but going only up to the IR-UV interpolating region. The mirror of this results in $D6$ branes in a deformed conifold. Now, the gravity dual of the above picture - which is what we work with - involves a resolved warped deformed conifold with a black hole and $\overline{D5}$, $D7$ branes and $\overline{D7}$ branes (plus fluxes) on the type IIB side and the delocalized mirror yields a non-K\"{a}hler warped resolved conifold with a black hole and $D6, \overline{D6}$ branes (plus fluxes) on the type IIA side.   The latter (warped resolved conifolds) are more easier to deal with computationally than  the former(resolved warped deformed conifolds).

\end{itemize}

(b) (related to (a) above) There is an intimate connection between Strominger-Yau-Zaslow mirror of resolved warped deformed conifolds and  thermal QCD at strong coupling and finite number of colors; hence, {\it delocalized Strominger-Yau-Zaslow mirror construction is an entirely new technique  used  for hadron spectroscopy}.

(c) The third is that a BH gravitational dual and a thermal gravity dual yield nearly isospectral spectra for the light vector mesons; the same is partially true for the lightest scalar mesons too.  Explicit computations reveal  that a BH type IIA gravity dual obtained by delocalized SYZ  mirror transform of the type IIB holographic dual of \cite{metrics} is not only able to provide a good match with PDG values for the lightest vector and scalar mesons, it is also able to thereby obtain an explicit temperature-dependence of the (pseudo-)vector masses as a bonus and realize the $\log r_h$-dependence in the same appears at the sub-dominant ${\cal O}\left(\frac{1}{(\log N)^2}\right)$ with the desired feature of a small large-$N$ suppressed  decrease with increase of temperature.

\end{enumerate}

\item
{\bf Future Directions: Glueball Decays into Mesons}

It will be interesting to look at the various glueball-to-meson decay modes. To that end, Performing a Kaluza-Klein reduction similar to \cite{Hashimoto-glueball}: $A_Z = \phi(Z)\pi(x^\mu), A_\mu = \psi (Z) \rho_\mu(x^\nu)$, and similar to \cite{Constable_Myers}, we can look at the following M-theory metric perturbations $h_{MN}(M,N=0,...,10;\mu=t,a, a=1,2,3)$:
\begin{eqnarray}
\label{M-theory-metric-perturbations}
& & h_{tt}(r,x^\mu) = q_1(r) G(x^\mu) G^{\cal M}_{tt}\nonumber\\
& & h_{rr}(r,x^\mu) = q_2(r)G(x^\mu) G^{\cal M}_{rr}\nonumber\\
& & h_{ra}(r,x^\mu) = g_3(r)\partial_a G(x^\mu) G^{\cal M}_{aa}\nonumber\\
& & h_{ab}(r,x^\mu) = G^{\cal M}_{ab}\left(q_4(r) + q_5 \frac{\partial_a\partial_b}{m^2}\right)G(x^\mu)\ {\rm no\ summation}\nonumber\\
& & h_{10\ 10}(r,x^\mu) = q_6(r) G(x^\mu)G^{\cal M}_{10\ 10}.
\end{eqnarray}
Using Witten's prescription of going from type IIA to M-theory
we could work back and using the aforementioned M-theory metric perturbations, work out the type IIA metric perturbations which hence would yield (in the following $\tilde{G}^{\rm IIA}_{\alpha\beta} = G^{\rm IIA}_{\alpha\beta} + h_{\alpha\beta}; \alpha,\beta=0,1...,9$ and $h_{\alpha\beta}$ being type IIA metric perturbations):
\begin{eqnarray*}
& & e^{\frac{4\phi^{\rm IIA}}{3}} = G^{\cal M}_{10\ 10} + h_{10\ 10},\nonumber\\
& & \frac{\tilde{G}^{\rm IIA}_{rr,tt}}{\sqrt{G^{\cal M}_{10\ 10} + h_{10\ 10}}} = G^{\cal M}_{rr,tt} + h_{rr,tt},\nonumber\\
& &  \frac{\tilde{G}^{\rm IIA}_{ra}}{\sqrt{G^{\cal M}_{10\ 10} + h_{10\ 10}}} = h_{ra},\nonumber\\
& & \frac{\tilde{G}^{\rm IIA}_{ab}}{\sqrt{G^{\cal M}_{10\ 10} + h_{10\ 10}}} = G^{\cal M}_{ab} + h_{ab}.
\end{eqnarray*}

Solving  (with a slight abuse of notation) the first order perturbation of the M-theory Einstein's EOM (assuming the flux term providing a cosmological constant):
$R_{MN}^{(1)} \sim \frac{G_4\wedge*G_4}{\sqrt{G}}h_{MN}$,
 for $q_{1,...6}$, one can obtain the glueball-meson interaction Lagrangian density (metric perturbation corresponding to glueballs and gauge field fluctuations corresponding to mesons), using which one can work out glueball decays into mesons.

\end{itemize}

\section*{Acknowledgements}

VY is supported by a Junior Research Fellowship (JRF) from the University Grants Commission, Govt. of India. KS is supported by a Senior Research Fellowship (SRF) from the Ministry of Human Re-source and Development (MHRD), Govt. of India. AM was partly supported by IIT Roorkee. He would also like to thank the theory group at McGill University (K.Dasgupta in particular), for the warm hospitality during the period of his stay wherein a part of this work was done. AM would like to thank K. Dasgupta for several useful discussions and M. Dhuria for suggesting this project several years ago.

\appendix
\section{Triple-T Duality Rules}
\setcounter{equation}{0} \seceqaa

In this section, we summarize the Buscher triple T-duality rules for T-dualizing first along $x$, then along $y$ followed by along $z$. The starting metric in the type IIB theory has the following components
\begin{eqnarray}
ds^2_{\rm IIB} & = &
g^{\rm IIB}_{\mu \nu}dx^\mu ~dx^\nu + g^{\rm IIB}_{x\mu} dx ~dx^\mu +  g^{\rm IIB}_{y \mu} dy~
dx^\mu +  g^{\rm IIB}_{z\mu} dz ~ dx^\mu +  g^{\rm IIB}_{xy} dx ~dy
  + g^{\rm IIB}_{xz} dx ~dz +  g^{\rm IIB}_{zy} dz ~ dy  \nonumber\\
 &&  +  g^{\rm IIB}_{xx} dx^2 +  g^{\rm IIB}_{yy}dy^2
 +  g^{\rm IIB}_{zz}~dz^2,
  \end{eqnarray}
where $\mu, \nu \neq x, y, z$.
As shown in \cite{SYZ 3 Ts}, the form of the metric of the mirror manifold after performing three T-dualities, first along $x$, then along $y$ and finally along $z$:
\begin{eqnarray}
\label{mirror_metric}
& & ds^2 =
\left( G_{\mu\nu} - {G_{z\mu}G_{z\nu} - {\cal B}_{z\mu} {\cal
B}_{z\nu} \over G_{zz}} \right) dx^\mu~dx^\nu +2 \left( G_{x\nu} -
{G_{zx}G_{z\nu} - {\cal B}_{zx} {\cal B}_{z\nu}
 \over G_{zz}} \right) dx~dx^\mu  \nonumber\\
& &  + 2\left( G_{y\nu} - {G_{zy}G_{z\nu} - {\cal B}_{zy} {\cal B}_{z\nu}
 \over G_{zz}}\right) dy~dx^\nu +
2\left( G_{xy} - {G_{zx}G_{zy} - {\cal B}_{zx} {\cal B}_{zy} \over
G_{zz}}\right) dx~dy  \nonumber\\
& &  + {dz^2\over G_{zz}} + 2{{\cal
B}_{\mu z} \over G_{zz}} dx^\mu~dz + 2{{\cal B}_{xz} \over G_{zz}}
dx~dz + 2{{\cal B}_{yz} \over G_{zz}} dy~dz \nonumber\\
& & +  \left( G_{xx}
- {G^2_{zx} - {\cal B}^2_{zx} \over G_{zz}} \right) dx^2 + \left(
G_{yy} - {G^2_{zy} - {\cal B}^2_{zy} \over G_{zz}} \right)
dy^2.
\end{eqnarray}
 The various components of the metric after three successive T-dualities along $x, y$ and $z$ respectively, can be written as \cite{SYZ 3 Ts}:
 {\small
\begin{eqnarray}
\label{G_munu}
& & G_{\mu\nu} = {g^{\rm IIB}_{\mu\nu}g^{\rm IIB}_{xx} -
g^{\rm IIB}_{x\mu}g^{\rm IIB}_{x\nu} + b^{\rm IIB}_{x\mu}b^{\rm IIB}_{x\nu} \over g^{\rm IIB}_{xx}} -
{(g^{\rm IIB}_{y\mu}g^{\rm IIB}_{xx} - g^{\rm IIB}_{xy} g^{\rm IIB}_{x \mu} + b^{\rm IIB}_{xy} b^{\rm IIB}_{x\mu})
(g^{\rm IIB}_{y\nu}g^{\rm IIB}_{xx}
 - g^{\rm IIB}_{xy} g^{\rm IIB}_{x \nu} + b^{\rm IIB}_{xy} b^{\rm IIB}_{x\nu}) \over
g^{\rm IIB}_{xx}(g^{\rm IIB}_{yy}g^{\rm IIB}_{xx}- g^{\rm IIB}_{xy}\ ^2 + b^{\rm IIB}_{xy}\ ^2)} \nonumber\\
&& + {(b^{\rm IIB}_{y\mu}g^{\rm IIB}_{xx} - g^{\rm IIB}_{xy} b^{\rm IIB}_{x \mu} + b^{\rm IIB}_{xy}
g^{\rm IIB}_{x\mu})(b^{\rm IIB}_{y\nu}g^{\rm IIB}_{xx}
 - g^{\rm IIB}_{xy} b^{\rm IIB}_{x \nu} + b^{\rm IIB}_{xy} g^{\rm IIB}_{x\nu})\over
g^{\rm IIB}_{xx}(g^{\rm IIB}_{yy}g^{\rm IIB}_{xx}- g^{\rm IIB}_{xy}\ ^2 + b^{\rm IIB}_{xy}\ ^2)},
\end{eqnarray}

\begin{eqnarray}
\label{Gmuz}
& & G_{\mu z} = {g^{\rm IIB}_{\mu z}g^{\rm IIB}_{xx} -
g^{\rm IIB}_{x\mu}g^{\rm IIB}_{xz} + b^{\rm IIB}_{x \mu}b^{\rm IIB}_{xz} \over g^{\rm IIB}_{xx}} - {(g^{\rm IIB}_{y\mu}g^{\rm IIB}_{xx}
- g^{\rm IIB}_{xy} g^{\rm IIB}_{x \mu} + b^{\rm IIB}_{xy} b^{\rm IIB}_{x\mu}) (g^{\rm IIB}_{yz}g^{\rm IIB}_{xx} - g^{\rm IIB}_{xy} g^{\rm IIB}_{x
z} + b^{\rm IIB}_{xy} b^{\rm IIB}_{xz}) \over g^{\rm IIB}_{xx}(g^{\rm IIB}_{yy}g^{\rm IIB}_{xx}- g^{\rm IIB}_{xy}\ ^2 +
b^{\rm IIB}_{xy}\ ^2)} \nonumber\\
& &  + {(b^{\rm IIB}_{y\mu}g^{\rm IIB}_{xx} - g^{\rm IIB}_{xy} b^{\rm IIB}_{x \mu} +
b^{\rm IIB}_{xy} g^{\rm IIB}_{x\mu})(b^{\rm IIB}_{yz}g^{\rm IIB}_{xx} - g^{\rm IIB}_{xy} b^{\rm IIB}_{x z} + b^{\rm IIB}_{xy}
g^{\rm IIB}_{xz})\over g^{\rm IIB}_{xx}(g^{\rm IIB}_{yy}g^{\rm IIB}_{xx}- g^{\rm IIB}_{xy}\ ^2 + b^{\rm IIB}_{xy}\ ^2)},
\end{eqnarray}

\begin{eqnarray}
\label{Gzz}
& & G_{zz} =  {g^{\rm IIB}_{zz}g^{\rm IIB}_{xx} - j^2_{xz} +
b^2_{xz}\over g^{\rm IIB}_{xx}} - {(g^{\rm IIB}_{yz}g^{\rm IIB}_{xx} - g^{\rm IIB}_{xy} g^{\rm IIB}_{xz} + b^{\rm IIB}_{xy}
b^{\rm IIB}_{xz})^2 \over g^{\rm IIB}_{xx}(g^{\rm IIB}_{yy}g^{\rm IIB}_{xx}- g^{\rm IIB}_{xy}\ ^2 + b^{\rm IIB}_{xy}\ ^2)} + {(b^{\rm IIB}_{yz}g^{\rm IIB}_{xx} - g^{\rm IIB}_{xy} b^{\rm IIB}_{x z} + b^{\rm IIB}_{xy} g^{\rm IIB}_{xz})^2 \over
g^{\rm IIB}_{xx}(g^{\rm IIB}_{yy}g^{\rm IIB}_{xx}- g^{\rm IIB}_{xy}\ ^2 + b^{\rm IIB}_{xy}\ ^2)},
\end{eqnarray}

\begin{eqnarray}
\label{Gymu}
& & G_{y \mu} = -{b^{\rm IIB}_{y \mu} g^{\rm IIB}_{xx} - b^{\rm IIB}_{x \mu} g^{\rm IIB}_{xy} + b^{\rm IIB}_{xy}
 g^{\rm IIB}_{\mu x} \over g^{\rm IIB}_{yy}g^{\rm IIB}_{xx}- g^{\rm IIB}_{xy}\ ^2 + b^{\rm IIB}_{xy}\ ^2},
~ G_{y z} = -{b^{\rm IIB}_{y z} g^{\rm IIB}_{xx} - b^{\rm IIB}_{x z} g^{\rm IIB}_{xy} + b^{\rm IIB}_{xy} g^{\rm IIB}_{z x}
\over g^{\rm IIB}_{yy}g^{\rm IIB}_{xx}- g^{\rm IIB}_{xy}\ ^2 + b^{\rm IIB}_{xy}\ ^2},
\end{eqnarray}

\begin{eqnarray}
\label{Gyy}
& & G_{yy} = {g^{\rm IIB}_{xx} \over g^{\rm IIB}_{yy}g^{\rm IIB}_{xx}- g^{\rm IIB}_{xy}\ ^2 +
b^{\rm IIB}_{xy}\ ^2},~ G_{xx} = {g^{\rm IIB}_{yy} \over g^{\rm IIB}_{yy}g^{\rm IIB}_{xx}- g^{\rm IIB}_{xy}\ ^2 +
b^{\rm IIB}_{xy}\ ^2}, ~G_{xy} = {-g^{\rm IIB}_{xy} \over g^{\rm IIB}_{yy}g^{\rm IIB}_{xx}- g^{\rm IIB}_{xy}\ ^2 +
b^{\rm IIB}_{xy}\ ^2},
\end{eqnarray}

\begin{eqnarray}
\label{Gmux}
& & G_{\mu x} = {b^{\rm IIB}_{\mu x} \over g^{\rm IIB}_{xx}} + {(g^{\rm IIB}_{\mu y} g^{\rm IIB}_{xx} -
 g^{\rm IIB}_{xy} g^{\rm IIB}_{x \mu} + b^{\rm IIB}_{xy} b^{\rm IIB}_{x \mu}) b^{\rm IIB}_{xy} \over
g^{\rm IIB}_{xx}(g^{\rm IIB}_{yy}g^{\rm IIB}_{xx}- g^{\rm IIB}_{xy}\ ^2 + b^{\rm IIB}_{xy}\ ^2)}
+ {(b^{\rm IIB}_{y \mu} g^{\rm IIB}_{xx} - g^{\rm IIB}_{xy} b^{\rm IIB}_{x \mu} + b^{\rm IIB}_{xy} g^{\rm IIB}_{x \mu}) g^{\rm IIB}_{xy}
 \over g^{\rm IIB}_{xx}(g^{\rm IIB}_{yy}g^{\rm IIB}_{xx}- g^{\rm IIB}_{xy}\ ^2 + b^{\rm IIB}_{xy}\ ^2)},
 \end{eqnarray}

\begin{eqnarray}
\label{Gzx}
& & G_{z x} = {b^{\rm IIB}_{z x} \over g^{\rm IIB}_{xx}} + {(g^{\rm IIB}_{z y} g^{\rm IIB}_{xx} -
g^{\rm IIB}_{xy} g^{\rm IIB}_{x z} + b^{\rm IIB}_{xy} b^{\rm IIB}_{x z}) b^{\rm IIB}_{xy} \over
g^{\rm IIB}_{xx}(g^{\rm IIB}_{yy}g^{\rm IIB}_{xx}- g^{\rm IIB}_{xy}\ ^2 + b^{\rm IIB}_{xy}\ ^2)}
 + {(b^{\rm IIB}_{y z} g^{\rm IIB}_{xx} - g^{\rm IIB}_{xy} b^{\rm IIB}_{x z} + b^{\rm IIB}_{xy} g^{\rm IIB}_{xz}) g^{\rm IIB}_{xy}
  \over g^{\rm IIB}_{xx}(g^{\rm IIB}_{yy}g^{\rm IIB}_{xx}- g^{\rm IIB}_{xy}\ ^2 + b^{\rm IIB}_{xy}\ ^2)}.
\end{eqnarray}}
In the above formulae we have denoted the type IIB
$B$ fields as $b^{\rm IIB}_{mn}$.  For the generic case we will switch on all the
components of the $B$ field:
{\small
\begin{eqnarray}
B^{\rm IIB} & = &
b^{\rm IIB}_{\mu\nu} ~ dx^\mu \wedge dx^\nu + b^{\rm IIB}_{x \mu} dx \wedge dx^\mu +  b^{\rm IIB}_{y \mu}
~ dy~\wedge dx^\mu + b^{\rm IIB}_{z \mu} ~ dz \wedge dx^\mu \nonumber\\
& & + ~ b^{\rm IIB}_{xy}
~ dx \wedge dy +
 b^{\rm IIB}_{xz} ~ dx  \wedge dz +  b^{\rm IIB}_{zy}~ dz  \wedge dy.
 \end{eqnarray}}
\noindent After applying again the T-dualities, the type IIA NS-NS $B$ field in the mirror set-up will take the form:
{\small
\begin{eqnarray}
\label{B}
B^{IIA} & = & \left( {\cal B}_{\mu\nu}
+ {2 {\cal B}_{z[\mu} G_{\nu]z} \over G_{zz}} \right) dx^\mu
\wedge dx^\nu + \left( {\cal B}_{\mu x} + {2 {\cal B}_{z[\mu}
G_{x]z} \over G_{zz}}\right)
 dx^\mu \wedge dx  \nonumber\\
& & \left( {\cal B}_{\mu y} + {2 {\cal B}_{z[\mu} G_{y]z} \over G_{zz}}
 \right) dx^\mu \wedge dy
+ \left( {\cal B}_{xy} + {2 {\cal B}_{z[x} G_{y]z} \over G_{zz}}
\right) dx \wedge dy \nonumber\\
& &  + {G_{z \mu} \over G_{zz}} dx^\mu
\wedge dz + {G_{z x} \over G_{zz}} dx \wedge dz + {G_{z y} \over
G_{zz}} dy \wedge dz.
\end{eqnarray}}
 Here the $G_{mn}$ components have been
given above, and the various ${\cal B}$  components can  be
written as:
{\small
\begin{eqnarray}
\label{Bmunu}
{\cal B}_{\mu\nu} & = &
{b^{\rm IIB}_{\mu\nu} g^{\rm IIB}_{xx} + b^{\rm IIB}_{x \mu} g^{\rm IIB}_{\nu x} - b^{\rm IIB}_{x \nu} g^{\rm IIB}_{\mu x}
\over g^{\rm IIB}_{xx}} \nonumber\\
& & +  {2 (g^{\rm IIB}_{y[\mu}g^{\rm IIB}_{xx} - g^{\rm IIB}_{xy}g^{\rm IIB}_{x[\mu} +
b^{\rm IIB}_{xy} b^{\rm IIB}_{x[\mu}) (b^{\rm IIB}_{\nu]y}g^{\rm IIB}_{xx} - b^{\rm IIB}_{\nu]x}g^{\rm IIB}_{xy} - b^{\rm IIB}_{xy}
g^{\rm IIB}_{\nu]x}) \over g^{\rm IIB}_{xx}(g^{\rm IIB}_{yy}g^{\rm IIB}_{xx}- g^{\rm IIB}_{xy}\ ^2 + b^{\rm IIB}_{xy}\ ^2)},
\end{eqnarray}

\begin{eqnarray}
\label{Bmuz}
{\cal B}_{\mu z} & = &  {b^{\rm IIB}_{\mu z} g^{\rm IIB}_{xx} +
b^{\rm IIB}_{x \mu} g^{\rm IIB}_{z x} - b^{\rm IIB}_{x z} g^{\rm IIB}_{\mu x} \over g^{\rm IIB}_{xx}} \nonumber\\
& & +  {2
(g^{\rm IIB}_{y[\mu}g^{\rm IIB}_{xx} - g^{\rm IIB}_{xy}g^{\rm IIB}_{x[\mu} + b^{\rm IIB}_{xy} b^{\rm IIB}_{x[\mu})
(b^{\rm IIB}_{z]y}g^{\rm IIB}_{xx} - b^{\rm IIB}_{z]x}g^{\rm IIB}_{xy} - b^{\rm IIB}_{xy} g^{\rm IIB}_{z]x}) \over
g^{\rm IIB}_{xx}(g^{\rm IIB}_{yy}g^{\rm IIB}_{xx}- g^{\rm IIB}_{xy}\ ^2 + b^{\rm IIB}_{xy}\ ^2)},
\end{eqnarray}

\begin{eqnarray}
\label{Bmuy}
{\cal B}_{\mu y} & = & {g^{\rm IIB}_{\mu y} g^{\rm IIB}_{xx} - g^{\rm IIB}_{xy} g^{\rm IIB}_{x \mu} + b^{\rm IIB}_{xy} b^{\rm IIB}_{x \mu}
 \over g^{\rm IIB}_{yy}g^{\rm IIB}_{xx}- g^{\rm IIB}_{xy}\ ^2 + b^{\rm IIB}_{xy}\ ^2},\nonumber\\
{\cal B}_{z y} & = & {g^{\rm IIB}_{z y} g^{\rm IIB}_{xx} - g^{\rm IIB}_{xy} g^{\rm IIB}_{x z} + b^{\rm IIB}_{xy} b^{\rm IIB}_{x z} \over g^{\rm IIB}_{yy}g^{\rm IIB}_{xx}-
 g^{\rm IIB}_{xy}\ ^2 + b^{\rm IIB}_{xy}\ ^2},
 \end{eqnarray}

\begin{eqnarray}
\label{Bmux}
{\cal B}_{\mu x} & = & {g^{\rm IIB}_{\mu x} \over g^{\rm IIB}_{xx}} - {g^{\rm IIB}_{xy} (g^{\rm IIB}_{\mu y} g^{\rm IIB}_{xx} -
g^{\rm IIB}_{xy} g^{\rm IIB}_{x \mu} + b^{\rm IIB}_{xy} b^{\rm IIB}_{x \mu}) \over
g^{\rm IIB}_{xx}(g^{\rm IIB}_{yy}g^{\rm IIB}_{xx}- g^{\rm IIB}_{xy}\ ^2 + b^{\rm IIB}_{xy}\ ^2)} + {b^{\rm IIB}_{xy} (b^{\rm IIB}_{x\mu}g^{\rm IIB}_{xy} - b^{\rm IIB}_{y\mu}g^{\rm IIB}_{xx} -
b^{\rm IIB}_{xy}g^{\rm IIB}_{xz})
 \over g^{\rm IIB}_{xx}(g^{\rm IIB}_{yy}g^{\rm IIB}_{xx}- g^{\rm IIB}_{xy}\ ^2 + b^{\rm IIB}_{xy}\ ^2)},
 \end{eqnarray}

\begin{eqnarray}
\label{Bzx}
{\cal B}_{z x} & = & {g^{\rm IIB}_{z x} \over g^{\rm IIB}_{xx}} - {g^{\rm IIB}_{xy} (g^{\rm IIB}_{z y} g^{\rm IIB}_{xx} -
g^{\rm IIB}_{xy} g^{\rm IIB}_{xz} + b^{\rm IIB}_{xy} b^{\rm IIB}_{x z}) \over
g^{\rm IIB}_{xx}(g^{\rm IIB}_{yy}g^{\rm IIB}_{xx}- g^{\rm IIB}_{xy}\ ^2 + b^{\rm IIB}_{xy}\ ^2)} + {b^{\rm IIB}_{xy} (b^{\rm IIB}_{xz}g^{\rm IIB}_{xy} - b^{\rm IIB}_{yz}g^{\rm IIB}_{xx} -
 b^{\rm IIB}_{xy}g^{\rm IIB}_{xz})
\over g^{\rm IIB}_{xx}(g^{\rm IIB}_{yy}g^{\rm IIB}_{xx}- g^{\rm IIB}_{xy}\ ^2 + b^{\rm IIB}_{xy}\ ^2)},
\end{eqnarray}

\begin{eqnarray}
\label{Bxy}
{\cal B}_{xy} & = & {-b^{\rm IIB}_{xy} \over g^{\rm IIB}_{yy}g^{\rm IIB}_{xx}- g^{\rm IIB}_{xy}\ ^2 + b^{\rm IIB}_{xy}\ ^2}.
\end{eqnarray}}

\section{Vector Meson Embedding Functions appearing in the DBI action for $D6$-branes}
\setcounter{equation}{0}\seceqbb

In this appendix we give the embedding functions $\Sigma_{0,1}(r;g_s,N_f,M,N)$ relevant to the embedding of $D6$-branes in the delocalized SYZ type IIA mirror of the type IIB construct of \cite{metrics} that appear in (\ref{DBI-det_i}) in Section {\bf 3}.
The same are given as under:
{\footnotesize
\begin{eqnarray}
\label{DBI-det_ii}
& & \hskip -0.2in \Sigma_0(r;g_s,N_f,M,N) \equiv -\frac{1}{97844723712 \pi
   ^{11} \alpha_{\theta_1}^8 \alpha_{\theta_2}^4 {g_s} N^{26/5} \left(9 a^2+r^2\right)}\Biggl\{r^6 \left(6 a^2+r^2\right)\nonumber\\
   & & \times \Biggl[81 \sqrt{2} 3^{5/6} \alpha_{\theta_1}^9 \sqrt[5]{N}-54 \sqrt[3]{3} \alpha_{\theta_1}^8 N^{2/5}+81 \sqrt{2} 3^{5/6} \alpha_{\theta_1}^7
   \alpha_{\theta_2}^2\nonumber\\
   & & -54 \left(2+\sqrt[3]{3}\right) \alpha_{\theta_1}^6 \alpha_{\theta_2}^2 \sqrt[5]{N}+24 \sqrt{6} \alpha_{\theta_1}^5 \alpha_{\theta_2}^2 N^{2/5}-8 \alpha_{\theta_1}^4 \alpha_{\theta_2}^2
   N^{3/5}\nonumber\\
   & & -24 \sqrt{6} \alpha_{\theta_1}^3 \alpha_{\theta_2}^4 \sqrt[5]{N}+16 \alpha_{\theta_1}^2 \alpha_{\theta_2}^4 N^{2/5}-8 \alpha_{\theta_2}^6 \sqrt[5]{N}\Biggr]\nonumber\\
   & & \times \left(3 {g_s} M^2
   \log (r) (-2 {g_s} {N_f} \log (\alpha_{\theta_1}  \alpha_{\theta_2})+{g_s} {N_f} {\log N}-6 {g_s} {N_f}+{g_s} {N_f} \log (16)-8 \pi )-36
   {g_s}^2 M^2 {N_f} \log ^2(r)+32 \pi ^2 N\right)^4\nonumber\\
   & & \times \left(3 {g_s} M^2 \log (r) (2 {g_s} {N_f} \log (\alpha  \alpha_{\theta_2})-{g_s}
   {N_f} {\log N}+6 {g_s} {N_f}-2 {g_s} {N_f} \log (4)+8 \pi )+36 {g_s}^2 M^2 {N_f} \log ^2(r)+32 \pi ^2 N\right)\Biggr\}\nonumber\\
& &\hskip -0.2in \Sigma_1(r;g_s,N_f,M,N) \equiv \frac{1}{165112971264 \pi ^{19/2} \alpha_{\theta_1}^{10} \alpha_{\theta_2}^6 {g_s}^{3/2} N^{49/10}}\Biggl\{r^4 \left(r^4-{r_h}^4\right)
\nonumber\\
& & \times \Bigg[-486 \sqrt{6} \alpha_{\theta_1}^{11} N+324 \alpha_{\theta_1}^{10} N^{6/5}+972 \sqrt{2} \sqrt[6]{3} \alpha_{\theta_1}^9 \alpha_{\theta_2}^2
   \sqrt[5]{N} \left(\left(-9-3 \sqrt[3]{3}+6 3^{2/3}\right) \alpha_{\theta_2}^2-\sqrt[3]{3} N^{3/5}\right)
 \nonumber\\
& &    +108 \alpha_{\theta_1}^8 \Biggl(27 \sqrt{2} \sqrt[6]{3}
   \left(3^{2/3}-3\right) \alpha_{\theta_2}^5 \sqrt[10]{N}+18 \left(\sqrt[3]{3}-1\right)^2 \alpha_{\theta_2}^4 N^{2/5}\nonumber\\
   & & +2 \left(3+3^{2/3}\right) \alpha_{\theta_2}^2
   N\Biggr)-9 \alpha_{\theta_1}^7 \Biggl(243 \sqrt{2} \sqrt[6]{3} \alpha_{\theta_2}^6-216 \sqrt[3]{3} \left(\sqrt[3]{3}-1\right) \alpha_{\theta_2}^5 N^{3/10}+54 \sqrt{6}
   \alpha_{\theta_2}^4 N^{3/5}\nonumber\\
   & & +16 \sqrt{2} \sqrt[6]{3} \alpha_{\theta_2}^2 N^{6/5}\Biggr)+2 \alpha_{\theta_1}^6 \Biggl(243 3^{2/3} \alpha_{\theta_2}^6 \sqrt[5]{N}+54 \left(3+2
   3^{2/3}\right) \alpha_{\theta_2}^4 N^{4/5}\nonumber\\
   & & +8 3^{2/3} \alpha_{\theta_2}^2 N^{7/5}\Biggr)-16 3^{2/3} \alpha_{\theta_1}^4 \alpha_{\theta_2}^4 N^{6/5}+144 \sqrt{2} \sqrt[6]{3} \alpha
   ^3 \alpha_{\theta_2}^6 N^{4/5} -16 3^{2/3} \alpha_{\theta_1}^2 \alpha_{\theta_2}^6 N+16 3^{2/3} \alpha_{\theta_2}^8 N^{4/5}\Biggr]\nonumber\\
   & & \times \left(3 {g_s} M^2 \log (r) (-2 {g_s}
   {N_f} \log (\alpha  \alpha_{\theta_2})+{g_s} {N_f} {\log N}-6 {g_s} {N_f}+{g_s} {N_f} \log (16)-8 \pi )-36 {g_s}^2 M^2 {N_f}
   \log ^2(r)+32 \pi ^2 N\right)^4\Biggr\}.\nonumber\\
   & &
\end{eqnarray}}

\section{Functions Appearing in the Vector and Scalar Mesons' Actions }
\setcounter{equation}{0}\seceqcc

\subsection{Vector Meson Action Functions}

The functions ${\cal V}_{1,2}$ appearing in equation (\ref{Ftildesq-i}) in Section {\bf 4} in the context of the vector meson action obtained by substituting a KK ansatz (\ref{AZ+Amu}) into the DBI action for $N_f=2$ $D6$-branes of (\ref{DBI-ii}) are given as under:
\begin{eqnarray*}
\label{V1}
& &  {\cal V}_1(Z) = e^{-\Phi^{IIA}}\sqrt{h}G^{ZZ}\sqrt{{\rm det}_{\theta_2,\tilde{y}}\left(i^*(G+B)\right)}\sqrt{{\rm det}_{\mathbb{R}^{1,3},Z}(i^*G)},\nonumber\\
& &  = \frac{1}{36 \sqrt{2} \pi ^{3/2} \alpha_{\theta_1}^2 \alpha_{\theta_2} {g_s}^{3/2}}\Biggl\{\sqrt[5]{N} e^{-4 |Z|} \left(e^{4 |Z|}-1\right)\nonumber\\
& &  \times\Biggl(3 a^2 {g_s} {N_f} \log N-9 a^2 {g_s} {N_f} \log
   ({r_h})-9 a^2 {g_s} {N_f} |Z|-9 a^2 {g_s} {N_f}+12 \pi  a^2+2 {g_s} {N_f} {r_h}^2 e^{2 |Z|} \log N
   \nonumber\\
& &    -6 {g_s} {N_f} {r_h}^2 e^{2 |Z|} |Z|-6 {g_s} {N_f} {r_h}^2 e^{2 |Z|} \log ({r_h})+8 \pi
   {r_h}^2 e^{2 |Z|}\Biggr)\Biggr\}\nonumber\\
   & & +\frac{1}{576 \pi ^{3/2} \alpha_{\theta_1}^4 \alpha_{\theta_2}^3
   {g_s}^{3/2}}\Biggl\{e^{-4 |Z|} \left(e^{4 |Z|}-1\right) \left(27
   \sqrt{2} \sqrt[3]{3} \alpha_{\theta_1}^6-24 \sqrt{3} \alpha_{\theta_1}^3 \alpha_{\theta_2}^2-8 \sqrt{2} \alpha_{\theta_2}^4\right)\nonumber\\
  & & \times  \Biggl(-3 a^2 {g_s} {N_f} \log
   \left(\frac{1}{N}\right)-9 a^2 {g_s} {N_f} \log ({r_h})-9 a^2 {g_s} {N_f} |Z|-9 a^2 {g_s} {N_f}+12 \pi  a^2-2
   {g_s} {N_f} {r_h}^2 e^{2 |Z|} \log \left(\frac{1}{N}\right)\nonumber\\
   & & -6 {g_s} {N_f} {r_h}^2 e^{2 |Z|} |Z|-6 {g_s} {N_f}
   {r_h}^2 e^{2 |Z|} \log ({r_h})+8 \pi  {r_h}^2 e^{2 |Z|}\Biggr)\Biggr\} + {\cal O}\left(\frac{1}{N^{6/5}}\right),
   \end{eqnarray*}
   \begin{eqnarray}
   \label{v1v2}
& & {\cal V}_2(Z) =  e^{-\Phi^{IIA}}\frac{h}{2}\sqrt{{\rm det}_{\theta_2,\tilde{y}}\left(i^*(G+B)\right)}\sqrt{{\rm det}_{\mathbb{R}^{1,3},|Z|}(i^*G)}\nonumber\\
& & =\frac{1}{18 \sqrt{2 \pi
   } \alpha_{\theta_1}^2 \alpha_{\theta_2} \sqrt{{g_s}} {r_h}^2}\Biggl\{N^{6/5} e^{-2 |Z|} \Biggl(3 a^2 {g_s} {N_f} \log \left(\frac{1}{N}\right)+9 a^2 {g_s} {N_f} \log ({r_h})+9 a^2 {g_s}
   {N_f} |Z|\nonumber\\
   & & -9 a^2 {g_s} {N_f}-12 \pi  a^2+2 {g_s} {N_f} {r_h}^2 e^{2 |Z|} \log N-6 {g_s}
   {N_f} {r_h}^2 e^{2 |Z|} |Z|-6 {g_s} {N_f} {r_h}^2 e^{2 |Z|} \log ({r_h})+8 \pi  {r_h}^2 e^{2 |Z|}\Biggr)\Biggr\}\nonumber\\
   & & +\frac{1}{288 \sqrt{\pi } \alpha_{\theta_1}^4 \alpha_{\theta_2}^3 \sqrt{{g_s}}
   {r_h}^2}\Biggl\{N e^{-2 |Z|} \Biggl(27 \sqrt{2} \sqrt[3]{3} \alpha_{\theta_1}^6-24 \sqrt{3} \alpha_{\theta_1}^3
   \alpha_{\theta_2}^2-8 \sqrt{2} \alpha_{\theta_2}^4\Biggr)\nonumber\\
   & & \times \Biggl(-3 a^2 {g_s} {N_f} \log N+9 a^2 {g_s} {N_f}
   \log ({r_h})+9 a^2 {g_s} {N_f} |Z|-9 a^2 {g_s} {N_f}-12 \pi  a^2+2 {g_s} {N_f} {r_h}^2 e^{2 |Z|} \log
  N \nonumber\\
  & & - 6 {g_s} {N_f} {r_h}^2 e^{2 |Z|} |Z| - 6 {g_s} {N_f} {r_h}^2 e^{2 |Z|} \log ({r_h})+8 \pi
   {r_h}^2 e^{2 |Z|}\Biggr)\Biggr\}  + {\cal O}\left(\frac{1}{N^{6/5}}\right).
\end{eqnarray}

\subsection{Scalar Meson Action Functions}

The scalar meson functions ${\cal S}_{1,2,3}$ appearing  in the DBI action (\ref{DBI-scalar}) for $N_f$ $D6$-branes and (\ref{mass-term-identification+EOM}) are given as under:
{\footnotesize
\begin{eqnarray}
\label{S123}
& & \hskip -0.5in {\cal S}_1 = \frac{1}{1296 \sqrt{2} \sqrt[3]{3}
   \pi  \alpha_{\theta_1}^24 \alpha_{\theta_2}^5 {g_s} Z^2}\Biggl\{{\cal C}^2 N^{11/10} {r_h}^2
    \left(27 \sqrt[3]{3} \alpha_{\theta_1}^26-12 \sqrt{6} \alpha_{\theta_1}^23 \alpha_{\theta_2}^2+8 \alpha_{\theta_1}^22 \alpha_{\theta_2}^2 \sqrt[5]{N}-8
   \alpha_{\theta_2}^4\right)\nonumber\\
   & &\hskip -0.5in \times \Bigg[-{g_s} {N_f} \log (N) \left(3 \left(\frac{4 {g_s} M^2 (\log ({r_h})+1)}{N}+0.6\right)^2-2 e^{2
   |Z|}\right)+3 {g_s} {N_f} (\log ({r_h})+|Z|) \left(3 \left(\frac{4 {g_s} M^2 (\log ({r_h})+1)}{N}+0.6\right)^2-2 e^{2 |Z|}\right)\nonumber\\
   & & \hskip -0.5in-3
   (3 {g_s} {N_f}+4 \pi ) \left(\frac{4 {g_s} M^2 (\log ({r_h})+1)}{N}+0.6\right)^2+8 \pi  e^{2 |Z|}\Biggr]\Biggr\},\nonumber\\
   & & \hskip -0.5in {\cal S}_2 = \frac{1}{15552 \sqrt{2}
   \pi ^2 \alpha_{\theta_1}^22 \alpha_{\theta_2}^5 {g_s}^2 Z^2}\Biggl\{{\cal C}^2 \sqrt[10]{N} {r_h}^4 e^{-2 |Z|} \left(e^{4 |Z|}-1\right) \left(81 \alpha_{\theta_1}^24-36 \sqrt{2} \sqrt[6]{3} \alpha  \alpha_{\theta_2}^2+8 3^{2/3}
   \alpha_{\theta_2}^2 \sqrt[5]{N}\right)\nonumber\\
   & &\hskip -0.5in \times \Biggl[{g_s} {N_f} \log (N) \left(3 \left(\frac{4 {g_s} M^2 (\log ({r_h})+1)}{N}+0.6\right)^2+2
   e^{2 |Z|}\right)-3 {g_s} {N_f} (\log ({r_h})+|Z|) \left(3 \left(\frac{4 {g_s} M^2 (\log ({r_h})+1)}{N}+0.6\right)^2+2 e^{2
   |Z|}\right)\nonumber\\
   & & \hskip -0.5in+(12 \pi -9 {g_s} {N_f}) \left(\frac{4 {g_s} M^2 (\log ({r_h})+1)}{N}+0.6\right)^2+8 \pi  e^{2 |Z|}\Biggr]\Biggr\}\nonumber\\
   & &  \hskip -0.5in {\cal S}_3 = \frac{1}{1944 \sqrt{2} \pi ^2 \alpha_{\theta_1}^22 \alpha_{\theta_2}^3 {g_s}^2
   \left(e^{4 |Z|}-1\right) Z^4}\Biggl\{\sqrt[5]{N} {r_h}^4 e^{-2 |Z|} \Biggl({g_s} {N_f} \log (N)\nonumber\\
   & & \hskip -0.5in\times \Biggl(3 \left(3^{2/3} {\cal C}^2 \sqrt[10]{N} \left(e^{4 |Z|}-1\right)^2-27
   \sqrt{\pi } \alpha_{\theta_2}^2 \sqrt{{g_s}} e^{4 |Z|} Z^2 \left(-2 |Z|+e^{4 |Z|} (6 |Z|+1)-1\right)\right) \nonumber\\
   & &\hskip -0.5in \times\left(\frac{4 {g_s} M^2 (\log
   ({r_h})+1)}{N}+0.6\right)^2+2 e^{2 |Z|} \left(27 \sqrt{\pi } \alpha_{\theta_2}^2 \sqrt{{g_s}} e^{4 |Z|} Z^2 \left(e^{4 |Z|} (4 |Z|+1)-1\right)+3^{2/3}
   {\cal C}^2 \sqrt[10]{N} \left(e^{4 |Z|}-1\right)^2\right)\Biggr)\nonumber\\
   & &\hskip -0.5in -3 {g_s} {N_f} (\log ({r_h})+|Z|) \Biggl[3 \left(3^{2/3} {\cal C}^2 \sqrt[10]{N}
   \left(e^{4 |Z|}-1\right)^2-27 \sqrt{\pi } \alpha_{\theta_2}^2 \sqrt{{g_s}} e^{4 |Z|} Z^2 \left(-2 |Z|+e^{4 |Z|} (6 |Z|+1)-1\right)\right)\nonumber\\
   & & \hskip -0.5in\times \left(\frac{4
   {g_s} M^2 (\log ({r_h})+1)}{N}+0.6\right)^2\nonumber\\
   & & \hskip -0.5in+2 e^{2 |Z|} \left(27 \sqrt{\pi } \alpha_{\theta_2}^2 \sqrt{{g_s}} e^{4 |Z|} Z^2 \left(e^{4 |Z|} (4
   |Z|+1)-1\right)+3^{2/3} {\cal C}^2 \sqrt[10]{N} \left(e^{4 |Z|}-1\right)^2\right)\Biggr]\nonumber\\
   & &\hskip -0.5in+3 \left(\frac{4 {g_s} M^2 (\log ({r_h})+1)}{N}+0.6\right)^2
   \Biggl[(3^{2/3} {\cal C}^2 \sqrt[10]{N} \left(e^{4 |Z|}-1\right)^2 (4 \pi -3 {g_s} {N_f})-27 \sqrt{\pi } \alpha_{\theta_2}^2 \sqrt{{g_s}} e^{4 |Z|} |Z|\nonumber\\
& & \hskip -0.5in\times   \left(3 {g_s} {N_f} |Z| \left(3 |Z|+e^{4 |Z|} (|Z|+1)-1\right)+4 \pi  |Z| \left(-2 |Z|+e^{4 |Z|} (6 |Z|+1)-1\right)\right)\Biggr]+2 \sqrt{\pi } e^{2 |Z|}\nonumber\\
& &\hskip -0.5in \times   \left(27 \alpha_{\theta_2}^2 \sqrt{{g_s}} e^{4 |Z|} Z^2 \left(4 \pi  \left(e^{4 |Z|} (4 |Z|+1)-1\right)-3 {g_s} {N_f} \left(e^{4 |Z|}-1\right)
   |Z|\right)+4 3^{2/3} \sqrt{\pi } {\cal C}^2 \sqrt[10]{N} \left(e^{4 |Z|}-1\right)^2\right)\Biggr)\Biggr\}.
\end{eqnarray}}

\end{document}